\documentclass[fleqn,usenatbib]{mnras}

\usepackage{amsfonts,amssymb}
\usepackage{newtxtext,newtxmath}
\usepackage[T1]{fontenc}
\DeclareRobustCommand{\VAN}[3]{#2}
\let\VANthebibliography\thebibliography
\def\thebibliography{\DeclareRobustCommand{\VAN}[3]{##3}\VANthebibliography}

\usepackage{graphicx}	% Including figure files
\usepackage{amsmath}	% Advanced maths commands
\usepackage{array}
\usepackage{tablefootnote}
\title[Radio Emission Beyond the Deathline]{Beyond the Rotational Deathline: Radio Emission from Ultra-long Period Magnetars}

\author[A. J. Cooper et al.]{
A. J. Cooper,$^{1}$\thanks{E-mail: alexander.cooper@physics.ox.ac.uk}
Z. Wadiasingh,$^{2,3,4}$
\\
$^{1}$Astrophysics, The University of Oxford, Keble Road, Oxford, OX1 3RH, UK\\
$^{2}$Department of Astronomy, University of Maryland, College Park, MD 20742-4111, USA\\
$^{3}$Astrophysics Science Division, NASA Goddard Space Flight Center, 8800 Greenbelt Road, Greenbelt, MD 20771, USA \\
$^{4}$Center for Research and Exploration in Space Science and Technology, NASA/GSFC, Greenbelt, Maryland 20771, USA  \\
}

\date{Accepted XXX. Received YYY; in original form ZZZ}

\pubyear{2024}

\begin{document}
\label{firstpage}
\pagerange{\pageref{firstpage}--\pageref{lastpage}}
\maketitle

\begin{abstract}
Motivated by the recent detection of ultra-long period radio transients, we investigate new models of coherent radio emission via low-altitude electron-positron pair production in neutron stars beyond rotationally-powered curvature radiation deathlines. We find that plastic motion (akin to `continental drift') and qualitatively similar thermoelectric action by temperature gradients in the crusts of slowly rotating, highly magnetized neutron stars could impart mild local magnetospheric twists. Regardless of which mechanism drives twists, we find that particle acceleration initiates pair cascades across charge-starved gaps above a mild critical twist. Cascades are initiated via resonant inverse-Compton scattered photons or curvature radiation, and may produce broadband coherent radio emission. We compute the pair luminosity (maximum allowed radio luminosity) for these two channels, and derive deathlines and `active zones' in $P-\dot{P}$ space from a variety of considerations. We find these twist-initiated pair cascades only occur for magnetar-like field strengths $B \gtrsim 10^{14}$~G and long periods: $P_{\rm RICS} \gtrsim 120 \; (T/10^{6.5} {\rm K})^{-5} \, {\rm sec}$ and $P_{\rm curv} \gtrsim 150 \; ({\rm v_{\rm pl}}/10^{3} {\, \rm cm \, yr^{-1}})^{-7/6} \, {\rm sec}$. Using a simplified geometric model, we find that plastic motion or thermoelectrically driven twists might naturally reproduce the observed luminosities, timescales, and timing signatures. We further derive `active zones' in which \textit{rotationally-powered} pair creation occurs via resonantly scattered photons, beyond standard curvature deathlines for pulsars. All cascades are generically accompanied by simultaneous (non-)thermal X-ray/UV counterparts which might be detectable with current instrumentation.

\end{abstract}


%%%%%%%%%%%%%%%%%%%%%%%%%%%%%%%%%%%%%%%%%%%%%%%%%%

%%%%%%%%%%%%%%%%% BODY OF PAPER %%%%%%%%%%%%%%%%%%

\section{Introduction}
The nature of recently reported Galactic long period pulsating radio sources (\citealt{hyman_powerful_2005,caleb_discovery_2022,hurley-walker_radio_2022, hurley-walker_long-period_2023,caleb_2024})\footnote{See also the growing population of bright, unassociated, image-domain Galactic radio transients \citep[e.g.,][]{2014ApJ...785...27O,2021ApJ...920...45W,2021MNRAS.504.4706K,2022MNRAS.516.5972W,2023arXiv231107394D,2024arXiv240612352D}} is uncertain. High brightness temperatures ($T_{\rm B} > 10^{12}$~K) and timing stability imply coherently emitting, relativistic magnetized plasmas within rotating compact object magnetospheres are likely sources. Highly magnetized neutron stars are obvious candidates, due to their large magnetic energy reservoir and known ability to produce coherent radio pulsations (e.g. \citealt{2006Natur.442..892C,rea_fundamental_2012}). The observed properties of these sources imply they may be much older than known magnetars, possibly pointing towards a common evolutionary channel where magnetic fields do not substantially decay within 10 kyr \citep{beniamini_evidence_2023,rea_long-period_2023}, or to magnetospheric conditions that deviate significantly from younger, rotationally-dominated sources towards `dead' pulsar solutions (e.g. \citealt{krause-polstorff_electrosphere_1985,petri_global_2002}). Alternative evolutionary channels may include enhanced spin-down through supernovae fall-back disks, giant flares or mass-loaded winds \citep{beniamini_periodicity_2020,ronchi_long-period_2022,2024arXiv240403882F}. Ultra-long period magnetars (henceforth ULPMs) might also evolve to attain long periods in tight binary systems, if strong magnetic or propeller torques enforce synchronisation with the orbital period \citep[e.g.,][]{2008ApJ...681..530P}. However, this binary synchronisation mechanism is likely ruled out for 1E 161348--5055 \citep{2017ApJ...841...11T} and GPM J1839--10 \citep{hurley-walker_long-period_2023}. Period derivative upper limits of strongly challenge traditional rotation-powered models of coherent radiation (e.g. \citealt{sturrock_model_1971,ruderman_theory_1975}), and may point towards a novel magnetically-powered coherent emission mechanism. Highly magnetized rotating white dwarfs (WDs) have also been suggested as plausible sources \citep{hurley-walker_radio_2022,katz_2022,rea_long-period_2023}, although this interpretation is disfavoured on energetic \& evolutionary grounds, and by X-ray upper limits \citep{beniamini_evidence_2023}. Similarly, bright radio emission from the large known population of nearby WDs is also largely ruled out \citep{2024arXiv240211015P}.
\par
Though interesting in their own right, ULPMs are of special interest due to a possible connection to the unknown sources of extragalactic millisecond-duration fast radio bursts (FRBs; \citealt{petroff_fast_2022, zhang_physics_2023}). FRBs are widely hypothesised to originate from highly magnetized neutron stars, through either impulsive magnetically-powered particle acceleration \citep[e.g.,][]{kumar_fast_2017,wadiasingh_repeating_2019,2019MNRAS.488.5887S,wang_coherent_2019,2020arXiv200505093L,cooper_coherent_2021} or synchrotron maser emission from relativistic flares \citep[e.g.,][]{lyubarsky_model_2014,metzger_fast_2019,margalit_implications_2020}. The former magnetospheric theories of FRBs appear to be preferred by recent observations of bursts with nanosecond variability (e.g., \citealt{snelders_microsecond-duration_2023,hewitt_2023}; see also \citealt{Beniamini+2020,Lu+2022}), a lack of optical afterglows \citep{kilpatrick_deep_2021, cooper_testing_2022,kilpatrick_limits_2023,hiramatsu_limits_2023}, and statistical similarity to earthquakes \citep{totani_2023,2024MNRAS.530.1885T}. Magnetospheric FRB theories often invoke sudden crustal disturbances such as starquakes \citep{wadiasingh_repeating_2019,2019MNRAS.488.5887S,kumar_frb_2020,2024Natur.626..500H} or oscillations \citep{wadiasingh_fast_2020,wadiasingh_fast_2020b} as triggers for particle acceleration and electron+positron cascades. Furthermore, two repeating FRB sources appear to show long-term periodicity: FRB 180916.J0158+65 (16.35 days; \citealt{collaboration_periodic_2020, pastor-marazuela_chromatic_2021}); and tentatively in FRB 201102 ($\approx$157 days; \citealt{rajwade_possible_2020, cruces_repeating_2020}). These discoveries led \cite{beniamini_periodicity_2020} to investigate ULPMs as candidate FRB progenitors, where the long observed periods is simply associated with the magnetar spin period. 
\par
Most X/$\gamma$-ray phenomena observed from magnetars exceed the spin-down luminosity budget, leading to the assumption that the large magnetic reservoir powers observed persistent and transient emission. In magnetars, the energy density of the field in the crust, $B^2$, is of similar order to the bulk shear modulus $\mu \sim 10^{30}$ erg cm$^{-3}$ of the crust's ion lattice \citep{1991ApJ...375..679S}, leading to a distinct and rich set of phenomena where the crust is pivotal \citep[e.g.,][]{2011ApJ...727L..51P,2015MNRAS.449.2047L,lander_game_2023}. Magnetars undergo outbursts characterised by the enhancement of persistent X-ray luminosity to $L_{\rm X} \approx 10^{31-36} \, {\rm erg \, s^{-1}}$ \citep{kaspi_magnetars_2017}, in addition to variable bursting activity on timescales of millisecond to second timescales \citep{2008A&ARv..15..225M}, often followed by a slow decay of persistent emission (e.g. \citealt{coti_zelati_systematic_2018}). The soft X-ray persistent spectrum of these sources is commonly characterised by either: two blackbodies (${\rm BB_{\rm i} + {\rm BB_{\rm rc}}}$) corresponding to intrinsic neutron star (NS) temperature with area $A_{\rm i} \approx 4 \pi R_{*}^2$ \footnote{$R_{*}$ denotes the radius of the NS throughout, although some of the mechanisms in this work may also be applicable to hybrid/quark stars.}, and hotspots associated with return currents from magnetospheric particle acceleration or heat transport from the interior crust $A_{\rm rc} \ll R_{*}^2$; or a blackbody and a non-thermal power-law component (${\rm BB_{\rm i}}$+PL), where the latter is attributed to Comptonization of thermal X-rays by mildly-relativistic and hot electrons in a low altitude corona. The physics and dynamics of crust is likely vital to observed activity.  Magnetic field footpoint displacement and shear associated crustal activity naturally twist the magnetosphere, resulting in a localized torodial magnetic field component $B_{\phi}$ and nonzero current $\nabla \times B \neq 0$. The gradual untwisting of the field lines dissipates energy, and is modulated by the current requirements of the twisted magnetic flux tubes. The resultant time-averaged voltage $\Phi$ across the length of the field lines is limited by runaway pair creation episodes, which screens the electric field on plasma timescales (commensurate with the inverse of the local plasma frequency). Accelerated relativistic charges may bombard the NS surface, resulting in hotspots which are observable as persistent but decaying signatures of untwisting field lines \citep{beloborodov_magnetar_2016,igoshev_2021_strong} if the charge deposition depth is sufficiently deep in the atmosphere (requiring high particle energies). This model has had some success in predicting the evolution persistent thermal X-ray emission following magnetar outbursts.  

\subsection{Ultra-long Period Coherent Radio Sources}
\label{sect:sources}
There are a handful of known astronomical sources which appear to be slowly rotating neutron stars including those in high-mass X-ray binary systems, periodic FRBs, long period $\gtrsim 12 \, {\rm s}$ but otherwise typical radio-loud pulsars. We list the properties of the three sources most relevant for this work in Table 1, but two other sources of specific note include: 1E 161348-5055, a 6.7 hour period central compact object associated with a supernova remnant RCW 103 (likely $880-4400$ years old; \citealt{braun_progenitors_2019}) which has displayed magnetar-like activity in X-rays \citep[e.g.,][]{2006Sci...313..814D,rea_magnetar-like_2016,2018MNRAS.478..741B}; and PSR J0901-4046, a radio pulsar with a 76 second period and spin-down age and dipole magnetic field of $5.3$ Myr and $1.3 \times 10^{14}$ G respectively \citep{caleb_discovery_2022}. We refer the reader to \cite{beniamini_evidence_2023} for an exhaustive list prior to the reported discovery of GPM J1839-10. These discoveries imply a large existing population and current bias against their detection. New observational identifications of this population are expected in the near future. 

\begin{table*}
\begin{center}
\setlength{\extrarowheight}{2pt}
% \captionof{Table to test captions and labels.}
\begin{tabular}{m{12em} | m{2cm}| m{2cm} | m{2cm} | m{2cm} }

\textbf{Source} & \textbf{GLEAM-X J1627} & \textbf{GPM J1839-10} & \textbf{GCRT J1745} & \textbf{ASKAP J1935} \\ 
\hline
P [min]  & 18.18 & 21 & 77 & 53.8\\  
\hline
$\dot{P}$ & < $1.2 \times 10^{-9}$ & $< 4.6 \times 10^{-13}$ & n/a & $< 1.2 \times 10^{-10}$ \\
\hline
Main Pulse [s] & 30-60 & 30-300 & $\sim$600 & 10-50 \\
\hline
Distance [kpc] & 1.3$\pm$0.5 & 5.7$\pm$2.9 & $\sim$ 8 & $\sim 4.75$ \\
\hline
$F_{\nu, \rm radio}$ [Jy] & 5-40 & 0.1-10 & $\sim$1 & $\sim$0.1 \\
\hline
$L_{\rm radio}$ [erg/s] & $\approx10^{28-31}$ & $\approx10^{28}$ & $\sim 10^{30}$ & $\sim 10^{30}$ \\
\hline
$L_{\rm spin-down}$ [erg/s] & $\lesssim1.2 \times 10^{28}$ & $\lesssim10^{25}$ & n/a & $\lesssim {\rm few} \times 10^{31}$\\
\hline
$L_{\rm X, 0.3-10 keV}$ [erg/s] & $<10^{32}$ & $<1.5
 \times 10^{32}$ & $\lesssim 3\times 10^{35}$ & $\lesssim 4\times 10^{30}$ \\
\hline
Activity duty cycle & $\approx$ 2 months & $\gtrsim$ 33 years & $\sim$7 hours & $\sim$2 weeks (main) \\
% \hline
\end{tabular}
\caption{Table 1: Properties of ultra-long period coherent radio sources \protect\citep{hyman_powerful_2005,hurley-walker_radio_2022,hurley-walker_long-period_2023}. Luminosities quoted are isotropic-equivalent values.}
\label{tab:source_properties}
\end{center}
\end{table*}

\subsubsection{GLEAM-X J1627}
GLEAM-X J162759.5-523504.3 (henceforth GLEAM-X J1627) was discovered as a serendipitous transient by the Murchison Widefield Array (MWA) as part of an extended image-plane observing program GLEAM-X \citep{hurley-walker_galactic_2022}. It has a period of 18.18 minutes, with bright, linearly polarized ($\approx 88 \%)$ coherent radio emission persisting for 30-60 seconds \citep{hurley-walker_radio_2022}. Emission is visible across the entire 72-231 MHz band with a variable temporal morphology (including sub-second spikes unresolved in MWA data), peaking at maximum flux densities of 5-40 Jy. Two separate periods of activity were observed, separated by around 20 days, in which the flux irregularly rises within days and decays within a month. In many epochs, two distinct sub-pulses appear to emerge; yet at times the lightcurve appears to have four or more distinct components. Pulse analysis supports favours $\dot{P} < 1.2 \times 10^{-9}$ s s$^{-1}$, meaning peak radio luminosities exceed the upper limit on the spin-down luminosity by orders of magnitude, suggesting a magnetic origin of emission. X-ray observations obtained with Swift XRT limit the $0.3-10$keV luminosity to $L_{\rm X} < 10^{32} \, {\rm erg \, s^{-1}}$ \citep{hurley-walker_radio_2022}. Deeper (but non-contemporaneous) X-ray limits were reported by \cite{rea_constraining_2022} \citep[see also Appendix F of][]{beniamini_evidence_2023} constraining the $0.3-8$~keV luminosity to $L_{\rm X} \lesssim 10^{30} \, {\rm erg \, s^{-1}}$ depending on the assumed spectral distribution.

\subsubsection{GPM J1839–10}
GPM J1839–10 was also discovered by MWA during a Galactic plane search for slow transients \citep{hurley-walker_long-period_2023}. It has a similarly long period of approximately 22 minutes, with greatly varying pulses lasting between 30 and 300 seconds, reaching maximum flux densities of 0.1-10 Jy. The pulses are rich in substructure, with observed variability resolved to timescales of milliseconds. Remarkably, archival searches discovered similar pulses reaching back to 1988 with the Very Large Array (VLA) and the Giant Metrewave Radio Telescope (GMRT). This led to a tighter constraint on the period derivative of $\dot{P} \lesssim 3.6 \times 10^{-13}$ at 1$\sigma$, implying stringent limits on the spin-down luminosity. The maximal dipole magnetic field as implied by spin-down upper limits is $B_{\rm s} \lesssim 2 \times 10^{15}\,$G. X-ray observations with XMM-Newton derive a limit to the persistent $0.3-10$ keV X-ray luminosity of $0.1-1.5 \times 10^{32}\, {\rm erg \,  s^{-1}}$ dependent on assumed spectral flux distribution; with limits of $\lesssim 2 \times 10^{33}\, {\rm erg \, s^{-1}}$ simultaneous with two bright radio pulses.

\subsubsection{GCRT J1745-3009}
GCRT J1745-3009 (the 'Galactic Burper', henceforth GCRT J1745) is a radio transient with a 77 minute period discovered by the VLA \citep{hyman_powerful_2005,spreeuw_new_2009}. Initial observations reported five 10-minute duration pulses observed at 330 MHz, with the high brightness temperature and minute-variability implying coherent emission from a compact source. One further burst was reported in \cite{hyman_new_2006}, but no new activity has been reported since. The authors note that despite a spatial location consistent with the Galactic centre, this source could be a local, less energetic flaring dwarf star ($D < 70$~pc; see also \citealt{zhang_gcrt_2005}). However this interpretation is disfavoured based on observed pulse properties, and the lack of an near-infrared counterpart \citep{kaplan_search_2008}. Serendipitous archival X-ray ($2-10$ keV) observations from the Rossi X-Ray Timing Explorer (RXTE) find no X-ray emission at a limit of $F_{\rm X} \approx 5 \times 10^{-12}\, {\rm erg \, cm^{-2}\, s^{-1}}$, although the X-ray epoch does not overlap with any radio activity.

\subsubsection{ASKAP J1935+2148}
ASKAP J193505.1+214841.0\footnote{ASKAP J1935 was reported during the review of this work. As such, we include discussion of this source in the introduction but not in later Sections of this work.} (henceforth ASKAP J1935) was discovered serendipitously by the Australian Square Kilometre Array Pathfinder (ASKAP) in the course of observing GRB 221009A \citep{caleb_2024}. It is a highly polarized ($\gtrsim 90\%$), coherent radio transient with a period of 53.8 minutes. It displays distinct emission modes similar to some radio pulsars: a the bright $F_{\nu} \sim 100 {\rm \, mJy}$ mode with pulses lasting 10-50 seconds, a fainter circularly polarized mode with $\sim 100 {\rm \, ms}$ pulses at higher frequencies, and a quiescent state. Archival Chandra and Swift observations rule out a persistent X-ray counterpart to below approximately $4 \times 10^{30} \; {\rm erg \, s^{-1}}$ for an assumed distance of 4.85 kpc. An isolated magnetized white dwarf interpretation of the source is disfavored owing to the extremely large required magnetic field $B_{\rm S, WD} \gtrsim 10^{10} \, {\rm G}$.

\subsection{Outline of This Work}
In this work, we explore models for radio emission from ULPMs from magnetospheric twists powered by plastic crustal motion, thermoelectric action sourced by temperature gradients, or prior crustal activity. In Section \ref{sect:model} we introduce the required concepts including crustal and magnetospheric evolution. In Section \ref{sect:plastic_motion} we show how plastic motion or thermoelectric action on a NS crust twists magnetic field lines, driving dissipation which can power ULPM pulsations. In Section \ref{sect:microphysics} we discuss the relevant physics for pair cascades, detailing the relevant regimes and new deathlines and active zones associated with this model. In Section \ref{sect:discussion}, we make predictions for multi-wavelength emission and timing signatures by which the model may be verified. We consider dissipation from pre-existing twists presumed to have arisen from past, transient crustal activity in Appendix \ref{sect:quake_induced}. We present our conclusions in Section \ref{sect:conclusions}.

\section{Magnetospheric Twists}
\label{sect:model}
\subsection{Crustal Evolution}
\label{sect:crustal_motion}

Unlike normal pulsars, the magnetic field energy density in the crust of magnetars is at a similar scale to the bulk shear modulus $\mu \sim 10^{30}$ erg cm$^{-3}$ ion lattice \citep{1991ApJ...375..679S}. As the lattice is fixed (with neutron fraction increasing with depth) but electrons are free, the field may evolve to states where mechanical stresses readily build in the crust. The variety of bursts, flares and persistent electromagnetic emission observed from magnetars are most readily explicable by disruptions and failures of the NS crust \citep[e.g.,][]{1992AcA....42..145P,thompson_soft_1995,1998ApJ...498L..45D,2008ApJ...680.1398T} which inject energy into the magnetosphere powering radiative processes. 
\par
NS evolution is only fully described by considering the co-dependent evolution of the core, crust, and external magnetosphere including possible superconductivity effects and field expulsion in the core \citep[e.g.,][]{2017PhRvC..96f5801H}. Although none of these individual elements are fully understood, magneto-thermal simulations have made substantial progress into our understanding of NS field evolution \citep[e.g.,][]{goldreich_magnetic_1992,2011ApJ...741..123P,vigano_unifying_2013,2016PNAS..113.3944G,vigano_simflowny-based_2019,vigano_magneto_2021,elfritz_simulated_2016,akgun_crustmagnetosphere_2018,igoshev_2021_strong,2019LRCA....5....3P,skiathas_combined_2024}. Recent theoretical studies suggest that in addition to canonical field evolution mechanisms of Hall drift and Ohmic decay \citep{goldreich_magnetic_1992,Pons2009}, plastic flow may dominate magnetic field evolution if $B_{\rm s} \gtrsim 10^{15}$ G \citep{lander_magnetar_2016,gourgouliatos_axisymmetric_2021}, and possibly stall magnetic field decay \citep{Suvorov_evolutionary_2023}. Plastic flow occurs when magnetically-induced crustal strain exceeds the elastic limit (i.e. the point at which the crust can return to the original unstressed configuration) such that the crust fails \citep{2009PhRvL.102s1102H,2012MNRAS.426.2404H}. Sudden supra-elastic failures, or excitation of magneto-elastic crustal normal modes could trigger FRBs in the magnetosphere (e.g. \citealt{wadiasingh_repeating_2019,wadiasingh_fast_2020,wadiasingh_fast_2020b}). After failures, the elastic-phase equations are no longer valid and the crust may enter a phase of slow deformation, known as plastic flow \citep{lander_magnetic-field_2019}.

\begin{figure}
  \centering
{\includegraphics[width=.5\textwidth]{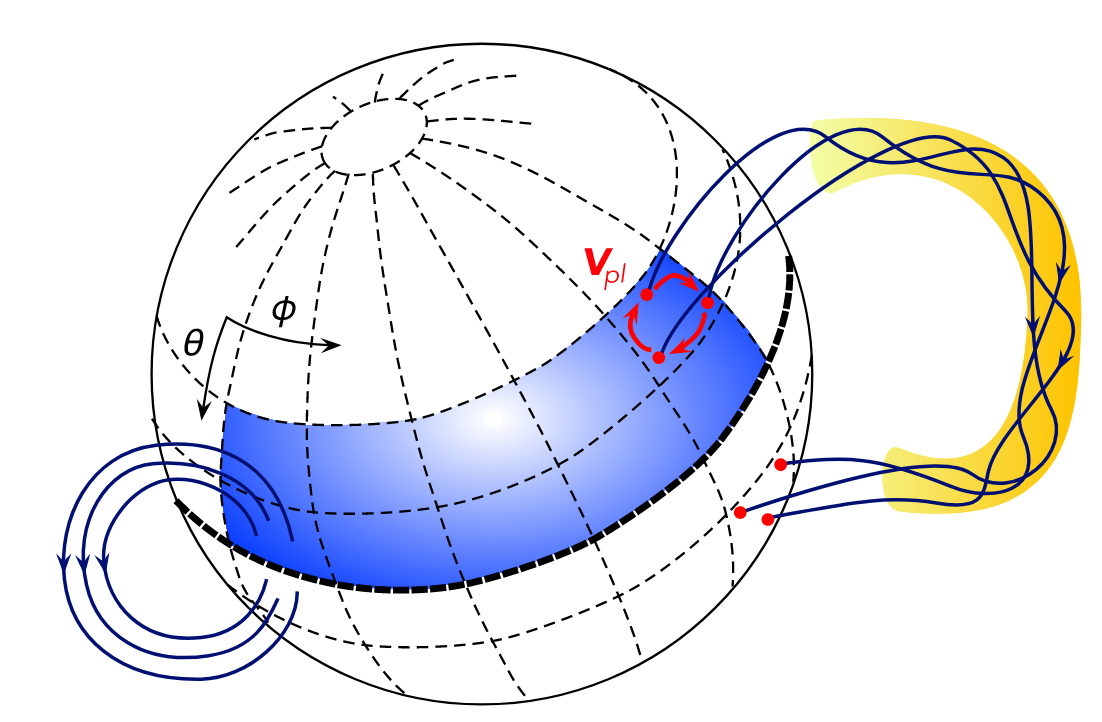}}
\caption{Illustration of twisted field lines via plastic flow from \protect\cite{lander_game_2023}. The depicted twist is exaggerated for clarity.}
\label{fig:illustration}
\end{figure}

Magnetic field lines are threaded through the crust, anchored at some point in the NS interior $r_{\rm c} < R_{*}$ which is strong enough to sustain the azimuthal Ampere forces, such that crustal motion imparts a mild twist in the magnetosphere (Fig. \ref{fig:illustration}) in regions where $\nabla \times B \neq 0$ generically far from force-free electrodynamics typically assumed for expediency in other works. Plastic flow induced twist may increase over long timescales and, if not dissipated via global field reconfiguration, will dissipate electromagnetically when magnetospheric charge density is insufficient. 

\par
The plastic flow velocity ${\rm v_{\rm pl}}$ depends on the unknown viscosity of the crust at the location driving the flow $\nu_{\rm crust}$. \cite{lander_magnetic-field_2019} use observationally motivated estimates of $\nu_{\rm crust} = 10^{36-38}\;$poise (${\rm cm^{-1}\, g^{-1}\, s}$), based on X-ray evolution timescales of thermal magnetars \citep{beloborodov_corona_2007}. \cite{lander_game_2023} suggest that the viscosity may be temperature-dependent, as in terrestrial liquids \citep{andrade_theory_1934}, and use values in the range $\nu_{\rm crust} = 10^{34-36}\;$poise\footnote{For reference, the measured viscosity of famously viscous pitch is between $10^{7-10}$ poise dependent on temperature, as measured by the Queensland pitch drop experiment \citep{edgeworth_pitch_1984}.}. The authors suggest that a modified Andrade equation may describe the viscosity of NS crusts as follows\footnote{We use convenient notation $X_{\rm n} \equiv X/10^n$ throughout.}:
\begin{equation}
    \nu(\rho,T) = 5 \times 10^{35} \, B_{\rm s, 15}^2/(4 \pi) \, \exp\bigg(\frac{5}{1 + T_{9}}\bigg)
\end{equation}
This might imply that twist-induced return current hotspots (e.g. \S \ref{sect:x-rays}) powered by plastic motion decrease local viscosities at magnetic footpoints creating a positive feedback loop enabling faster local plastic flow. 
\par
Numerical simulations suggest if  $B \approx 10^{15}$~G, estimated viscosities permit for plastic flow velocities\footnote{For clarity, ${\rm v}_{\rm pl}$ is quoted in units of ${\rm cm \, yr^{-1}}$ throughout, whereas all other quantities are in centimetre-gram-second convention.} of $1-100 \; {\rm cm \, yr^{-1}}$, which increase with $B$ and depend inversely on the $\nu_{\rm crust}$, as expected \citep{lander_magnetic-field_2019}. \cite{younes_pulse_2022} suggest that observed pulse migration from SGR 1830-0645 may be explained by plastic flow velocities of ${\rm v}_{\rm pl} \approx 10^{6} \; {\rm cm \, yr^{-1}}$, plausibly attainable through a relatively shallow field line anchor point in the outer crust. We adopt $1 \; {\rm cm \, yr^{-1}} \lesssim {\rm v}_{\rm pl} \lesssim 10^{6} \; {\rm cm \, yr^{-1}}$ as a feasible range throughout. 
\par
Using cellular automata to model self-organised criticality behaviour \citep{1987PhRvL..59..381B} likely realized in magnetar crusts, \cite{lander_game_2023} find that a range of magnetar activity including giant flares, short bursts and periods of quiescence can be, at least qualitatively, explained by Hall drift and plastic flow of crustal cells with length scales of approximately $L = 1\, {\rm km}$ in size, in broad agreement with magneto-thermal evolution simulations \citep[e.g.,][]{2016PNAS..113.3944G,igoshev_2021_strong}. Given this, we adopt throughout a characteristic plastic flow footpoint area of $A_{\rm fp} \approx 1\, {\rm km}^2$. 

\subsection{Field Twists}
\label{sect:field_lines_belo}
Crustal motion perturbs magnetic footprints on the surface resulting in local torodial twists $B_{\phi}$ in the magnetosphere, which is otherwise approximately dipolar embedded in curved spacetime. Consider one such twist along a field line $\textbf{B}$ which runs from two points on the NS surface $[Q_{1}, Q_{2}]$ corresponding to footpoints at co-latitudes $\theta_{\rm fp}$ in separate hemispheres. In all cases $B_\phi \ll B_0$ such that the twist and magnetic helicity can be treated linearly. The associated free energy in MHD equilibria is \citep{beloborodov_untwisting_2009}\footnote{See also \cite{wolfson_shear_induced_1995} for a seminal mathematical treatment of force-free magnetospheric twists. }:
\begin{equation}
    E_{\rm tw} = \int_{\rm r > r_c} \frac{B_\phi^2}{8 \pi} dV
    \label{eq:energy_free_def}
\end{equation}
The twist angle $\Psi$ is the integrated twist along $\textbf{B}$ \citep{thompson_electrodynamics_2002}. If the twist is uniformly distributed along the current loop according to local values of $B_0$:
\begin{equation}
    \Psi = \int d \phi = 2 \int_{\rm \theta_{\rm fp}}^{\pi/2} \frac{B_\phi}{B_0} \frac{d\theta}{\sin(\theta_{\rm fp})} \approx - 2 \ln(\theta_{\rm fp}) \frac{B_\phi}{B_0} \sim \frac{B_\phi}{B_0}
    \label{eq:twist_def}
\end{equation}
Where we integrate from footpoints to the equator, resulting in the factor two. The twist demands and is by supported by a current which is maintained by a time-averaged longitudinal voltage $\Phi$ by an electric field parallel to $\textbf{B}$: 
\begin{equation}
    \Phi = \int_{Q_1}^{Q_2} E_{\parallel} dl
    \label{eq:voltage_def}
\end{equation}
\cite{beloborodov_untwisting_2009} show that the luminosity associated with twist dissipation is directly correlated to this voltage and the current $I$:
\begin{equation}
  L_{\rm diss} = \int_{Q_1}^{Q_2} \textbf{E} \cdot \textbf{L}(l) d\textbf{l} \sim I \Phi
  \label{eq:diss_l}
\end{equation}

\subsection{Pair Production}
\label{sect:pair_production_brief}
Electrons and ions may be pulled from the atmosphere of the neutron star, which defines a minimum voltage along the twist which in the limit $R \ll R_{\rm max}$: 
\begin{equation}
     q \Phi_{\rm crust} = \frac{G M_{\rm NS} m_{\rm i}}{R_{*}} \approx
  \begin{cases} 
  10^{5} \, {\rm eV} & \text{electrons}  \\
   2 \times 10^{8} \, {\rm eV}       & \text{ions}
  \end{cases}
     \label{eq:crust_thres}
\end{equation}
Where $M_{\rm NS}$ is the NS mass, and $m_{\rm i}$ is the mass of the particle, depending on which footpoint is considered. The inevitability of pair production is an open question. \cite{beloborodov_corona_2007} argue charges lifted from the star accelerated along the twisted field lines cannot describe a realistic steady-state magnetosphere, as this would imply an unacceptable departure from neutrality or a violation of the conservation of particle momenta due to the gravitational force acting similarly on charges of both signs; where as the $E_{\parallel}$ accelerating charges in opposite directions. \cite{beloborodov_corona_2007} claim, supported by toy models and numerical experimentation, that a large $\Phi$ should persist along field lines, growing until rampant pair creation cascades screen the electric field and prevent twist dissipation. Such pair production can occur through one-photon magnetic pair production of either curvature photons or the resonant inverse-Compton scattering (RICS) of soft X-rays from the stellar surface by accelerated electrons \citep{2007Ap&SS.308..109B}. For high temperatures, it may be the case that the critical voltage for RICS pair production to occur lies intermediate to the voltages in Eq. \ref{eq:crust_thres}. This, along with directionality of electrons (ingoing or outgoing) influences the RICS scattering rate and may mean pair production is favored at just one footpoint (i.e. the footpoint where crust ions are pulled). Both mechanisms of pair creation are detailed in \S\ref{sect:microphysics}, but directionality influences and tenability of pair production at one or both footpoints is not considered in this work. If pair creation occurs it will regulate the dissipation rate, establishing relative stability at the pair production threshold. \footnote{The plasma response time is very short ($\tau \approx \omega_{p}^{-1} = (m_e/4 \pi q^2 n)^{1/2}$), and variability due pair creation occurs on microsecond timescales associated with the gap size or plasma skin depth.} Despite this, it is conceivable that specific magnetospheric conditions may allow lifted crustal charges to wholly account for current requirements of magnetospheric twists. We do not consider this further, but note coherent radio emission with comparable luminosities may be present without pair production; e.g. similar to the mechanism in AR Sco \citep{Marsh_2016}, albeit with lower observed brightness temperatures.

\subsection{Radio Emission and Frequency}
\label{sect:radio}
The production of coherent radio emission in radio pulsars (and FRBs) involves tiny portion of the total energetics, and is not fully understood. For decades, it has been thought that the production of new electron-positron pairs which supplies the plasma is necessary for coherent radio emission within the magnetosphere, with much theoretical work devoted to the study of magnetospheric pair creation in stationary (time-independent) gaps \citep[e.g.,][]{sturrock_model_1971,1973ApJ...183..625T,ruderman_theory_1975,1982ApJ...252..337D,daugherty_pair_1983,1997ApJ...476..246H,1998ApJ...508..328H,2000ApJ...531L.135Z,2001ApJ...547..929B,2001ApJ...554..624H,2001ApJ...560..871H,2002ApJ...576..366H,2011ApJ...726L..10H}. In \cite{timokhin_time_2010} \& \cite{timokhin_current_2013}, the authors use 1D particle-in-cell simulations to show that pair creation close to the stellar surface is intrinsically non-stationary, resulting in limit-cycles of pair creation and acceleration. They find that bursty pair creation cycles are a necessary and sufficient condition for generation of superluminal electrostatic waves, and conjecture that similar superluminal electromagnetic waves may be realized beyond their 1D construct in pulsars. This was borne out in 2D numerical experiments by \cite{philippov_origin_2020} (see also \citealt{2024arXiv240520866B}), albeit with parameters far from realistic pulsars, where non-stationary pair creation occurs across field lines resulting in superluminal electromagnetic waves. Such waves might be capable of escaping the magnetosphere and propagating to the observer as coherent radio emission. Therein, the curvature of field lines and tilt of the pair formation front is essential for generation of such superluminal waves which, if the plasma were homogeneous, would map to superluminal O-modes.
\par
It is common in pulsar physics to express observed radio luminosities as a fraction of dissipated (spin-down) luminosity. In \S\ref{sect:microphysics}, we study the microphysical processes which govern the acceleration gap, and thus we can adopt a two-stage efficiency convention: 
\begin{equation}
    L_{\rm r} \approx \eta_{\rm r} L_{\rm e^+e^-} = \eta_{\rm r} f_{\rm pair} L_{\rm diss}
    \label{eq:radio_efficiency}
\end{equation}
Where radio efficiency parameter $\eta_{\rm r} < 1$ is the radio efficiency from the pair luminosity (see \S\ref{sect:pair_luminosity}), and $f_{\rm pair} \leq 1$ is the fraction of twist dissipation luminosity which accelerates primary electrons. Generically, the pair luminosity scales similarly to the voltage ($\propto B$) rather than the total dissipation luminosity ($\propto B^2$), and $\eta_r$ approaches unity in rotational-powered pulsars near `death' zones \citep{2002ApJ...568..289A}.

The emission plausibly generated by the Timokhin-Arons acceleration mechanism, considered in \cite{philippov_origin_2020}, is broadband in nature as the sum of a spectrum of non-stationary pair discharges across a range of wavenumbers. The radio emission is produced at the local electron plasma frequency in discharges $\omega_{\rm e} \sim \sqrt{4 \pi q \mathcal{M}_\pm \rho_{\rm twist}/m_{\rm e}}$ and its harmonics ($\mathcal{M}_\pm$ is a pair multiplicity). Thus, lower pair multiplicities and less chaotic limit cycles are likely produce more narrow band radio emission. The current density associated with pair creation is $\rho_{\rm diss} = \frac{q \Phi_{\rm pp}}{4 \pi h_{\rm gap}^2}$ and the minimum frequency of radio emission is therefore:
\begin{equation}
\begin{split}
    \nu_{\rm min} &= \frac{1}{2 \pi} \sqrt{\frac{q \mathcal{M} \Phi}{m_{\rm e} h_{\rm gap}^2}} = 8 \times 10^{7} \; {\rm Hz} \, \Phi_{\rm pp, 12}^{1/2}  \, \mathcal{M}_{2}^{1/2} \, h_{\rm gap, 6}^{-2} 
\end{split}
\end{equation}
For typical gap heights and $\mathcal{M} \lesssim 10^{5}$, emission is generally predicted at observed frequencies of ultra-long period sources, consistent with theoretical maximum multiplicities for pulsars \citep{timokhin_multiplicity_2019}. 
\par
We note that alternative radiation mechanisms may power coherent emission, following a similar efficiency convention. This may occur, for example, via the bunching of leptons accelerated by the large $E_{\parallel}$ to produce coherent curvature radiation \citep{kumar_fast_2017, wang_coherent_2019}, where the maximum allowed luminosity \citep{cooper_coherent_2021} is consistent with observations. However, charge bunches have not yet been demonstrated in kinetic simulations in the pulsar context.

\section{Plastic motion induced twists}
\label{sect:plastic_motion}

\subsection{Time-dependent Dissipation Toy Models}
Assuming one static footpoint and one undergoing plastic motion with a velocity ${\rm {\rm v}_{\rm pl}}$, magnetospheric twist builds up based on the differential angular velocity:
\begin{equation}
    \dot{\Psi}_{\rm pl} = \frac{{\rm v}_{\rm pl}}{R_{*}} = 10^{-4} \; {\rm v}_{\rm pl, 2} \, R_{*, 6}^{-1} \: {\rm year^{-1}}
    \label{eq:twist_rate}
\end{equation}
 
The current demands of mild magnetospheric twists can be dealt with by the local excess Goldreich-Julian charge density ($\rho_{\rm GJ} \approx \frac{2 B }{c P}$) present due to the rotation of the neutron star. \footnote{Ultra-long period magnetars' magnetospheres may admit inactive `disk-dome electrosphere' states \citep{krause-polstorff_electrosphere_1985,petri_global_2002} if pair production ceases. In this regime, the magnetosphere is charge-separated with an equatorial belt and polar cap dome. These specific regions may have charge densities slightly above $\rho_{\rm GJ}$ (by a factor of a few), but the remainder of the magnetosphere could obey near-vacuum conditions. A more detailed treatment, with co-latitude-dependence $\rho_{\rm GJ} \sim -\boldsymbol{\Omega} \cdot \boldsymbol{B}/(2\pi c)$ or of the disk-dome solutions could be explored but is not expected to qualitatively alter the conclusions of this work. Twists may also be prograde or retrograde to rotation, but do not meaningfully alter considerations here except in contrived scenarios (N.B. a non-axisymmetric local twist in an oblique rotator will have both prograde and retrograde components).} By equating $j_{\rm GJ}$ and $j_{\rm twist}$ (see, e.g., Eq. 2 in \citealt{wadiasingh_repeating_2019}), we can write down the critical value above which twist charge requirements are not satisfied and twist dissipation occurs:
\begin{equation}
\begin{split}
    \Psi_{\rm crit} &= \frac{8 \pi R_{*}}{c P \sin^2(\theta_{\rm fp})} = 8 \times 10^{-5} \; R_{*, 6} \, P_{\rm NS, 3}^{-1} \, \sin^{-2}(\theta_{\rm fp, -1})
\end{split}    
    \label{eq:critical_twist}
\end{equation}
Where $P$ is the NS period. Assuming quasi-constant plastic motion the timescale for the twist to reach a critical value is:
\begin{equation}
\begin{split}
    \tau_{\rm tw, c} &= \frac{\Psi_{\rm crit}}{\dot{\Psi}_{\rm pl}} = \frac{8 \pi R_{*}^2}{c P {\rm v}_{\rm pl} \sin^2(\theta_{\rm fp})} \\
    &\approx 1 \; P_{3}^{-1} \, {\rm v}_{\rm pl, 2}^{-1} \,  R_{*, 6}^{2} \, \sin^{-2}(\theta_{\rm fp, -1})\: {\rm year} 
     \end{split}
    \label{eq:timescale_critical}
\end{equation}
As twist is applied to field lines through crustal plastic motion, the critical twist (Eq. \ref{eq:critical_twist}) is regularly exceeded. The dissipation rate is proportional to the time-averaged longitudinal voltage $\Phi_{\rm pp}$, the value of which depends on the microphysics of pair production (see \S\ref{sect:microphysics}). The untwisting rate can be expressed to first order by consideration of the timescales of the dissipation of energy and twist. Using Eqs. 46 \& 48 in \citep{beloborodov_untwisting_2009}:
\begin{equation}
\begin{split}
    \frac{\Psi}{\dot{\Psi}_{\rm untwist}} &= \frac{E_{\rm twist}}{L_{\rm untwist}} \\
    \dot{\Psi}_{\rm untwist} &= \Psi \frac{L_{\rm untwist}}{E_{\rm twist}} = \frac{c \Phi(\Psi)}{B_{\rm s} R_{*}^2}
\end{split}
\label{eq:twistdown}
\end{equation}
The associated untwisting timescale for twists mildly in excess of the critical value is:
\begin{equation}
\begin{split}
    \tau_{\rm untwist} &= \frac{\Psi_{\rm crit}}{\dot{\Psi}_{\rm untwist}} = \frac{8 \pi B_{\rm s} R_{*}^3}{\Phi_{\rm pp} c^2 P \sin^2(\theta_{\rm fp})} \\
    &\approx 7 \; B_{\rm s, 15} \, R_{*, 6}^{3} \, \Phi_{\rm pp, 12} \, P_{3}^{-1} \, \sin^{-2}(\theta_{\rm fp, -1}) \: {\rm days}   
\end{split}
    \label{eq:untwisting_timescale_pl}
\end{equation}
In the following, we present two 1D simplified models of $\Psi(t)$ from plastic motion.

\subsubsection{$\big(\dot{\Psi}_{\rm untwist} > \dot{\Psi}_{\rm pl}\big)$: Step function voltage }
\label{sect:efficient_dissipation}
As discussed in \S\ref{sect:pair_production_brief}, $\Phi$ is expected to grow rapidly for super-critical twists, to produce pairs which screen $E_{\parallel}$ \citep[e.g.,][]{beloborodov_corona_2007,beloborodov_untwisting_2009}. Once the twist above the critical value is dissipated and returns to $\Psi_{\rm crit}$, pair production ceases as the field lines are not charge-starved. Based on this notion, we explore twist evolution where $\big(\dot{\Psi}_{\rm untwist} > \dot{\Psi}_{\rm pl}\big)$ in a simplified one-dimensional model. This approach is valid if quasi-stable plastic motion occurs and the following hierarchy of timescales are met:
\begin{equation}
    \tau_{\Phi} \ll \tau_{\rm untwist} \ll \tau_{\rm tw, c}
    \label{eq:timescales_belo}
\end{equation}
Where $\tau_{\rm \Phi}$ is the timescale at which the voltage increases to the pair production threshold when $j_{\rm twist} > j_{\rm GJ}$ (i.e. $h_{\rm gap}/c$ on the order microseconds), $\tau_{\rm untwist}$ is the timescale of untwisting (Eq. \ref{eq:untwisting_timescale_pl}), and $\tau_{\rm tw, c}$ is the timescale at which plastic motion increases the field line twist (Eq. \ref{eq:timescale_critical}). If this timescale hierarchy is met, then the twist evolution may be described by:
\begin{equation}
    \frac{d \Psi}{d t} = \frac{{\rm v}_{\rm pl}}{R_{*}} - \frac{c \Phi_{\rm pp}}{B_{\rm s} R_{*}^2} H(j_{\rm twist})
    \label{eq:odeint_pl}
\end{equation}
Where the $H$ is the Heaviside function:
\begin{equation}
    H(j_{\rm twist}) = 
    \begin{cases} 
    1 & \text{if} \: \: j_{\rm twist} > j_{\rm GJ}  \\
    0 & \text{if} \: \: j_{\rm twist} \leq j_{\rm GJ}
    \end{cases}
\end{equation}
Here, $j_{\rm twist} \approx \frac{c B \sin^2(\theta_{\rm fp}) \Psi}{4 \pi R_{*}}$ is the local current density demanded by the twist, and $j_{\rm GJ} \approx \rho_{\rm GJ} c = \frac{2 B}{q P}$ is the local Goldreich-Julian current density. Numerical integration of this differential equation shows that, as expected, the twist slowly increases to the critical value described by Eq. \ref{eq:critical_twist} before rapidly oscillating about this value as sparks of pair production occur, signalling quasi-stable radio emission. The dissipation of energy and twist occurs efficiently, such that any long timescale observable electromagnetic emission scales with $\dot{\Psi}_{\rm pl}$, not $\dot{\Psi}_{\rm untwist}$. In this regime, where $\big(\dot{\Psi}_{\rm untwist} > \dot{\Psi}_{\rm pl}\big)$, radio emission persists for as long as crustal motion remains active, expected to be on timescales of months to decades \citep{lander_game_2023}. One could think of the observed dissipation as more akin to terrestrial `continental drift'\footnote{Incidentally, neutron star plate tectonics was considered for neutron star models of gamma-ray bursts by \cite{1991ApJ...366..261R,1991ApJ...382..576R,1991ApJ...382..587R}. } of the crust driven by supra-elastic stresses rather than crustal `quakes', which may power observed magnetar short X-ray bursts and/or FRBs.

\subsubsection{Net excess current demand voltage $\big(\dot{\Psi}_{\rm untwist} < \dot{\Psi}_{\rm pl}\big)$}
\label{sect:voltage_net_excess}

Now we consider an alternative scenario in which the voltage value across twisted field lines is set exclusively by the excess current demand of the twist. In this case, the voltage associated with the net charge requirement can be rewritten as via Eq. \ref{eq:j_twist}:
\begin{equation}
\Phi_{\rm net} \approx 4 \pi (\rho_{\rm twist} - \rho_{\rm gj}) l_{\rm tw}^2  = 4 \pi l_{\rm tw}^2 B \bigg( \frac{\Psi \sin^2(\theta)}{4 \pi R_{*}} - \frac{2}{c P}\bigg)
\label{eq:net_phi}
\end{equation}
where $l_{\rm tw}$ is a length scale across which the voltage acts, which we show in Eq. \ref{eq:length_scale_dominate} to be $\approx R_{*}$. In this scenario, we can compute a stable $\Phi_{\rm max}$ as the time at which twist dissipation (\& Eq. \ref{eq:twistdown}) equals twist increment via plastic motion (Eq. \ref{eq:twist_rate}):
\begin{equation}
    q \Phi_{\rm max} = \frac{q {\rm v}_{\rm pl} B_{\rm s} R_{*}}{c} \approx 10^{11} \; {\rm eV} \; {\rm v}_{\rm pl, 5} \, B_{\rm s, 15}
    \label{eq:phi_max}
\end{equation}
To produce pairs and thus coherent radio emission, Eq. \ref{eq:phi_max} should exceed production of $10^{10-11}$ eV (see \S\ref{sect:pair_production_brief}). We can substitute this maximum voltage into Eq. \ref{eq:net_phi} to examine the maximum twist: 
\begin{equation}
\begin{split}
    \frac{{\rm v}_{\rm pl} B_{\rm s} R_{*}}{c} = 4 \pi l_{\rm tw}^2 B \bigg( \frac{\Psi \sin^2(\theta)}{4 \pi R_{*}} - \frac{2}{c P}\bigg) \\
    \Psi_{\rm max} = \frac{{\rm v}_{\rm pl} R_{*}^2}{l_{\rm tw}^2 c \sin^2(\theta_{\rm fp})} + \frac{8 \pi R_{*}}{c P \sin^2(\theta_{\rm fp})}
    \end{split}
\end{equation}
For most reasonable values of $P$ and ${\rm v}_{\rm pl}$, the maximum twist is dominated by the second term. This corresponds exactly with the critical twist in Eq. \ref{eq:critical_twist}, meaning that efficient dissipation of twist occurs as soon as the twist reaches a value above the value dealt with by the Goldreich-Julian current. For large values of ${\rm v}_{\rm pl} > 10^{5} \, {\rm cm \, yr^{-1}}$ within the plausible range, it is possible that the conditions required by Eq. \ref{eq:phi_max} are met without aforementioned step function voltage jumps in the super-critical twist regime. This results in a twist that cannot be efficiently dissipated and so $\Psi$ increases as plastic flow continues to act. For the remainder of the paper, we assume that the former case $\big(\dot{\Psi}_{\rm untwist} > \dot{\Psi}_{\rm pl}\big)$ holds, although we note that the main conclusions of the work are unaffected in either case if fast plastic motion is realized. The primary difference is that the dissipation luminosity (and with a weaker dependence, the radio luminosity set by the microphysics), and timescale for which emission persists will depend on the dissipation timescale rather than the time for which plastic motion persists.

\subsection{Twist Dissipation Luminosity}
\label{sect:radio_pl}
\begin{figure}
  \centering
{\includegraphics[width=.5\textwidth]{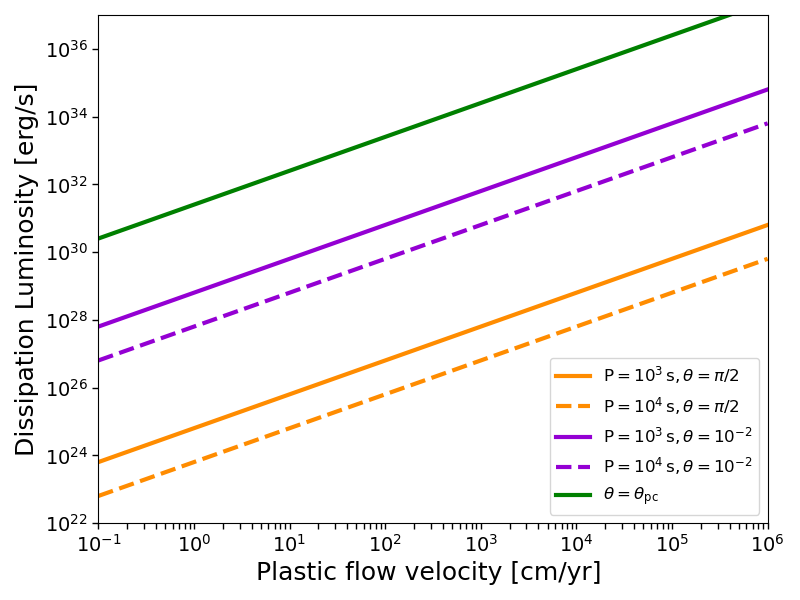}}
\caption{Dissipation power of twist induced plastic-flow assuming $B = 10^{15}$ G, $A_{\rm fp} = 10^{10}$ cm$^2$ via Eq. \ref{eq:luminosity_untwisting_global}. Results are shown for different values of NS period $P$, plastic flow velocity ${\rm v}_{\rm pl}$ ($L_{\rm diss, pl} \propto {\rm v}_{\rm pl} P^{-1}$) and co-latitude of the slab footpoints $\theta_{\rm fp}$. Dependence on $P$ disappears if emission occurs at maximal co-latitude at the polar cap boundary.}
\label{fig:plastic_flow}
\end{figure}

The maximum observable electromagnetic luminosity can be estimated as a fraction of the liberated twist energy which, if twist is efficiently dissipated, is the same as the power transferred to the torodial magnetic field by plastic motion. The total power (dissipation rate) above the critical twist may be estimated by:
\begin{equation}
    L_{\rm diss, pl} = \dot{E}_{\rm tw} \approx  \frac{d}{dt} \int_{r = r_{\rm c}} \frac{B_{\phi}^2}{8 \pi} dV
    \label{eq:L_diss_pl}
\end{equation}
The magnetic field lines flare to a maximal cross section at $R_{\rm max}$ of $A_{\rm max} = A_{\rm fp} \big(\frac{R_{\rm max}}{R_{*}}\big)^{3} \approx A_{\rm fp}/\sin(\theta_{\rm fp})^6$, which one can easily understand by the conservation of magnetic flux; i.e. decrease in $B(r)$ is commensurate with an increase in flux tube cross-section $A(r)$. This increases cross section is offset by the decreasing magnetic field $B(r) \propto \big(\frac{R_{\rm max}}{R_{*}}\big)^{3}$. Therefore the integrated free energy is:
\begin{equation}
    \begin{split}
    E_{\rm tw} &= \int_{r = r_{\rm c}} \frac{B_{\phi}^2}{8 \pi} dV = 2 \Psi^2 B_{\rm s}^2 A_{\rm fp} \int_{R_{*}}^{R_{\rm max}} \bigg(\frac{R}{R_{*}}\bigg)^{-3} dl
    \end{split}
\end{equation}
Where $dl$ is the line element along the field loop. One can express $dl$ in terms of $R_{\rm max}$ (Eq. 22 in \citealt{wadiasingh_resonant_2018}) as $dl = \frac{\sqrt{ 4 R_{\rm max} - 3 r}}{2 \sqrt{R_{\rm max} - r}} dr$ such that the free energy is:
\begin{equation}
\begin{split}
    E_{\rm tw} &= 2 \Psi^2 B_{\rm s}^2 A_{\rm fp} \int_{R_{*}}^{R_{\rm max}} \bigg(\frac{r}{R_{*}}\bigg)^{-3} \frac{\sqrt{ 4 R_{\rm max} - 3 r}}{2 \sqrt{R_{\rm max} - r}} dr \\
    &\approx \Psi^2 B_{\rm s}^2 A_{\rm fp} R_{*}
    \end{split}
\end{equation}
Where we have used the fact that the right hand side integral is:
\begin{equation}
    \begin{split}
   &\int_{R_{*}}^{R_{\rm max}}  \bigg(\frac{r}{R_{*}}\bigg)^{-3} \frac{\sqrt{4 R_{\rm max} - 3 r}}{2 \sqrt{R_{\rm max} - r}} dr\\
   &= \frac{R_{*}^3}{64 r^2 R_{\rm max}^2}  \Bigg[15 r^2 {\rm arctanh} \Bigg(\frac{2 \sqrt{R_{\rm max} - r}}{\sqrt{4 R_{\rm max} - 3r}}\Bigg) \\
   & \: \: \: \: \: \: \: \: \:   + 2 \sqrt{R_{\rm max} - r} \sqrt{4 R_{\rm max} - 3 r} \bigg(9 r + 8 R_{\rm max} \bigg) \Bigg] \\
    &\approx \frac{R_{*}}{2} \: \: \: \: \: \: \: \: \: \: \: \: \: \:{\rm \big( if \: \: R_{*} \ll R_{\rm max} \big)} 
    \end{split}
    \label{eq:length_scale_dominate}
\end{equation}
Where the appropriate limit is used corresponding to higher latitude footpoints. This result implies that most of the free energy is stored, and thus dissipation occurs, in the high $B$ region within one NS radii of the surface. Therefore using Eqs. \ref{eq:twist_rate}, \ref{eq:critical_twist} \& \ref{eq:L_diss_pl}, and assuming $\frac{d A_{\rm fp}}{dt} \approx 0$:
\begin{equation}
\label{eq:luminosity_untwisting_global}
    \begin{split}
        L_{\rm diss, pl} &\approx \frac{\Psi_{\rm crit} \dot{\Psi}_{\rm pl} B_0^2 R_{*} A_{\rm fp}}{8 \pi} 
         % &\approx \frac{4 \pi R_{*}}{c P \sin^2(\theta)} \frac{{\rm v}_{\rm pl}}{R_{*}} \frac{ B_0^2 L_{\rm fp}^2 R_{*}}{8 \pi} \\
         = \frac{{\rm v}_{\rm pl} B_{\rm s}^2 A_{\rm fp} R_{*}}{c P \sin^2(\theta_{\rm fp})} \\
         &\approx 1.5 \times 10^{31} \; {\rm erg \, s^{-1}}\; {\rm v}_{\rm pl, 4} \, B_{\rm s, 15}^2 \, A_{\rm fp, 10} \, P_{3}^{-1} \, \sin^{-2}(\theta_{\rm fp, -1})
    \end{split}
\end{equation}
The same result can be obtained via Eq. \ref{eq:diss_l} by the product of the current $I = j_{\rm twist} A_{\rm fp}$ and a voltage along the field line bundle $\Phi = (v_{\rm pl}/c) B R_{\rm max}$. The twist dissipation luminosity is shown in Fig. \ref{fig:plastic_flow}.
\par
Twists must occur within the closed field line region, bounded at the poles by polar cap half-angle $\sin(\theta) > \sin(\theta_{\rm PC}) \approx \big(\frac{R_{*}}{R_{\rm LC}}\big)^{1/2} = \big(\frac{2 \pi R_{*}}{c P}\big)^{1/2} \approx 5 \times 10^{-4} \; P_{3}^{-1/2} \, R_{*, 6}^{1/2}$. Given this, we can define the maximal luminosity where the twist co-latitude is located close to the polar cap boundary $\theta = \theta_{\rm PC}$:
\begin{equation}
    \begin{split}
        L_{\rm diss, pl, max}&= \frac{{\rm v}_{\rm pl} B_{\rm s}^2 A_{\rm fp}}{4 \pi} \approx 3 \times 10^{33} \, {\rm v}_{\rm pl,0} \, B_{\rm s, 15}^2 \, A_{\rm fp, km^2} \; {\rm erg \,s^{-1}}
    \end{split}
    \label{eq:diss_pl}
\end{equation}

We note that at latitudes much in excess of $\sin(\theta) < 0.1$, the footpoint area $A_{\rm fp}$ located in the closed field line region will necessarily decrease, resulting in slightly lower dissipation luminosities. 
\par

A fraction $f_{\rm pair}$ of dissipated luminosity powers the acceleration of pairs across the gap, of which a fraction $\eta_{\rm r} \ll 1$ will be radiated as coherent radio emission. To explain the (isotropic-equivalent) radio luminosities of GLEAM-X J1627 \& GPM J1839-10 with typical radio efficiencies $10^{-6} \lesssim f_{\rm pair} \eta_{\rm r} \lesssim 10^{-3}$, relatively fast plastic flow ${\rm v}_{\rm pl} > 10 \, {\rm cm \, yr^{-1}}$, or relatively high co-latitude footpoints ($\theta_{\rm fp} \lesssim 0.1$) are required. As these sources are the first detected within a new parameter space, they are expected to be the brightest of their class. Within this model, weaker emission should exist from plastic motion from lower co-latitudes, lower plastic motion velocities, and more slowly rotating magnetars.

\subsection{Alternative Sources of Field Twist}
Thus far we have considered crustal plastic motion as the dominant driver of twist. In Appendix \ref{sect:quake_induced} we discuss radio emission from twist induced by past starquakes, which is qualitatively different due to discontinuous and sporadic twist quakes may impart. However, temperature and density gradients in the NS crust can drive the production of horizontal magnetic field components in the crust perpendicular to both the temperature and density gradients through thermoelectric effects (e.g., \citealt{1980SvA....24..425U,Blandford_1983,1986MNRAS.219..703U,1991A&A...245..331W,2024arXiv240214911G} and references therein). This mechanism may introduce a small twist at a single footpoint with a hotspot in a quasi-steady state over the lifetime of the temperature gradient, in a similar nature to crustal plastic flow. Presumably, such a temperature gradient may arise from late-time core field evolution/expulsion, as suggested in \cite{beniamini_evidence_2023}.

\par
Including thermoelectric action, but here neglecting plastic flow, the full NS magnetic induction equation reads (Eq. 7 in \citealt{2024arXiv240214911G}):
\begin{equation}
    \frac{\delta \vec{B}}{\delta t} = - c \nabla \times \bigg( \frac{(\nabla \times \vec{B}) \times \vec{B}}{4 \pi q n_{\rm e}} + \frac{c \nabla \times \vec{B}}{4 \pi \sigma_{\rm con}} - \frac{S_{\rm e}}{q} \nabla T \bigg)
\end{equation}
Where the first term expresses the Hall effect, the second expresses Ohmic dissipation and the final term is the thermal battery term of interest, where $\sigma_{\rm con}$ is the electric conductivity, and the electron entropy $S_{\rm e}$ is defined as:
\begin{equation}
    S_{\rm e} = \bigg(\frac{\pi^4}{3 n_{\rm e}}\bigg)^{1/3} \frac{k_{\rm B}^2 T}{c \hbar}
\end{equation}

Magnetic field components driven by thermoelectric effects are expected to be perpendicular to both density gradients normal to NS surface and thermal gradients parallel to surface: thus primarily in contributing to the $B_{\rm \phi}$ twist component for high co-latitude footpoints. To estimate the effect, we assume two length scales across which the $T$ and $n_{\rm e}$ vary as $L_{\rm n}$ and $L_{\rm T}$, and associate an efficiency $\eta_{\phi}$ of order unity to encapsulate the uncertainty regarding the portion of $\frac{\delta \vec{B}}{\delta t}$ which results in twisted components. Dimensionally, therefore, the twist rate due this thermoelectric action sourced by temperature gradients is:
\begin{equation}
    \dot{\Psi}_{\rm batt} = \frac{\delta \vec{B}}{\delta t} \eta_{\phi} B_0^{-1} \approx \frac{c \eta_{\phi} S_{\rm e} T}{ B_0  q L_{\rm n} L_{\rm T}} = \frac{\pi^{4/3} \eta_{\phi} k_{\rm B}^2 T^2}{3^{1/3} n_{\rm e}^{1/3} q L_{\rm n} L_{\rm T} B_0 \hbar} 
\end{equation} 
We note that here, $T$ refers to the inner crust temperature and therefore is higher than typical surface temperatures considered in this work. This temperature is likely limited by neutrino cooling. Using a typical value of the electron density from \cite{2024arXiv240214911G}, we find that:
\begin{equation}   
    \dot{\Psi}_{\rm batt} \approx 1.2 \times 10^{-12} \; {\rm s^{-1}} \: \eta_{\phi, 0} \, T_{9}^2 \, n_{\rm e, 36}^{-1/3} \, L_{\rm n, 4}^{-1} \, L_{\rm T, 4}^{-1} \, B_{0,15}^{-1} 
\end{equation}
Which is roughly $\dot{\Psi}_{\rm batt} = 4 \times 10^{-5} \, {\rm yr^{-1}}$. We can compare this to typical plastic motion values: $\dot{\Psi}_{\rm pl} \approx 10^{-4} \, {\rm v}_{\rm pl,2} \; {\rm yr^{-1}}$, to see that these effects may compete particularly for where $B < 10^{15}$ G. In the remainder of the work, we principally consider plastic motion but an equivalent $\dot{\Psi}$ may arise from thermoelectric action. One or both may play an important role in powering radio emission via inducing small local field twists in long period sources. The effect on the magnetosphere is qualitatively and geometrically similar, and thus predictions made in this work generally hold in either case.

\section{Acceleration gap microphysics}
\label{sect:microphysics}

Coherent radio emission produced close to the surface of pulsars is thought to stem from non-stationary pair creation across acceleration gaps in the Timokhin-Arons mechanism \citep{timokhin_time_2010,timokhin_current_2013}. The existence of such gaps are essential to pair creation powered by twisted magnetospheres, and they are required to maintain the time and space averaged $\Phi_{\rm pp}$ discussed in \S\ref{sect:plastic_motion}. Leptons are accelerated across a $E_{\parallel} \neq 0$ gap to form a pair formation front (which screens the gap) at a distance $\sim h_{\rm gap}$. This total distance is split into an acceleration length, the distance electrons are accelerated before emitting photons capable of pair production; and a photon propagation length, the distance these photons travel before they produce pairs. 

In the Timokhin-Arons model, pair creation is non-stationary and undergoes limit-cycle behavior: bursts of pair production are followed by quiescent phases in which pairs screen the electric field. As freshly produced pairs leave the active region, the local charge density drops below that required by both the corotation and twist and a charge-starved gap reappears. The gap grows until the energy of particles accelerated across the gap becomes large enough to initiate electron-positron cascades through either RICS or curvature photons scattering with the magnetic field, restarting the cycle.

Four features of the cascade physics are qualitatively different in high magnetic fields compared to canonical radio pulsars, leading to different collective effects and radio emission. These are: (i) radiative losses for particles may be catastrophic (and quantized, with step changes in electron energy) in the accelerating beam of plasma, particularly for the RICS channel; (ii) photon splitting cascades may suppress pair production and increase the coherence length or duty cycle for sustained and unscreened electric fields (since photons have to reach threshold to pair produce); (iii) pair production occurs largely in the ground state at threshold at low altitudes, and is not as efficient as implied by extrapolation of the semi-classical Erber formulae to high fields \citep{1966RvMP...38..626E}; (iv) non-resonant Thomson/Compton scatterings are suppressed, unlike the normal pulsar case \citep[e.g.,][]{2001ApJ...554..624H}. As many pairs are produced in the ground state, little secondary synchrotron radiation occurs during the cascade, which largely proceeds through a single channel (e.g. RICS or curvature radiation). Other than `parasitic' losses to photon splitting, in high magnetic fields, less energy is lost to the creation of lower energy photons which do not pair produce or otherwise contribute to the cascade. 

In this Section we discuss the microphysics of acceleration gaps present in low-twist magnetar magnetospheres relevant for this work. We identify the relevant gap regimes, detailing the dominant physical processes, and derive the pair luminosity (power transferred to primary accelerated pairs) for these cases. This is compared to the twist luminosity (Eq. \ref{eq:luminosity_untwisting_global}) in \S \ref{sect:comparing_luminosities} to derive the necessary conditions for the production of coherent radio emission powered by plastic motion (or possibly thermoelectric effects), and associated death lines or active regions.
 
\subsection{Relevant Scales and their Hierarchy}
For the consideration of pair cascades and gaps in highly-magnetized compact objects, one must appraise at least five relevant micro-physical length scales and their hierarchy\footnote{These also must be compared to macro-physical length scales, such as the plastic motion footpoint length scale (i.e. $\sim1$ km) or polar cap radius in the case of a rotation-powered compact object.}. In the context of magnetized compact objects \citep[see, e.g.][ for a pulsar treatment]{2001ApJ...554..624H}, they are: (1) the free acceleration length scale $\ell_{\rm acc} \sim c \gamma_{e}/\dot{\gamma}_{e, \rm acc}$, which is also (within factors of order unity) the skin depth of the relativistic plasma beam with pair multiplicity of unity; (2) the (magnetic) photon pair creation length scale $\ell_{\rm pp}$, in either above threshold for low B, pair production in high landau states or at threshold ($B\gtrsim B_{\rm cr}$, ground state pair production $\ell_{\rm pp} \sim 2 \rho_c/\varepsilon_{\gamma}$) pair creation regimes \citep{daugherty_pair_1983}; (3) the primary particle energy loss timescale in a given channel $\ell_{\rm loss} = c \gamma_{e}/\dot{\gamma}_{e}$;  (4) the length scale to emit a photon quanta of energy $\varepsilon$, $\ell_{\rm emit} = c\varepsilon_{\gamma}/\dot{\gamma}_{e}$; and (5) the photon splitting $\gamma \rightarrow \gamma \gamma$ scale $\ell_{\rm split}$, which may operate below 1 MeV and split photons before they have the opportunity to produce pairs \citep{1997ApJ...476..246H,2001ApJ...547..929B}. For standard rotationally-powered pulsars which sample the curvature channel, $\ell_{\rm loss} > \ell_{\rm acc} > \ell_{\rm pp} \gg \ell_{\rm emit}$ and $\ell_{\rm pp} \ll \ell_{\rm split}$, which implies particles are in the free acceleration regime (i.e. never attain radiation reaction) before gap termination which is largely set by the minimized free acceleration length scale. 
\par
These processes are discussed in detail in the following subsections, and are summarized in Fig. \ref{fig:lengthscales_fud} (see also Figs \ref{fig:lengthscales_higherrhocc}, \ref{fig:lengthscales_lowB}, \ref{fig:lengthscales_small_rho_c}). The figures can be read from bottom to top as follows. Blue lines follow the acceleration of primary electrons and positrons to Lorentz factors ($\gamma_{\rm e}$, x-axis) as a function of distance traversed (y-axis), for various mild twists. If these acceleration profiles intersect with solid magenta curves, RICS scattering dominates resulting in high-energy photons which (following the intersection vertically from the intersection point) may split (dashed) or pair produce (dotted) depending on the photon polarization. If however the particle acceleration paths `miss' the magenta curved lines, they continue to accelerate until they intersect with dashed/dotted green lines, referring curvature photon splitting or pair production respectively. The solid green line refers to the curvature loss length scale (i.e. radiation-reaction) which is not generally important, and the dot-dash line is the length scale to emit curvature photons. These intersections and the summations of relevant length scales enables gap height calculations in either regime, for various magnetospheric parameters.

\subsection{Gap Electrodynamics and Free Acceleration}
The current demanded globally sets the physical scales the pair cascade gap must satisfy locally \citep{timokhin_time_2010,timokhin_current_2013}. Let us assume global current density required by an effective twist $\Delta \Psi$ is:
\begin{equation}
  \label{eq:j_twist}
  j_{\rm{twist}}=\frac{c}{4\pi} \left| \boldsymbol{\nabla}\times\boldsymbol{B}  \right|\sim
  \frac{c}{4\pi}\frac{B}{R_{*}}\,f(\theta_{\rm 0}) \,\Delta\Psi\,
\end{equation}
where $f(\theta_{\rm 0})>0$ is a geometric factor associated with the footpoint co-latitude of the twisted flux tubes. For axisymmetric dipolar force-free twists, $f(\theta_{\rm 0}) \sim \sin^{2}(\theta_{\rm fp})$ which we will adopt henceforth for simplicity. Note that this specific form will not hold in generality for arbitrary crust shear and multipolar fields, including potential locally helical, kinked or fluted geometries constructions one may propose \cite[see, e.g.,][]{parfrey_dynamics_2013,2023arXiv231204620R}. One can, however, subsume these geometric and topological considerations into the effective twist  $\Delta \Psi$, which is otherwise $\approx \Psi_{\rm crit}$.

In a one-dimensional gap construct, $d E_{\rm gap}/dt \sim -4 \pi (j- j_{\rm twist})/c$ from the induction Maxwell equation where $j$ is realized current density locally and $j_{\rm twist}$ is the one demanded by Eq.~(\ref{eq:j_twist}). 
The requisite charge density is
\begin{equation}
\rho_{\rm twist} \sim  j_{\rm twist}/c.
\label{eq:rho_twist}
\end{equation}
Omitting factors of unity, it follows that the gap electric field is $d E_{\rm gap}/d\ell \sim 4 \pi (\rho-\rho_{\rm twist})$. Now since $(\rho - \rho_{\rm twist}) \sim \rho_{\rm twist}$ to leading order\footnote{The departure from the global current/charge demand is defined by the $\xi = \rho/\rho_{\rm twist}$ parameter in \cite{timokhin_current_2013}. In fact, a spectrum of this charge density is realized as the gaps are intrinsically non-stationary and involves limit cycle behavior. This is conjectured to result in broadband power-law radio spectra \citep{philippov_origin_2020}.}, it follows the required vacuum gap electric field is

\begin{equation}
E_{\rm gap} \sim 4 \pi \rho_{\rm twist} \ell_{\rm acc} \sim  \frac{B}{R_{*}}\,\sin^{2}(\theta_{\rm fp}) \,\Delta\Psi\, \ell_{\rm acc} 
\label{Egap}
\end{equation}
where $\ell_{\rm acc}$ is the free acceleration length to Lorentz factor $\gamma_{e}$ for a primary electron/positron prior to gap termination (in the ultrarelativistic limit), i.e., $ m_{e} c \,d(\beta_{e}\gamma_{e})/dt  \approx m_{e} c^{2} d\gamma_e/d\ell_{\rm acc}= q E_{\rm gap}$:
\begin{equation}
\gamma_{e, \rm acc} \equiv \frac{q \, B}{2 R_{*} m_e c^2 }\sin^{2}(\theta_{\rm fp})\,\Delta\Psi\, \ell_{\rm acc}^2 \quad , \quad \ell_{\rm acc} \sim \left( \frac{2 R_{*} m_{e} c^{2} \gamma_{e}}{q B \sin^{2}(\theta_{\rm fp}) \Delta \Psi} \right)^{1/2}.
\label{eq:gammaacc}
\end{equation}
Consequently, the free acceleration length scale increases as $\ell_{\rm acc} \propto \sqrt \gamma_e$ \citep[see also the ultrarelativistic $\xi >1$ limit of Appendix B in ][]{timokhin_current_2013}. We note that the skin depth (and Debye length) for a relativistic beam $\ell_{\rm skin} =c/\omega_{e}$, where $\omega_{e} \sim \sqrt{4 \pi \rho_{\rm twist} q/(\gamma_{e} m_{e})}$, is by construction equivalent to the free acceleration length scale, $\ell_{\rm skin} \sim \ell_{\rm acc}$ (for primaries, prior to the pair formation front).

\subsection{Relevant Scales for the Curvature Radiation Channel}

 \def\lambar{\lambda\llap {--}}

We now detail the relevant lengthscales in the curvature photon pair production channel for our problem. Classically, curvature radiation is the low pitch-angle limit of synchrotron emission in a curved field \citep[e.g.,][]{2015AJ....149...33K}. It is broadband and peaks at a characteristic dimensionless photon energy, $\varepsilon_{\rm \gamma, CR}  \sim 3 (\lambar/\rho_c)  \gamma_e^3/2$ where $\lambar \equiv \hbar/(m_e c)$ is the reduced Compton wavelength and $\rho_c$ is the local field curvature radius\footnote{We omit departures due to the quantum regime of curvature radiation \citep[e.g.,][]{2002MNRAS.335...99H,2006RPPh...69.2631H,2017PhRvD..95j5008V} -- these departures are small in the ultrarelativistic parallel momentum and high field limit where particles largely occupy the ground perpendicular state, and the emitted curvature dimensionless photon energy is large compared to both $\lambar/\rho_{c}$ and the energy of perpendicular momenta (Landau) states. Electron recoil (and large gamma-ray quanta) effects are negligible provided that $\varepsilon_{\rm \gamma, CR} /\gamma_{e} \ll 1$ or $\gamma_{e}^{2} \lambar/\rho_{c} \ll 1$ which implies $\gamma_{e} \ll 4\times 10^{8} \rho_{c,7}^{1/2}$ \citep{1991ApJ...371..265N}.}, noting that mild twists (see Eq. \ref{eq:critical_twist}) contribute negligibly to $\rho_{\rm c}$. The single particle classical energy loss rate is $\dot{\gamma}_{\rm CR} = (2/3) \alpha_{\rm f} \lambar c \gamma_{e}^{4}/\rho_{c}^{2}$ where $\alpha_{\rm f} \approx 1/137$ is the fine structure constant\footnote{CR and RICS subscripts denote curvature radiation and resonant inverse-Compton scattering quantities respectively.}.

The energy loss length scale $\ell_{\rm loss} = (3/2) \rho_{c}^{2} / (\alpha_{\rm f} \lambar \gamma_{e}^{3})$ while the ground state pair production length scale is $ \ell_{\rm pp} \sim 2 \rho_c/\varepsilon_{\gamma} \sim (4/3) \rho_{c}^{2} / (\lambar \gamma_e^{3})$. For the curvature channel for threshold pair creation in the ground state, regardless of the driver (twists or corotation), it is apparent that,
\begin{equation}
\frac{\ell_{\rm pp}}{\ell_{\rm loss, CR}} \sim \alpha_{\rm f} \ll 1
\label{eq:ellgamm_ellloss}
\end{equation}
which implies radiation reaction is unlikely to be realized prior to gap termination in the CR channel in high fields.

\begin{figure}
  \centering
{\includegraphics[width=.5\textwidth]{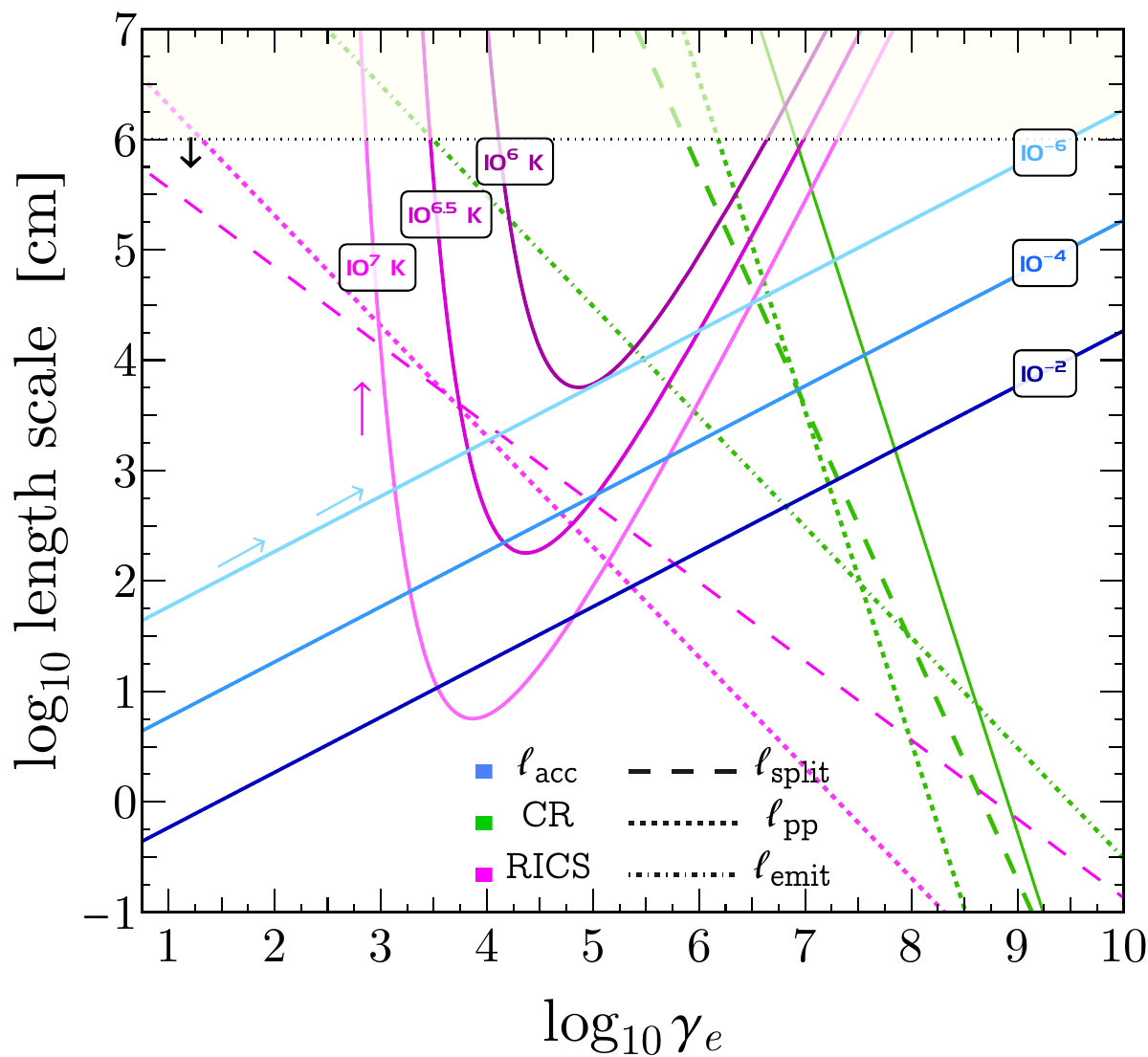}}
\caption{Schematic length scale plot for relevant processes as detailed in the text. Blue lines (dark to light) are acceleration length scales $\ell_{\rm acc}$ for a vacuum gap with $\Delta \Psi =\{10^{-2},10^{-4},10^{-6}\}$. Solid magneta lines (dark to light) are RICS cooling lengthscales for $T=\{10^6,10^{6.5},10^7\}$ K. Long dashed magenta is the splitting lengthscale for RICS photons. Short-dashed magenta is RICS photon mean free path for at-threshold pair creation. Green is for the curvature channel. Dash-dotted green is curvature energy loss length scale. Short-dashed green is CR photon mean free path for at-threshold pair creation. Solid green is $\ell_{\rm loss, CR}$. Dotted green is the mean free path for emitting a photon. Long dashed is the photon splitting length scale for CR photons.
Parameters: $B=10^{15}$ G,  $\rho_c = 10^7$ cm,  $\theta_0 = 0.1$. As $B$ is assumed constant and RICS curves are computed for interactions near the surface, plotted quantities are not valid for length scales $R_* \gtrsim 10^6$ cm.
}
\label{fig:lengthscales_fud}
\end{figure}

The length scale for the accelerated primary to emit a curvature gamma-ray quanta of energy $\varepsilon_{\gamma}$ is $\ell_{\rm emit} = c\varepsilon_{\gamma}/\dot{\gamma}_{e} \sim (9/4) \rho_{c}/(\alpha_{\rm f}\gamma_{e})$. For such emission length to impact the gap structure, it must be large compared to $\ell_{\rm pp}$ and $\ell_{\rm acc}$. However, the ratio,
\begin{equation}
\frac{\ell_{\rm emit}}{\ell_{\rm pp}} = \frac{27}{16} \frac{\gamma_{e}^{2}}{\alpha_{\rm f}} \left(\frac{\lambar}{\rho_{c}}\right) \sim \alpha_{\rm f}^{-1} \left(\frac{\varepsilon_{\gamma}}{\gamma_{e}} \right)
\end{equation}
is small except in the extreme radiation reaction limit, which, as Eq.~(\ref{eq:ellgamm_ellloss}) implies, will generally not be realized. These are plotted in Fig. \ref{fig:lengthscales_fud} and \ref{fig:lengthscales_higherrhocc}--\ref{fig:lengthscales_small_rho_c}.

\subsection{Relevant Scales for the RICS Channel}
\label{sec:rics}
Resonant inverse-Compton scattering is a scattering process of lower energy photons by relativistic electrons in a strong magnetic field. It is effectively cyclotron absorption followed by immediate remission \citep[e.g,][]{1979PhRvD..19.2868H,1986ApJ...309..362D,1986PhRvD..34..440B,2000ApJ...540..907G,2014PhRvD..90d3014G,2016PhRvD..93j5003M}, as the lifetime of Landau states are extremely short in supercritical magnetic fields realized in magnetars. Unlike classical non-magnetic Compton scattering, RICS is effectively first order in $\alpha_{\rm f}$ and can exceed the Thomson cross section by $\sim 2-3$ orders of magnitude at resonance. For ultrarelativistic electrons/positrons, angles of initial photons are nearly parallel to the electron motion (and magnetic field) in the electron rest frame -- in this limit, the cross section exhibits a single resonance at the cyclotron fundamental. The resonance is finite capped by an cyclotron lifetime from intermediate virtual Landau and spin electron states \citep[e.g.,][]{2005ApJ...630..430B}.

For such ultrarelativistic electrons, resonant scatterings are always accessed in a thermal bath with a peak in the electron energy loss rate (and scattering rate) at $\gamma \Theta \sim {\cal B}$ where $\Theta = k_{b} T/(m_{e} c^{2})$ is the dimensionless temperature and ${\cal B} = B/B_{\rm cr}$. A nuance is that the correct relativistic treatment of the spin dependence of the cross section and RICS must be formulated in the Sokolov-Ternov basis \citep[see discussion in][]{2014PhRvD..90d3014G} rather than the magnetic Thompson or Johnson-Lippmann formulations \citep{1949PhRv...76..828J,1990ApJ...360..197D,1995ApJ...445..736S}, inappropriate for the supercritical regime relevant for regions close to the surface of a magnetar. \cite{baring_cooling_2011} computed the energy loss rates in a fully relativistic treatment in the Sokolov \& Ternov basis. The lengthscale found by \cite{baring_cooling_2011} is $\ell_{\rm loss,RICS} \sim 2 \lambar {\cal B} /(\alpha_{\rm f} \Theta^{3})$. This supercritical regime is suitable for this work, as twist dissipation occurs close the magnetar surface. The RICS cooling rate, on the surface of the NS, retaining its Lorentz factor dependence (with $\beta_e = \sqrt{\gamma_e^2-1}/\gamma_e$) is given by the asymptotic resonant approximation of Eq. (50) in \cite{baring_cooling_2011}:

\begin{equation}
\dot{\gamma}_{\rm RICS} \approx \frac{3}{8\pi} \frac{\sigma_T c}{\lambar^3} \frac{\Theta}{\gamma_e \Gamma} {\cal R}({\cal B}) \left( g_+ - g_- \right) \quad, \quad \gamma_{e} \Theta \gtrsim {\cal B}
\label{eq:RICS_BWG11}
\end{equation}
where ${\cal R}(B)$ is an involved function of ${\cal B}$ detailed in the Appendices of \cite{baring_cooling_2011}, $\Gamma$ is the relativistic spin-averaged width (also a function of ${\cal B}$), $\sigma_T$ is the Thomson cross section, and 
\begin{equation}
g_\pm = -\ln (1-\exp(-\chi_\pm))
\end{equation}
with 
\begin{equation}
   \chi_\pm = \frac{{\cal B}}{\gamma_e \Theta (1 + \beta_e \mu_\mp) }
\end{equation}
Here $\mu_\pm$ are the angle cosines of the incoming photon distribution (in the lab frame), where for ingoing electrons/positrons on the surface $\{\mu_-,\mu_+\} =\{0, 1\} $ and for outgoing ones $\{\mu_-,\mu_+\} =\{-1, 0\} $. For arbitrary interaction points at higher altitudes, the form of these angles is more involved. For outgoing electrons close to the surface, Eq.~(\ref{eq:RICS_BWG11}) peaks at $\dot{\gamma}_{\rm RICS, peak} \sim \alpha_{\rm f}c \Theta^{2}/(2 \lambar)$ when ${\cal B} \gg 1$; this peak is realized when $\gamma_{e} \Theta \sim {\cal B}$. In a gap, as the cascade ensues, electrons and positrons will in general move in opposite directions (i.e. ingoing versus outgoing, sampling different target photon fields). This will result in different scattering and loss rates, modifying the collective effects and gap dynamics beyond the scope of this work. 

The width $\Gamma$ here is for the ground-state to first excited state transition and is spin-averaged. It may be computed from the approximation of \cite{2005ApJ...630..430B} retaining terms only a few ($\sim 6$) orders in the series
\begin{equation}
\Gamma \equiv \overline{\Gamma}_{1\rightarrow0} = \frac{\alpha_{\rm f} {\cal B}}{{\cal E}_1} I_1({\cal B}) \quad , \quad  I_1 \approx \sum_{k=0}^{\infty} \frac{(-1)^k}{k!} \left[ Q_k (z) - z Q_{k+1} (z)  \right]
\end{equation}
where $Q_k$ are Legendre functions of the second kind, $z = 1 + 1/{\cal B}$, and ${{\cal E}_1} \approx 1 + {\cal B}$ at resonance.

For a thermal bath, RICS photons form a flat power-law with significant emission anisotropy \citep{wadiasingh_resonant_2018}. Scattered photons are narrowly beamed in the forward direction. The maximum energy of photons attained for head-on collisions in the forward direction (tangent to the local magnetic field) is
\begin{equation}
\varepsilon_{\rm RICS,max} \approx \frac{ \gamma_{e}(1+\beta_e) {\cal B}}{1+2{\cal B}}
\label{eq:ergmax}
\end{equation}
%\ 
and the luminosity (energy loss) is dominated by these photons given the intrinsically hard spectrum of scatterings \citep{wadiasingh_resonant_2018}. In the relativistic regime ${\cal B}\gtrsim 1$, it follows $\varepsilon_{\rm RICS,max} \approx \gamma_{e}$, i.e. electron recoil is large and energy losses are catastrophic. Thus, by nature RICS exhibits $\ell_{\rm emit} \approx \ell_{\rm loss}$, similar to the Klein-Nishina domain of Compton scattering.

Non-resonant scatterings, pertinent at lower $\gamma_e$ below the resonance cool electron/positrons as $\dot{\gamma}_e \propto \gamma_e^4 \Theta^6 {\cal B}^{-2}$ or $\ell_{\rm loss} \propto \gamma_e^{-3} \Theta^{-6} {\cal B}^{2} $ \citep{baring_cooling_2011}. In lower-field pulsars non-resonant scatterings can be important \citep{2001ApJ...554..624H}, however the suppression by ${\cal B}^2$, and lower temperatures and photon densities in ULMPs means these scattering are subdominant and are not considered in this work.

\subsection{Photon Splitting}

In strong fields ${\cal B} \gtrsim 0.1$ relevant here, photon splitting $\gamma \rightarrow \gamma\gamma$ \citep{1971AnPhy..67..599A,1979JPhA...12.2187S,1991A&A...249..581B,1997ApJ...482..372B} could influence the gap physics. It is a third-order QED process forbidden when $B=0$, but is permissible in when $B>0$ from virtual pairs radiating when interacting with the field with a splitting probability or rate $\propto \alpha_{\rm f}^3 \varepsilon^{5} {\cal B}^{6}$. Photon splitting may operate below 1 MeV ($\varepsilon < 2$) thereby quenching pair production, acting as a mechanism to increase the gap height \citep{1983SvAL....9..212U,1997ApJ...476..246H,2001ApJ...547..929B}. In regimes of large curvature radius, it may out-compete pair production significantly, overruling the tenability of large toriodal current demands in twists approaching the split monopole force-free solution \citep[see discussion in][]{hu_high-energy_2022} with large local curvature radii.

In the weakly dispersive limit ${\cal B} \ll 1$ of vaccum polarization, only the extraordinary photon state (electric field vector $\perp$ to momentum and B) may split into two ordinary modes, $\perp \rightarrow \parallel \parallel$ limiting splitting cascades. However, in stronger fields where nonlinear effects come into play, and more modes could split. By charge-parity symmetry, splitting modes which could be permitted include $\parallel \rightarrow \perp \parallel$ and $\perp \rightarrow \perp \perp$. The allowance of these additional modes when ${\cal B}$ is high permits a full splitting cascade to develop, significantly impacting pair production and the gap \citep{2001ApJ...547..929B}. As a scattering process, RICS produces mostly photons in the $\perp$ state \citep{2007Ap&SS.308..109B,wadiasingh_resonant_2018} and as such its photons (at the highest energies) can readily split \citep{2019BAAS...51c.292W}. Likewise, curvature radiation also mostly produces photons in the $\perp$ state.

For footpoints close to the magnetic pole, \cite{2019MNRAS.486.3327H} computed polarization averaged spitting lengths $\ell_{\rm split}$ (Eq. (40) in that work) which we adopt here,
\begin{equation}
\ell_{\rm split, ave} \approx \left(\frac{2 {\cal M}_1^2}{3{\cal M}_1^2+{\cal M}_2^2}\right)^{1/7} \ell_{\perp \rightarrow \parallel \parallel}    
\end{equation}
where on the surface of the star
\begin{equation}
\ell_{\perp \rightarrow \parallel \parallel} \approx R_* \left(\frac{573440\pi^2}{243\alpha_{\rm f}^3} \frac{\lambar}{{\cal M}_1^2 {\cal B}^6 \theta_0^6 R_*} \right)^{1/7} \varepsilon^{-5/7}
\end{equation}
Here ${\cal M}_{1,2}$ are splitting amplitude integrals which depend solely on ${\cal B}$, and are detailed in Eq.~(21)--(22) of \cite{2019MNRAS.486.3327H}, first derived by \cite{1971AnPhy..67..599A}.

\subsection{RICS Acceleration Gaps}

The RICS gap will be set by the Lorentz factor which primaries encounter the energy-loss ``wall'' and enter radiation reaction evolution. For RICS to be of relevance, $\ell_{\rm acc} = \ell_{\rm loss, RICS}$, i.e. the interaction of the blue and magenta curves in Fig. \ref{fig:lengthscales_fud},  must be realized. When primary electrons/positrons encounter this cooling ``wall", they lose energy catastrophically; the beam will then comprise of a piled-up population in momentum space just below the resonant Lorentz factor. The collective plasma dynamics and other implications of this inverted population is interesting but beyond the scope of this work. As the target photons constitute a thermal photon distribution of a variety of incoming angles, this resonant wall is extensive and not fine-tuned \citep[see discussion in ][]{baring_cooling_2011}. Free (re)acceleration of lower energy primaries will continue until termination of gap, which will largely be set by $\ell_{\rm pp}$\footnote{\cite{2000ApJ...531L.135Z} inappropriately assume gap height for RICS is always set by radiation reaction, rather than $\ell_{\rm pp}$ \citep[see also discussion in][]{2007MNRAS.382.1833M}.}. As $\ell_{\rm pp} \gg \ell_{\rm acc}$ after encountering the resonant wall, this sets the gap height. In fact, the pair multiplicity here (pseudo-multiplicity, given our stationary gap approximation) is approximately ${\cal M} \sim \ell_{\rm loss, RICS}/\ell_{\rm pp}$ since subsequent generations of pairs are mostly created in the ground state and $\ell_{\rm emit} \approx \ell_{\rm loss}$. In general, this multiplicity is expected to be lower than the CR channel, and as such, the radio emission could be intrinsically more narrow band in the RICS channel.

Note that \cite{2007MNRAS.382.1833M} considered RICS gaps for rotation-powered pulsars and found longer periods were favored for cascades (as we do here, see below). However, they did not use the correct relativistic forms of the angular-dependent scattering kinematics or the appropriate RICS cross section, and did not capture the kinematics where resonant scatterings are first accessed in the Wien tail of the target photon distribution. The analysis here supersedes it for both rotation and magnetically (e.g. plastic motion or thermoelectric) powered cases.

The intersection Lorentz factor $\gamma_{\rm int}$ may be found, via approximation to the Wien tail of the RICS cooling rate from \cite{baring_cooling_2011},  as a solution of a transcendental equation, expressible in terms of the Lambert W function $W_{-1}(z)$,
\begin{equation}
\gamma_{\rm int} \approx \frac{2 \cal{B}}{3 \Theta |W_{-1}({\cal Z})|}
\label{eq:gam_rics_int}
\end{equation}
for outgoing electrons $\{\mu_{-},\mu_{+}\} = \{-1,0\}$, where ${\cal Z}$ is a dimensionless argument 
\begin{equation}
{\cal Z}_{\rm twists} = -\frac{2^{2/3}}{3} \left(\frac{   {\cal B}^{4} \Delta \Psi \sin^{2}(\theta_{\rm fp}) \Gamma^{2} }{\alpha_{\rm f}^{4}  \Theta^{5} R({\cal B})^{2}  }\right)^{1/3} \left(\frac{\lambar}{R_{*}}\right)^{1/3}.
\label{eq:arg_W}
\end{equation}
For corotation or rotationally-driven accelerating fields, one may adopt the identification $4 \pi R_* /(c \sin^{2}(\theta_{\rm fp}) \Delta \Psi) \equiv P_{\rm crit}$ in Eq.~(\ref{eq:arg_W}), corresponding to equivalent twist and rotational charge requirements.  This yields an analogous expression,
\begin{equation}
{\cal Z}_{\rm rot} = -\frac{2}{3} \left(\frac{   {\cal B}^{4}  \Gamma^{2} }{\alpha_{\rm f}^{4}  \Theta^{5} R({\cal B})^{2}  }\right)^{1/3} \left(\frac{2 \pi \lambar}{c P_{\rm crit}}\right)^{1/3}.
\label{eq:arg_Wrot}
\end{equation}
where note that $2 \pi/(c P_{\rm crit}) = R_L$ is the light cylinder of the rotator. Similar corresponding expressions for ingoing electrons/positrons are readily derivable but not shown here.

The function $W_{-1}(z)$ is negative and real-valued provided that $z \geq -1/e$ with $W_{-1}(-1/e) = -1$. For this limiting value, the acceleration curve narrowly intersects the cooling curve at cusped base of the magenta curves in Fig.~\ref{fig:lengthscales_fud} (see also Figs \ref{fig:lengthscales_higherrhocc} -- \ref{fig:lengthscales_small_rho_c} which consider different magnetospheric regions and conditions). This constitutes one boundary of for the regime of validity of RICS gaps. Numerical evaluation yields $W_{-1}({\cal Z}) \sim -10 $ to $-3$ over a wide parameter range of interest, as its dependence is logarithmic. Consequently, the RICS gap height is $h_{\rm gap, RICS} \sim \ell_{\rm pp} \sim 2 \rho_{c} /\varepsilon_{\rm RICS,max}$ evaluated at $\gamma_{\rm int}$. 
Thus, to zeroth order\footnote{For $B \gtrsim 10^{13}$ G, the approximation in Eq. \ref{eq:zero_order_Rics_gap} differs from the full calculation (Eq. \ref{eq:hgap_rics}) by only $W_{-1}({\cal Z})$, roughly a factor $\sim 6$.}, the RICS gap height when ${\cal B} \gg1$ is:
\begin{equation}
    h_{\rm gap, RICS} \sim 30 \rho_{c} \Theta/{\cal B}
    \label{eq:zero_order_Rics_gap}
\end{equation}

More accurately, the total gap size is $h_{\rm gap, RICS} \sim \ell_{\rm loss, RICS} + \ell_{\rm pp}$ subject to the constraints (1) a viable intersection $\ell_{\rm loss, RICS} \approx \ell_{\rm acc}$ (principally in the Wien tail), and (2) the pair luminosity bounded by the dissipation power, (3) gap scale much less than $R_*$ and (4) the pseudo pair multiplicity $\ell_{\rm pp}/\ell_{\rm int}$ (see below) greater than unity. These four requirements define the boundaries of `active zones' for pair production via RICS in $P-B$ or $P-\dot{P}$ space.

For a more accurate gap estimate, substitution of Eq.~(\ref{eq:gam_rics_int}) into Eq.~(\ref{eq:RICS_BWG11}) readily gives $\ell_{\rm loss, RICS}$ approximated to the Wien tail, or more simply by the substitution into $\ell_{\rm acc}$ in Eq.~(\ref{eq:gammaacc}). This yields,
\begin{eqnarray}
\ell_{\rm int, twists} &\approx& 2 \left[-\frac{\lambar R_*}{3 \Delta \Psi \sin^{2}(\theta_{\rm fp}) \Theta \left| W_{-1}({\cal Z}_{\rm twist}) \right|} \right]^{1/2} \\
\ell_{\rm int, rot} &\approx&  \left[\frac{\lambar c P_{\rm crit}}{3 \pi \Theta \left| W_{-1}({\cal Z}_{\rm rot})\right|} \right]^{1/2}
\end{eqnarray}
for the two cases, twist or corotationally-driven gaps. Then the gap height $\ell_{\rm int} + \ell_{\rm pp}$ is,
\begin{equation}
    h_{\rm gap, RICS} \approx \ell_{\rm int} + 3 \rho_c \Theta \frac{1+2{\cal B}}{2 {\cal B}^2} |W_{-1}({\cal Z})|
    \label{eq:hgap_rics}
\end{equation}
using expression Eq.~(\ref{eq:ergmax}) for the photon energy. 

A viable intersection corresponds to the bound  $z \geq -1/e$ for $W_{-1}(z)$, which implies,
\begin{equation}
\frac{4 \pi R_*}{c \sin^{2}(\theta_{\rm fp}) \Delta \Psi} =  P_{\rm crit} \gtrsim \frac{16 {\cal B}^4 \pi e^3}{27 \alpha_{\rm f}^4 \Theta^5} \frac{\lambar}{c} \left(\frac{\Gamma}{R({\cal B})} \right)^2
\label{eq:Lambertcondition}
\end{equation}
In the limit of ${\cal B} \gg1$, $R({\cal B}) \rightarrow  (1 - 2/e) {\cal B}^2 $ and $\Gamma \rightarrow \alpha_{\rm f} (1-1/e)$. It is readily apparent that in this limit (${\cal B} \gg1$), the ${\cal B}$ dependence of Eq.~(\ref{eq:Lambertcondition}) vanishes, with 
\begin{equation}
P_{\rm crit} \gtrsim 214 \left(\frac{\lambar}{c \alpha_{\rm f}^2 \Theta^5}\right) \approx  120 \, \, T_{6.5}^{-5} \quad \rm sec \quad ({\cal B} \gg 1)
\label{eq:Lambertcondition2}
\end{equation}
yielding a minimum period that solely and strongly depends on surface temperature.
For RICS to be of relevance in screening of the gap, ${\cal M} \gg 1$. This RICS pseudo pair multiplicity scales as ${\cal M} \sim \ell_{\rm pp}/\ell_{\rm int} \propto \rho_c \Theta^{3/2} {\cal B}^{-1}$ to leading order when  ${\cal B} \gg1$, i.e. pair cascades and gaps in polar locales with large curvature radii and hotter objects with middling field strengths will be determined by the RICS rather than CR channel. Explicitly,
\begin{eqnarray}
{\cal M}_{\rm twists} &\sim& \frac{3 \sqrt{3}(1+2{\cal B}) \Theta^{3/2}}{4 {\cal B}^2 } \frac{\rho_c \sqrt{\Delta \Psi \sin^{2}(\theta_{\rm fp})}}{\sqrt{\lambar R_*}} |W_{-1} ({\cal Z}_{\rm twist})|^{3/2} \nonumber \\
{\cal M}_{\rm rot} &\sim&  \frac{3 \sqrt{3 \pi}(1+2{\cal B}) \Theta^{3/2}}{2 {\cal B}^2 } \frac{\rho_c }{\sqrt{\lambar c P_{\rm crit}}} |W_{-1} ({\cal Z}_{\rm rot})|^{3/2}.
\label{eq:multiplicity_requirement}
\end{eqnarray}
Note that Eqs.~\ref{eq:hgap_rics}--\ref{eq:multiplicity_requirement} neglect photon splitting suppression of pair cascades in high $B \sim 10^{15}$ G fields. As it occurs for photons below pair threshold, this could reduce the pair multiplicity drastically, and would likely reduce the true radio luminosity well below the gap luminosity (see below), i.e. driving $\eta_{\rm r} \ll 1$. From Figures \ref{fig:lengthscales_fud} and \ref{fig:lengthscales_higherrhocc}--\ref{fig:lengthscales_small_rho_c} it is apparent that for some parameter regimes, the splitting length is shorter than $\ell_{\rm pp}$ for the highest energy photons in the RICS channel but not the CR channel. This would suggest RICS may be more favored in moderate fields $B \sim 10^{14} - 10^{13}$~G where $\ell_{\rm pp} < \ell_{\rm split}$ and pair creation slightly above threshold, see Fig.~\ref{fig:lengthscales_lowB}. A detailed Monte Carlo cascade treatment is required to assess the pair yield with splitting, and is beyond the scope of this work.
 
\subsection{$e^+e^-$ Pair Luminosities, Deathlines, and Active Regions}
\label{sect:pair_luminosity}

In the following we calculate the gap height and pair luminosity as a function of NS parameters for curvature radiation and RICS channels. We assume that the magnetospheric twist is close to the critical value outlined in Eq. \ref{eq:critical_twist}. The gap voltage $\Phi_{\rm gap}$ may be higher than the time-averaged pair production voltage $\Phi_{\rm pp}$, as the pair creation initiates and ceases on short timescales, and had a `time on' duty cycle less than $100 \%$. The pair luminosity, $L_{\rm e^+e^-}$, is the total power in accelerated primary pairs and constitutes the maximum realisable persistent radio luminosity:

\begin{equation}
L_{\rm e^+e^-} \sim q \Phi_{\rm gap} \dot{N}_{\rm twist}  
\end{equation}

where the gap voltage is  
\begin{equation}
 \Phi_{\rm gap} \sim  2 \pi \rho_{\rm twist} h_{\rm gap}^2
 \end{equation}
 while 
 \begin{equation}
 q \dot{N}_{\rm twist} \sim \rho_{\rm twist} A_{\rm fp} c
\end{equation}
Thus the twist pair luminosity is
 \begin{equation}
L_{\rm e^+e^-} \sim 2 \pi \rho_{\rm twist}^2 h_{\rm gap}^2 A_{\rm fp} c
\label{twistpairlum1}
\end{equation}
Where $\rho_{\rm twist}$ is given by Eqs. \ref{eq:j_twist} \& \ref{eq:rho_twist}.

Some caveats for this $L_{\rm e^+e^-}$ construct are: this luminosity is only for particles accelerated in the gap, i.e. primaries and the first generation of pairs. The gap is assumed to be screened beyond the scale $h_{\rm gap} \sim \ell_{\rm pp} + \ell_{\rm acc}$. In this approximation, secondary and subsequent pair generations do not tap much energy from the vacuum gap electric field Eq.~\ref{Egap} over lengthscales much larger than the computed $h_{\rm gap}$. This approximation is manifest in the equivalent field-energy-density derivation of Eq.~\ref{twistpairlum1}: $L_{\rm e^+e^-} \sim E_{\rm \parallel}^2/(8 \pi) \times A_{\rm fp} h_{\rm gap} \times c/h_{\rm gap}$. 

\subsubsection{Curvature radiation pairs} 

For at-threshold pair creation, $\varepsilon_{\rm CR} \ell_\gamma /\rho_c = 2$. The total gap size is $\ell_{\rm tot} = \ell_{\rm acc} + \ell_\gamma$ minimized for variations in $\ell_{\rm acc}$. From this minimization procedure (as in \citealt{1998ApJ...508..328H,timokhin_polar_2015,cooper_pulsar_2023}), and $h_{\rm gap, CR} \sim 2\ell_{\rm tot}$, we obtain the threshold gap:

\begin{equation}
\begin{split}
h_{\rm gap, CR} &= \bigg(\frac{4}{\sin^2(\theta_{\rm fp}) \Delta \Psi}\bigg)^{3/7} \left(\frac{B_{\rm cr}}{B}\right)^{3/7} \left(\lambar^2 \rho_{c}^{2} R_{*}^{3} \right)^{1/7} \\
&\approx 2 \times 10^{4} \; {\rm cm} \; R_{*, 6}^{3/7} \, P_{\rm NS, 3}^{3/7} \, B_{15}^{-3/7} \, \sin(\theta_{\rm fp, 0})^{-2/7}
\label{eq:h_gap_curv}
\end{split}
\end{equation}
Where we have assumed $\Delta \Psi = \Psi_{\rm crit}$ as in Eq. \ref{eq:critical_twist} and a dipolar magnetic field such that $\rho_{\rm c} \approx R_{*}/\sin(\theta_{\rm fp})$. Based on the above we obtain the gap voltage:

\begin{equation}
\begin{split}
\frac{q \Phi_{\rm gap, CR}}{m_e c^{2}} &= \frac{2^{19/7} q}{m_{\rm e} c^2 R_{*}} \big( \sin^2(\theta_{\rm fp}) \Delta \Psi\big)^{1/7} \big(B B_{\rm cr}^6\big)^{1/7} \bigg( 
\frac{\lambar^2 R_{*}^5}{\sin^2(\theta_{\rm fp})}\bigg)^{2/7} \\
% &= \frac{2^{19/7} q}{m_{\rm e} c^2 R_{*}} \big( \frac{8 \pi \kappa R_{*}}{c P}\big)^{1/7} \big(B B_{\rm cr}^6\big)^{1/7} \bigg( 
% \frac{\lambar^2 R_{*}^5}{\sin^2(\theta_{\rm fp})}\bigg)^{2/7} \\
&\approx 10^{12} \; {\rm eV} \, R_{*, 6}^{3/7} \, P_{\rm NS, 3}^{-1/7} \, B_{15}^{1/7} \, \sin(\theta_{\rm fp, 0})^{-4/7}
\end{split}
\end{equation}

Using Eqs. \ref{eq:j_twist}, \ref{twistpairlum1} \& \ref{eq:h_gap_curv}, the pair luminosity is thus,

\begin{equation}
\begin{split}
L_{\rm e^+e^-, CR} &\sim 2 \pi A c \rho_{\rm twist}^2 h_{\rm gap, CR}^2 \\
&= \bigg(\frac{A_{\rm fp} c}{2^{9/7} \pi R_{*}^2}\bigg) \big(\sin^2(\theta_{\rm fp}) \Delta \Psi\big)^{8/7} \big(B^4 B_{\rm cr}^3\big)^{2/7} \bigg(\frac{\lambar^2 R_{*}^5}{\sin^2(\theta_{\rm fp})}\bigg)^{2/7} \\
&\approx 10^{32} \; {\rm erg \, s^{-1}} \;  A_{\rm fp, 10}\, R_{*, 6}^{-4/7}  \, P_{\rm NS, 3}^{-8/7} \, B_{15}^{8/7} \, \sin(\theta_{\rm fp, 0})^{-4/7}
\label{eq:twist_luminosity_curv_free}
\end{split}
\end{equation}
This luminosity estimate corresponds to only those primaries produced required to screen the gap, as most of the pairs are produced above the gap \citep{timokhin_polar_2015}. We show the fiducial radio luminosity $L_{\rm r} = \eta_{\rm r} L_{\rm e^+e^-, CR}$ in Fig. \ref{fig:radio_luminosity_curv}. Rearranging Eq. \ref{eq:twist_luminosity_curv_free} in terms of observable properties allows new observations to constrain key parameters:
\begin{equation}
\begin{split}
L_{\rm r, 28} \, P_{\rm NS, 3}^{4/7} \, \dot{P}_{-12}^{-4/7}  \approx \eta_{-4}^{-1} \, A_{\rm fp, 10}  \, \sin(\theta_{\rm fp, 0})^{-4/7}
\label{eq:observer_luminosity_curv}
\end{split}
\end{equation}

\subsubsection{Resonant Inverse-Compton Scattering Pairs}

Using Eqs.~\ref{eq:hgap_rics} \&~\ref{twistpairlum1},
\begin{equation}
\begin{split}
    &L_{\rm e^+e^-, RICS, twists} \sim \frac{A_{\rm fp}}{\pi R_*^2} \frac{ m_e c^3 \Delta \Psi \sin^{2}(\theta_{\rm fp})}{288 \alpha_{\rm f} {\cal B}^3 \lambar^2 }  \\
    &\times \Bigl( 4\sqrt{3 R_*} {\cal B}^{5/2} [\Theta |W_{-1}({\cal Z}_{\rm twist})|]^{-1/2}  \\
    &+ 9\rho_c \lambar^{-1/2} (1+2{\cal B}) \left[\Delta \Psi \sin^{2}(\theta_{\rm fp}) {\cal B}\right]^{1/2} \Theta \left|W_{-1}({\cal Z}_{\rm twist})\right|   \Bigr)^2 
\label{eq:pair_production_twist_exc}
\end{split}
\end{equation}
Similarly, for the rotation-powered case, the appropriate area is the polar cap size, $A \rightarrow A_{\rm pc} \sim 4 \pi^2 R_*^3/(P c)$ and 
\begin{equation}
\begin{split}
    L_{\rm e^+e^-, RICS, rot} &\sim \frac{ m_e c^2  A_{\rm pc} }{18 \alpha_{\rm f} {\cal B}^3 \lambar^2 P  }  \times \Bigl( 2\sqrt{3} {\cal B}^{5/2} [\Theta |W_{-1}({\cal Z}_{\rm rot})|]^{-1/2}  \\
    &+ 9 \rho_c \sqrt{\frac{\pi {\cal B}}{P c\lambar}} (1+2{\cal B}) \Theta |W_{-1}({\cal Z}_{\rm rot})|    \Bigr)^2
    \label{eq:pair_production_rot_exc_1}
\end{split}
\end{equation}
In both cases, the first term dominates at ${\cal B} \gg 1$ due to ${\cal B}^{3/2}$ scaling.

\subsection{Comparison of Twist Dissipation and Pair Luminosities}
\label{sect:comparing_luminosities}
The twist dissipation luminosity (Eq. \ref{eq:luminosity_untwisting_global}) can be considered to be the persistent macroscopic luminosity qualitatively similar in nature to the Poynting flux spin-down power $\dot{E}_{\rm sd}$ of isolated pulsars. The pair luminosity  (Eq. \ref{eq:twist_luminosity_curv_free} and Eq. \ref{eq:pair_production_twist_exc}) corresponds to the microphysical luminosity transferred to the primary pairs accelerated across the resultant acceleration gap stemming from the twist itself. The pair luminosity cannot be greater than the dissipated twist luminosity: $L_{\rm e^+e^-} < L_{\rm diss, pl}$. In the following we examine the parameter space for which this inequality holds, constituting a deathline for this mechanism of magnetic coherent radio emission from twist dissipation. Similarly, the constraint $L_{\rm e^+e^-} < \dot{E}_{\rm sd}$ must hold for the rotation-powered case.

\subsubsection{Curvature radiation channel}
The ratio between Eq. \ref{eq:twist_luminosity_curv_free} \& Eq. \ref{eq:luminosity_untwisting_global} may be denoted as the efficiency ${f_{\rm pair}}$ by which twist luminosity is transferred to accelerated pairs:
\begin{equation}
    \begin{split}
        f_{\rm pair, CR} &= \frac{L_{\rm e^+e^-, CR}}{L_{\rm diss, pl}} \\
        &= \frac{4 c}{{\rm v}_{\rm pl}} \bigg(\frac{2 \pi}{c P}\bigg)^{1/7} \bigg(\frac{B_{\rm cr}}{B}\bigg)^{6/7} \bigg(\frac{\lambar^4 \sin^{10}(\theta)}{R_{*}^3}\bigg)^{1/7} 
    \end{split}
    \label{eq:ratio_pair_diss}
\end{equation}
Requiring $f_{\rm pair} \leq 1$ implies a minimum period of magnetically produced radio emission corresponding to $f_{\rm pair} = 1$, for twisted footpoints at the colatitude of the polar cap ($\theta = \theta_{\rm pc}$), as persisting twists must be confined to the closed field line region. Rearranging Eq. \ref{eq:ratio_pair_diss}: 
\begin{equation}
    \begin{split}
        P &> \frac{4^7 c^7}{{\rm v}_{\rm pl}^7} \bigg(\frac{2 \pi}{c}\bigg) \bigg(\frac{B_{\rm cr}}{B}\bigg)^{6} \bigg(\frac{\lambar^4 \sin^{10}(\theta)}{R_{*}^3}\bigg) \\
        P &> \frac{2^{24/6} c^{7/6}}{3^{2/6} {\rm v}_{\rm pl}^{7/6}} \bigg(\frac{8 \pi }{c}\bigg)^{1/6} \bigg(\frac{B_{\rm cr}}{B}\bigg) \bigg(\frac{\lambar^4}{R_{*}^3}\bigg)^{1/6} \bigg(\frac{2 \pi R_{*}}{c}\bigg)^{5/6} \\
        P &\gtrsim 10 \, {\rm seconds} \; {\rm v}_{\rm pl, 4}^{-7/6} \, B_{\rm 15}^{-1} \, R_{*, 6}^{1/3} 
    \end{split}
    \label{eq:period-limits}
\end{equation}

One can replace $B$ with the characteristic dipolar magnetic field naively assuming spin-down $\dot{P}$ is fully described by magnetic dipole radiation: $B_{\rm char} \sim 6.4 \times 10^{19} \, P^{1/2} \, \dot{P}^{1/2}$, where we assume a NS moment of inertia $I \approx 10^{45} \, {\rm g \, cm^{2}}$. The resulting deathlines in $P-\dot{P}$ space are shown as orange lines in Fig. \ref{fig:p-pdot_hurley_walker}.
\par
\begin{figure}
  \centering
{\includegraphics[width=.5\textwidth]{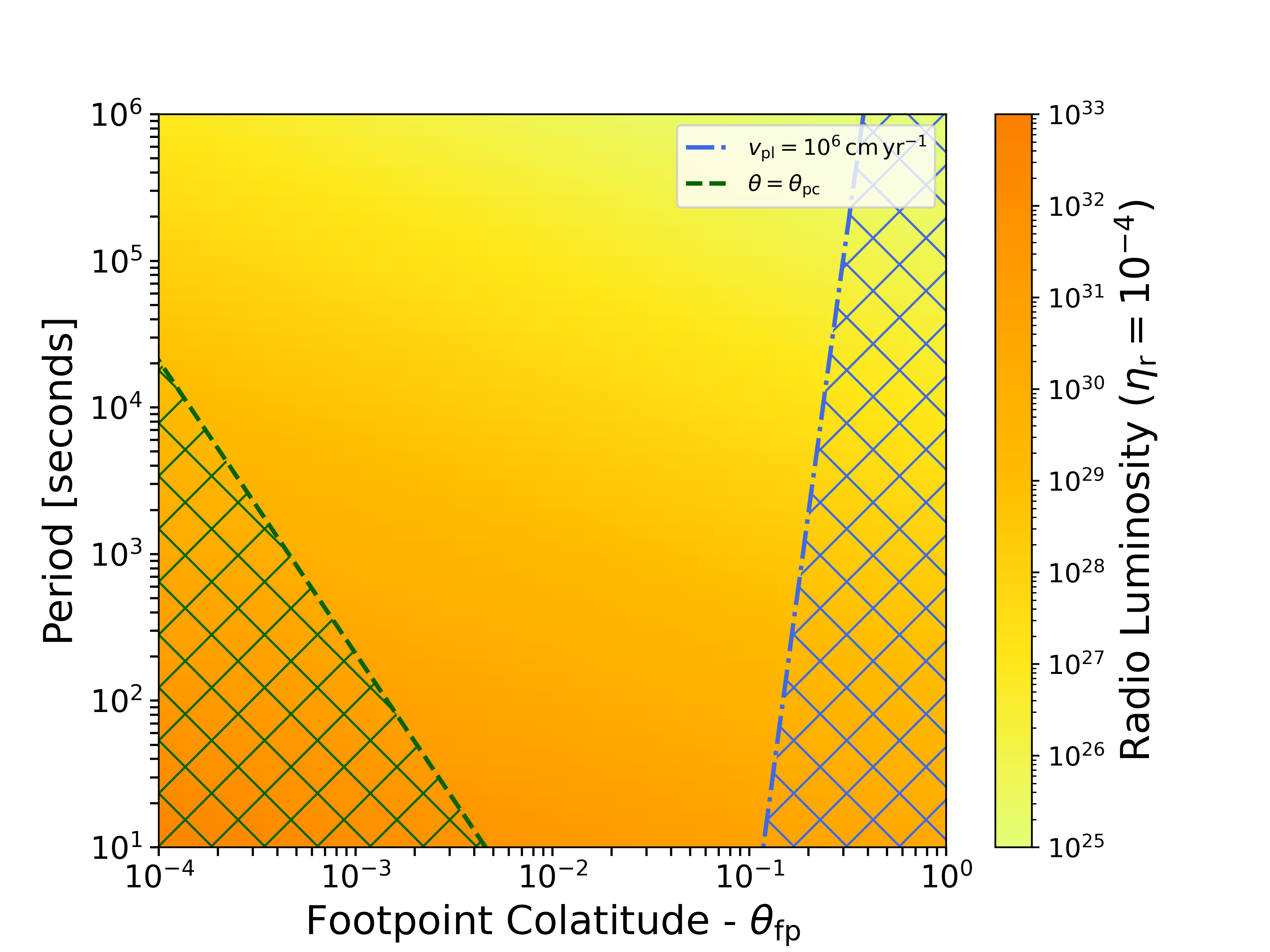}}
\caption{Fiducial model radio luminosity for curvature radiation initiated cascades at the critical twist. We adopt $L_{\rm r} = \eta_{\rm r} L_{\rm e^{+} e^{-}, curv}$ for $\eta_{\rm r} = 10^{-4}$ and ${\rm v}_{\rm pl} = 10^{6} \, {\rm cm \, yr^{-1}}$. Hatched regions are excluded - green denotes regions where footpoints are anchored above the polar cap, and the blue denotes the region where $f_{\rm pair} \geq 1$. }
\label{fig:radio_luminosity_curv}
\end{figure}

Interesting, this result also implies a minimum and maximum radio luminosity for a given $P$, ${\rm v}_{\rm pl}$, shown in Fig. \ref{fig:radio_luminosity_curv} as occurring at footpoint co-latitudes corresponding to $f_{\rm pair } = 1$ and $\theta = \theta_{\rm pc}$ respectively. This range of luminosities is small, such that we do not expect many less luminous sources for fixed $A_{\rm fp}$. The unknown source distributions of $P$, $A_{\rm fp}$, ${\rm v}_{\rm pl}$ prevent us from making firm predictions of ULPM luminosity function.

\subsubsection{RICS channel}
\label{sect:RICS_lum_comparisonsss}
\begin{figure*}
  \centering
{\includegraphics[width=1.\textwidth]{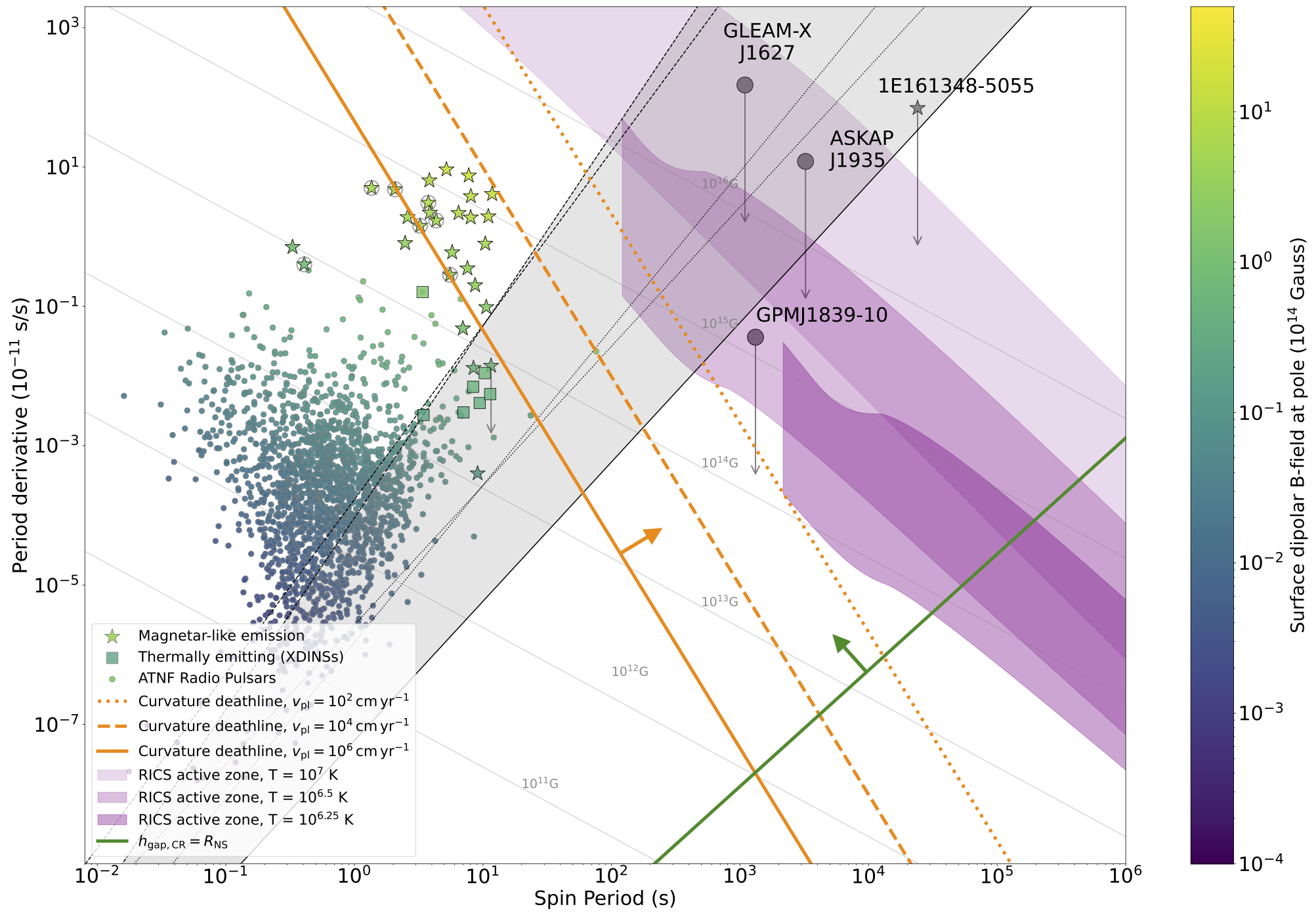}}
\caption{Deathlines and active zones for coherent radio emission stemming from plastic motion (${\rm v}_{\rm pl}$) or thermoelectric induced local twists. Purple active zones take into account all constraints to RICS pair production ($\theta_{\rm fp} = 0.1$, $\rho_{\rm c} = 10^{7}$ cm, $v_{\rm pl} = 10^{5}$ cm yr$^{-1}$) discussed in \S\ref{sect:microphysics} including a viable intersection (defining a lower limit on the period) and pair multiplicity and luminosity thresholds, bounding the regions at high and low $B$ respectively. Orange deathlines constrain parameter space for pair production from curvature channel (Eq. \protect\ref{eq:period-limits}) assuming $\theta_{\rm fp} = \theta_{\rm pc}$. In green we show the limit at which, for the curvature radiation regime, the gap height $h_{\rm gap, CR}$ exceeds the NS radius beyond which our microphysical analysis becomes unreliable. This figure is adapted (with permission) from \protect\cite{hurley-walker_long-period_2023}; making use of open-source code available at: \protect\url{https://github.com/nhurleywalker/GPMTransient}. Included are $P$-$\dot{P}$ locations of pulsars, magnetars and central compact objects \citep{2005AJ....129.1993M,2006Sci...313..814D,2014ApJS..212....6O,2016MNRAS.463.2394D,rea_magnetar-like_2016,coti_zelati_systematic_2018}, and various rotational-powered deathlines \citep{1993ApJ...402..264C,2000ApJ...531L.135Z} spanning the diagonal grey band. We remind the reader that $B_{\rm char}$ (light grey background contours) is an approximate but possibly unreliable estimate of the true magnetic field at the twist dissipation and pair production site. Finally, note that in the lower B regime $B \lesssim B_{\rm crit}$, magnetic pair production will occur above threshold (e.g. Fig. \ref{fig:lengthscales_lowB}), so equivalent contours will differ from those depicted.}
\label{fig:p-pdot_hurley_walker}
\end{figure*}

The RICS pair luminosity is an involved equation, however we can equate Eq. \ref{eq:luminosity_untwisting_global} \& Eq. \ref{eq:pair_production_twist_exc} to find the limiting case at which $f_{\rm pair} = 1$ (e.g. $L_{\rm e^+e^-, RICS} = L_{\rm diss, pl}$). This requirement defines the bottom left (low $B$) bound of purple active zones in Fig. \ref{fig:p-pdot_hurley_walker}; where $\dot{P} \propto P^{-9/7}$ to $P^{-1}$ depending on $\cal{B}$ and additional numerical factors that scale with $\cal{B}$ in Eq. \ref{eq:pair_production_twist_exc}. 
\par
A similar calculation is performed for the rotation case using Eq. \ref{eq:pair_production_rot_exc_1} (assuming $A = A_{\rm pc}$) for Fig. \ref{fig:P_Pdot_rotation_case}, instead comparing to the spin-down luminosity $L_{\rm sd} \approx 4 \pi^2 I \dot{P} P^{-3}$. Here the critical boundary $L_{\rm e^+e^-, RICS} = L_{\rm sd}$ scales as $\dot{P} \propto P^{3/5}$ for $\cal{B} \gg$ 1 and $\dot{P} \propto P^{-1/3}$ for $\cal{B} \ll$ 1, assuming a characteristic spin-down magnetic field.

\subsection{Rotationally-powered RICS Pairs in Pulsars and Magnetars}
\begin{figure}
  \centering
{\includegraphics[width=0.5\textwidth]{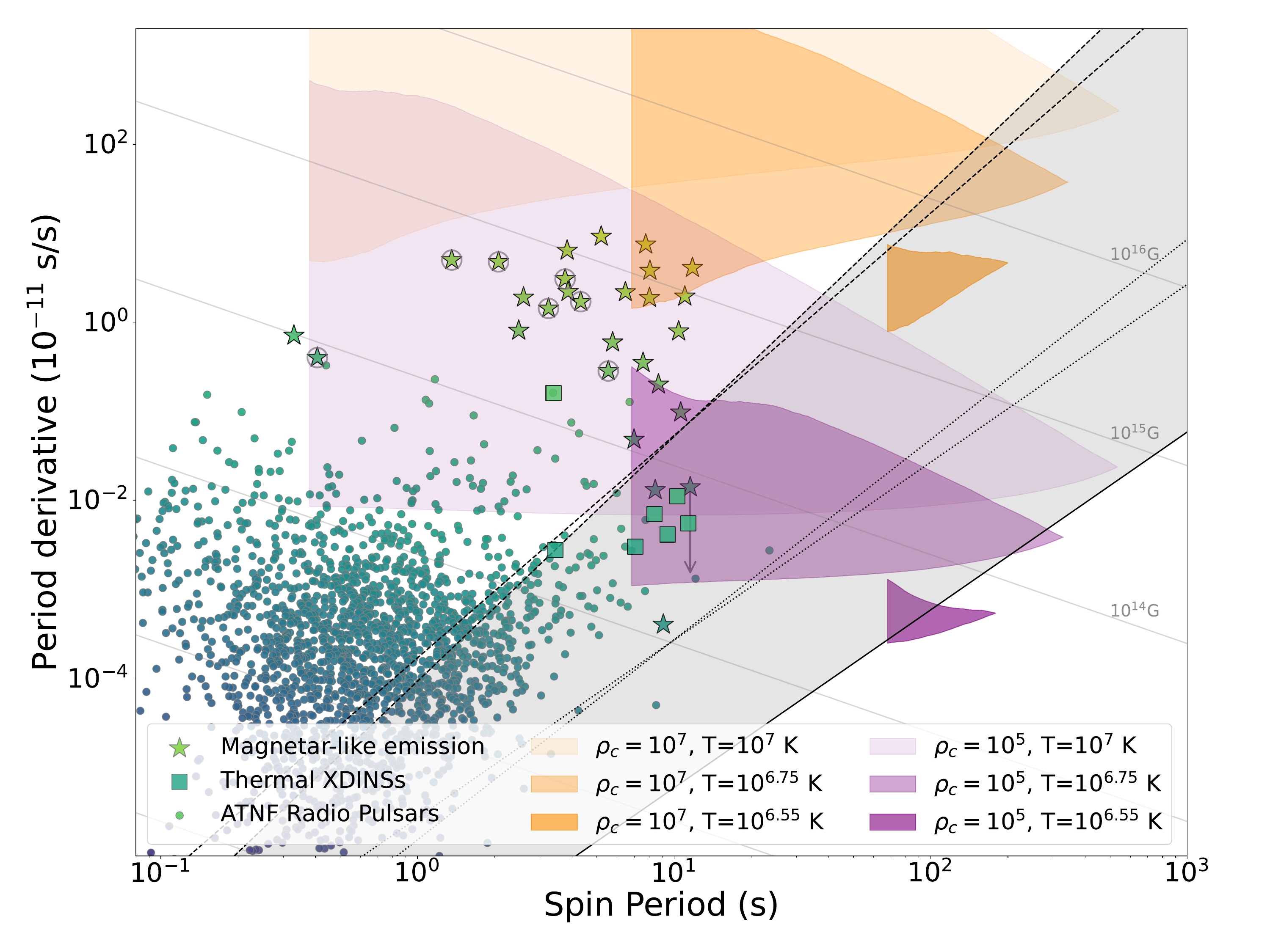}}
\caption{Active zones for \textit{rotationally-powered} pair production via resonant inverse-Compton scattering, assuming a polar cap area $A_{\rm pc} = \frac{4 \pi^2 R_*^3}{P c}$. Orange and purple zones respectively correspond to curvature radii of $\rho_{\rm c} = 10^{7}$ cm and $\rho_{\rm c} = 10^{5}$ cm; for temperatures ranging from $10^{6.55-7}$ K. Each zone is bounded on the left by the viable intersection constraint approximation (Eq. \protect\ref{eq:Lambertcondition2}), on the bottom by the requirement that the spin-down luminosity exceeds the pair production luminosity (\S \protect\ref{sect:RICS_lum_comparisonsss}; where boundary scales as $\dot{P} \propto P^{3/5}$ for $\cal{B} \gg$ 1 and $\dot{P} \propto P^{-1/3}$ for $\cal{B} \ll$ 1) and on the top/right by the pseudo-multiplicity requirement $\mathcal{M} > 1$ (Eq. \protect\ref{eq:multiplicity_requirement}).}
\label{fig:P_Pdot_rotation_case}
\end{figure}

As aforementioned, the RICS pair production channel can be generalised to the rotationally powered case, by identification of the critical twist with the period $P_{\rm crit}$ corresponding to the equivalent current density requirements. In Fig. \ref{fig:P_Pdot_rotation_case} we show the region in $P-\dot{P}$ space in which pair production may occur via rotationally-powered RICS for given $\rho_{\rm c}$ and $T$ given the following constraints: i.e. the psuedo pair multiplicity $\mathcal{M} > 1$, the gap height $h_{\rm gap} < R_{*}$, the spin-down luminosity should exceed the pair luminosity $\dot{E}_{\rm sd} > L_{\rm e^+e^-, RICS}$, and where primaries resonant interactions are accessible (see Eq.~\ref{eq:Lambertcondition}). 

\par
\begin{figure}
  \centering
{\includegraphics[width=0.5\textwidth]{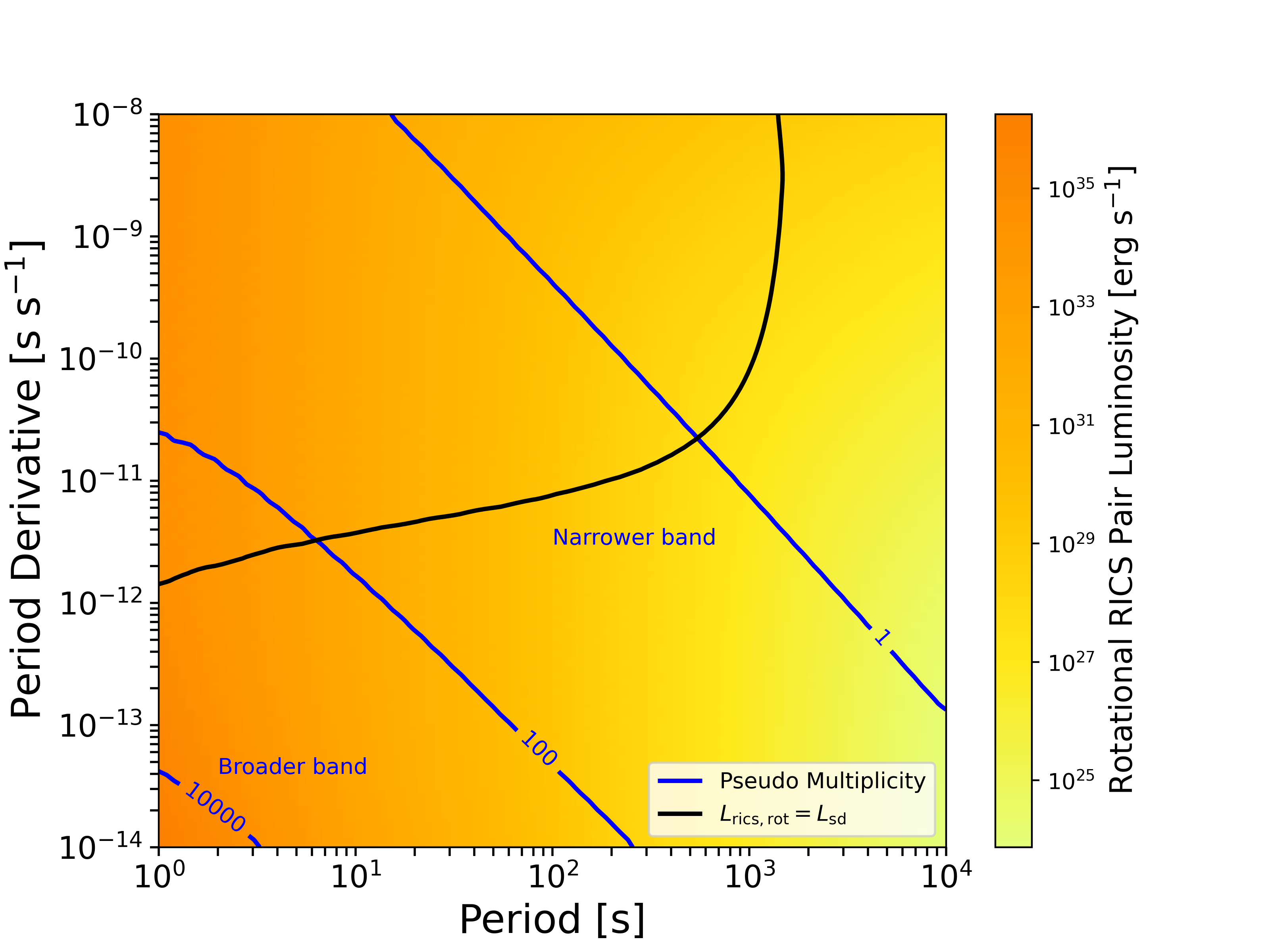}}
\caption{The rotationally-powered RICS pair luminosity (Eq. \protect\ref{eq:pair_production_rot_exc_1}) as a function of $P-\dot{P}$ for $T = 10^{7}\,$K and $\rho_{\rm c} = 10^{6}\,$cm. Blue contours represent pseudo pair multiplicity (Eq. \protect\ref{eq:multiplicity_requirement}), where we expect no emission in regions where $\mathcal{M} < 1$ and potentially progressively more broadband emission for higher expected multiplicities. The black contour represents the boundary at which $L_{\rm sd} = L_{\rm e^+e^-, RICS, rot}$, such that no emission is expected below this line. The intersection of this line and the $\mathcal{M} = 1$ contour can be identified with the rightmost point of active regions in Fig. \ref{fig:P_Pdot_rotation_case}}
\label{fig:luminosity_rotation_rics}
\end{figure}

Importantly, for sufficiently high temperatures $T \gtrsim 10^{6.5}$ K , large regions in $B-P$ parameter space may viably produce pairs via RICS, seeding plasma so that the global magnetospheric solution is never a `dead' pulsar e.g., the disk-dome solution \citep{krause-polstorff_electrosphere_1985,petri_global_2002}, albeit with low $\dot{E}_{\rm sd}$ and radio luminosity\footnote{However, for higher $T$, some of these could be considered as dim AXPs as the X-ray luminosity will generally exceed $\dot{E}_{\rm sd}$.}. In Fig. \ref{fig:luminosity_rotation_rics} we show the rotationally-powered RICS pair luminosity, overlaid with the psuedo-multiplicity contours (assuming $\eta_{\rm r} \approx 10^{-4}$). There is large parameter space where potentially undiscovered rotationally-powered pulsars that may viably produce pairs with $P \gtrsim 10$ s and relatively large magnetic fields $B \gg 10^{13}$ G. These could be perhaps similar in nature to PSR J0901-4046 \citep{caleb_discovery_2022}. Finally, it is a qualitative expectation that lower multiplicities, e.g., larger values of both $P$ and $\dot{P}$, should be commensurate with narrower band radio emission (see also \citealt{2024HEAD...2110725H} for a detailed Monte Carlo calculation of pair multiplicities of RICS casacades).

\section{Discussion: The Model's Generic Features}
\label{sect:discussion}
\subsection{X-ray Emission}
\label{sect:x-rays}
A generic prediction of the plastic motion or a thermoelectrically driven mechanism is that radio emission could have a thermal X-ray component associated with twist dissipation hotspot. Heating by back-flowing pairs from cascades will proceed by Bremsstrahlung and magneto-Coulomb interactions (e.g., \citealt{2019MNRAS.483..599G} and references therein). The surface heat deposition depends on particle luminosity and Lorentz factors: lower particle energies results in shallow penetration depths which will generally not efficiently heat the atmosphere. Lower Lorentz factors and multiplicities are generally realized in the RICS channel, possibly resulting in less efficient heating compared to the CR channel. 
\par
\begin{figure}
  \centering
{\includegraphics[width=.5\textwidth]{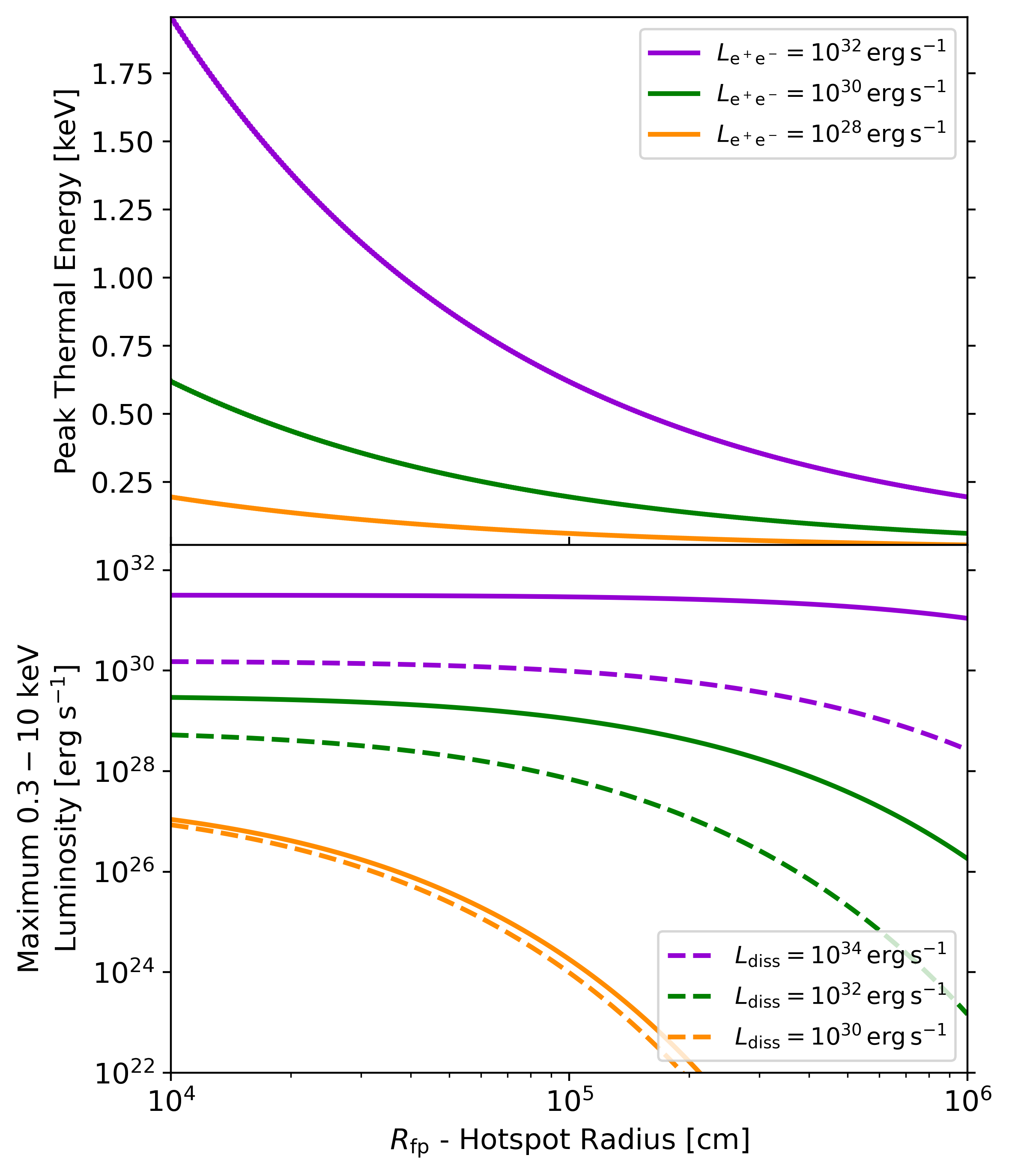}}
\caption{Maximum X-ray emission (from return currents only) as a function of blackbody hotspot radius, for different pair luminosities. The peak of the spectrum is shown in Panel 1. In Panel 2 we show the unabsorbed, integrated $0.3-10$ keV thermal luminosity for the $L_{\rm X} = L_{\rm e^+e^-}$ case (solid lines) and $L_{\rm X, 30} \sim L_{\rm diss, 33}^{0.88}$ (dashed lines) following \protect\cite{chang_observational_2023}. Our fiducial parameters ($L_{\rm e^+e^-} = 10^{31} \, {\rm erg \, s^{-1}}$, $R_{\rm fp} = 1 \, {\rm km}$) imply an unabsorbed X-ray luminosity of approximately $L_{\rm 0.3 - 10 keV} \approx 2 \times 10^{30} \, {\rm erg \, s^{-1}}$}
\label{fig:x_rays_thermal_rc}
\end{figure}

Detailed calculations are deferred for a future work, but here we use two approximations: a coarse upper limit assuming $L_{\rm X} = \sigma T_{\rm eff} \approx L_{\rm e^+e^-}$ (solid lines in Fig. \ref{fig:x_rays_thermal_rc}) and an empirical relationship extrapolated from high-energy pulsars of $\frac{L_{\rm X, therm}}{L_{\rm diss,pl}} \sim 10^{-3} L_{\rm diss, pl}^{-0.12}$ (dashed lines; see Fig. 11 in \citealt{chang_observational_2023}). In Fig. \ref{fig:x_rays_thermal_rc} we show maximum X-ray peak frequency and integrated luminosity for $L_{\rm e^+e^-}$, where solid purple lines explain observed radio sources assuming $\eta_{\rm r} = 10^{-4}$. X-ray fluxes fall slightly below the limits obtained for ULPMs thus far, nonetheless sensitive follow-up at keV energy ranges during radio-loud phases is clearly demanded particularly for sources with favourable absorption.
\par
Particle acceleration in gaps may lead to observable non-thermal emission either directly from RICS or secondary synchrotron pairs from the CR or RICS channels, as well as photon splitting cascades reprocessing higher energy emission. Primary (high-energy) gamma-ray CR is not observable due to the high magnetospheric opacity \citep{2019MNRAS.486.3327H}. Most of the magnetosphere at low altitudes is opaque below above several hundred keV to splitting. Hard X-ray non-thermal emission, from RICS, is observed in conventional magnetars with luminosities comparable or larger than thermal surface emission \citep[e.g.,][]{2004ApJ...613.1173K,2008A&A...489..245D,2008A&A...489..263D,2017ApJS..231....8E}. RICS is highly anisotropic, pulsed, and generically gives rise to a hard spectrum \citep[e.g.,][]{2007Ap&SS.308..109B,wadiasingh_resonant_2018, 2019BAAS...51c.292W} with a cutoff determined by kinematics or splitting opacities. {\color{blue} Harding et al. (in prep)} find secondary synchrotron from pair cascades occurring above the ground state is generally sub-dominant to RICS except at very low energies (< 50 keV). Thus a generic feature expected from ULPMs, if cascades are occurring in the RICS channel, is an expectation of power-law emission above $\sim$~few keV up to potentially an MeV with potentially high polarization degree.

\subsection{UV/Optical Emission}
\begin{figure}
  \centering
{\includegraphics[width=.5\textwidth]{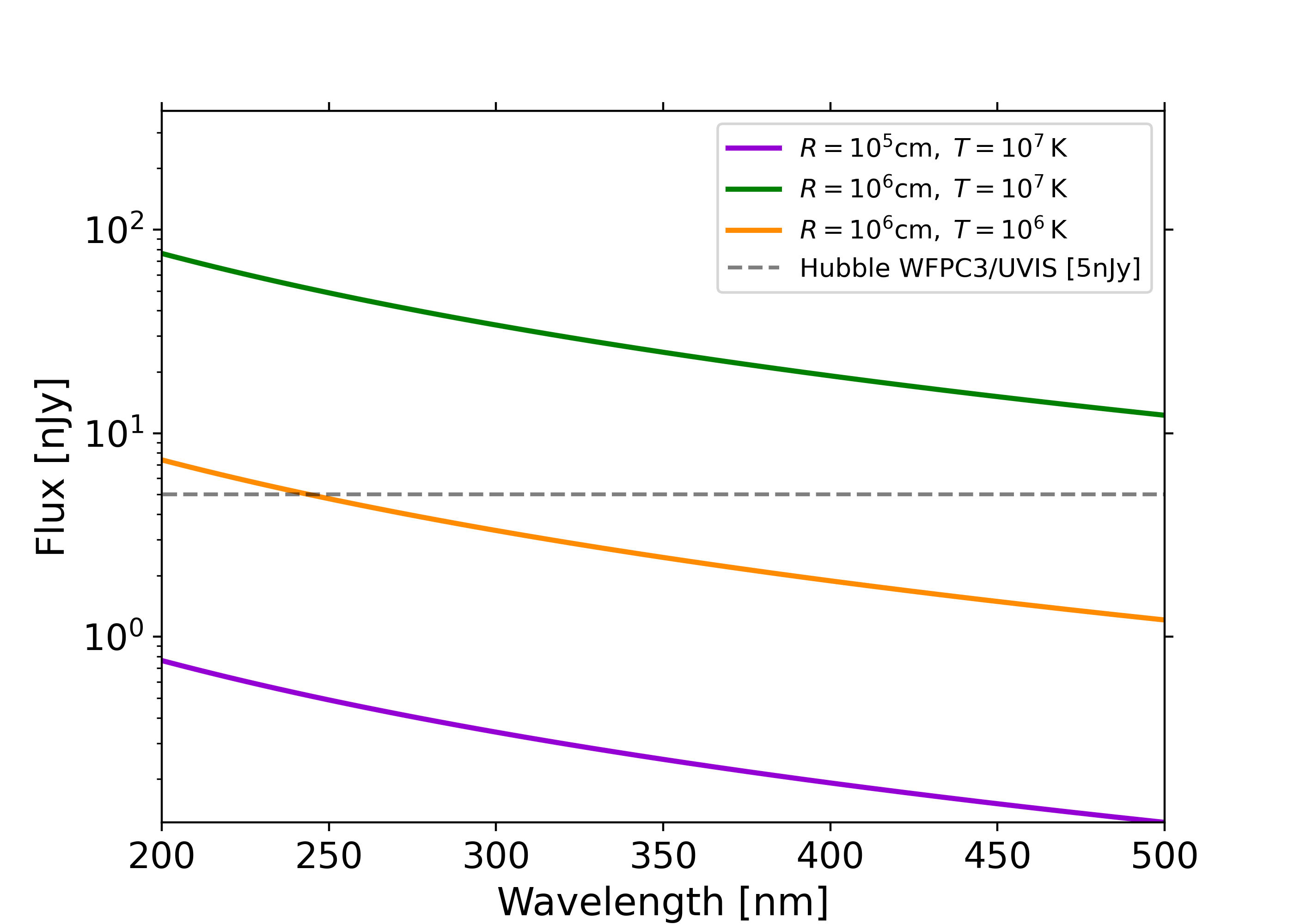}}
\caption{Predictions for unabsorbed thermal return current UV emission as a function of blackbody radius and temperature, assuming $D = 1$ kpc. A typical limiting HST sensitivity of 5 nJy (29.5 mag) is shown with a grey dashed line.}
\label{fig:optical}
\end{figure}

Although potential thermal hotspot emission peaks $\sim$keV, it may be observable with more sensitive observations at lower frequencies. In Fig. \ref{fig:optical} we show predictions for the UV spectra of emission for different blackbody radii and temperatures, assuming a distance of $1$ kpc. Observations with the Hubble Space Telescope may constrain larger area hotspot temperatures above $T>10^{7}$ K, corresponding to the RICS channel of pair production and inefficient radio emission (i.e. high pair luminosities). Such observations may also probe non-thermal emission components \citep{2024arXiv240503947H}.

ULPMs formed via magnetic synchronisation torques in tight binaries might have an optical counterpart with an orbital period similar to the radio pulse period \citep[e.g.,][]{2008ApJ...681..530P}. Additional periodic nulling may also be present due to stellar or plasma eclipses. However, unless the spin and orbital periods are significantly different in a tight binary evolved in this manner, the radio emission should be powered by magnetar activity rather than binary interaction.

\subsection{Radio Variability}
Plastic flow (or thermoelectric) powered radio emission will exhibit variability on a number of timescales. As mentioned likely short duration variability will occur on sub-millisecond timescales ($\omega_{\rm p}^{-1}$ and $h_{\rm gap}/c$) due to the gap microphysics. As in rotational-powered pulsars, we also expect significant pulse-to-pulse variability and jitter (e.g., \citealt{drake_second_1968,jenet_radio_1998}) or nulling (e.g., \citealt{backer_pulsar_1970,camilo_psr_2012,lyne_two_2017}) due to magnetospheric ``weather" regardless of specific emission mechanism.
The observer's viewing angle is also changing with time (uncorrelated with the magnetospheric weather), and samples different times and regions of the magnetosphere where waves decouple from the plasma and escape. In contrast, there is the magnetospheric ``climate" which is the time-averaged pulse profile (and polarization, see below) character of radio pulsars \citep[e.g.,][]{2024arXiv240410254J} and stability of gamma-ray emission in pulsars \citep[e.g.,][]{2022ApJ...934...30K} implying stability of the magnetospheric plasma on large length scales. Thus, for the ULPM context, polarization stacked and averaged pulses are might be more securely interpreted than variable single pulses. Yet, as slow rotators, the observer viewing direction changes relatively slowly in ULPMs compared to normal radio pulsars with much larger fluence per pulse. Thus, ULPMs might offer a powerful opportunity to scrutinize localized regions of the magnetosphere and plasma ``weather" on timescales much longer than the light crossing time of the emitting region than afforded by radio pulsars. 
\par
The plastic motion model predictions additional intrinsic variation depending primarily on the time-varying ${\rm v}_{\rm pl}(t)$ or footpoint morphology. Cellular crustal evolution models presented in \cite{lander_game_2023} find a bimodal distribution of crustal quiescence peaking on timescales of a few months and a 1-3 decades, therefore we may expect radio emission to be quenched on these timescales too. Interestingly, this appears to match the two observed ULPM candidates which have duty cycles of 2 months and $\gtrsim$30 years respectively, despite the very small number statistics. Variations in the plastic flow morphology (${\rm v}_{\rm pl}(t)$, $A_{\rm fp}(t)$ or slab shape) will result in pulse profile variability. The relatively gradual changes in pulse profile and luminosity in GLEAM-X J1627 could be explained by a combination single pulse variability, and deviations in one (or possibly two) slabs of non-uniform plastic flow. The extremely variability observed in GPM J1839-10 is more challenging to explain, requiring multiple active slabs and/or extreme non-uniformity in plastic flow across single slabs. This is also supported by the relatively high pulse fraction, which is consistent with multiple kilometre-sized slabs. These rapid changes to the plastic flow may be attributed crustal heating feedback leading to reduced viscosity or larger thermoelectric gradients.

\subsection{Observed Radio Polarization}
As in rotationally-powered pulsars, a secure expectation is high levels of linear polarization, and generically lower levels of circular polarization, attributable to propagation effects \citep[e.g.,][]{wang_polarization_2010} in a magnetically-dominated plasma with a highly ordered magnetic field. Polarization swing across pulses may broadly follow the Rotating Vector Model \citep{radhakrishnan_magnetic_1969} if the wave modes adiabatically follow as plasma modes, with small departures from the low magnetic twists invoked in this work. However, propagation effects of waves and their eigenmodes are uncertain as the plasma distribution function is unknown and could vary wildly. As aforementioned, observer's viewing angle is also changing with time and samples different portions of the magnetosphere where waves decouple from the plasma and escape. The near-surface pair production across gaps is non-stationary and generically results in inhomogeneous plasmas. This stirring is on the scale of $h_{\rm gap}$, but could potentially cascade to much larger length scales higher up in the magnetosphere. In such a situation (i.e. a time-dependent and inhomogeneous plasma), the notion of eigenmodes of propagation in a plasma are ill-posed.
\par
For longer period magnetars, lower plasma densities (i.e. $n_{\rm GJ} \propto P^{-1}$) suggests that vacuum polarization may be important for wave propagation at low altitudes. As the plasma is inhomogenous, the wave may encounter pockets of low or high plasma density for particular propagation directions even after it decouples from the plasma which produced it. Vacuum polarization could compete with plasma birefringence, especially if there is a significant proton component in the plasma (the modes of vacuum birefringence and a hydrogenic plasma are opposite for a guide B field). For a stationary hydrogenic plasma, the density at which the contributions to the dielectric tensor are compensating (the ``vacuum resonance") is \cite{1997MNRAS.288..596B,2002ApJ...566..373L}:
\begin{equation}
    n_{\rm res} \approx 1.8 \times 10^{12} \; {\rm cm^{-3}} \, B_{15}^2 \, \nu_{\rm GHz}^{2} \, Y_{\rm e,0}^{-1}.
\end{equation}
Here $Y_{\rm e}$ is the plasma electron fraction and we have omitted slowly varying factors of order unity. Below this hydrogenic plasma density, vacuum birefringence governs propagation modes over plasma modes. We can compare this to typical magnetospheric plasma densities:
\begin{equation}
    n_{\rm GJ} \approx \frac{2 {\cal M} B}{q c P} = 1.3 \times 10^{11} \; {\rm cm^{-3}} \, B_{15} \, P_{3}^{-1} \, {\cal M}
\end{equation}
which drops with altitude slower ($B^2$ versus $B$) than vacuum polarization density scale. Thus at lower altitudes vacuum modes may dominate over plasma modes for a high proton-to-positron ratios (or low ${\cal M}$). For large ${\cal M}$, plasma modes will likely always dominate. In fact, for a pair plasma, the vacuum resonance effect does not occur as the contributions to the dielectric tensor are in the same direction as vacuum polarization. If the proton fraction is locally high, the O/X modes for vacuum polarization and plasma effects are opposite and thus may result in mode switching observable as sharply varying polarization signatures, including intra-pulse (and pulse-to-pulse) jumps in the levels of linear and circular polarization. This is very complex situation -- localized regions may have wildly different pair fractions or plasma density. Polarmetric signatures of plasma/vacuum mode switching may be observed as the observer line of sight through the magnetosphere changes, or due to changes gap altitude due to the (non-stationary) plasmas.

\subsection{A Simplified Geometric Model of Pulse Morphology}
\subsubsection{Stratification of twist and dissipation}
Although the exact geometry of the crustal region undergoing plastic motion is unknown, here we construct a greatly simplified toy model (motivated by \citealt{lander_magnetar_2016, lander_magnetic-field_2019}, similar to Fig. \ref{fig:illustration}) in which plastic motion occurs for a circular slab of size $A_{\rm fp} \sim 1 \, {\rm km^2}$ with no net component (i.e. corresponding to a rotation on the surface). The circular slab has a radius $l$ and ${\rm v}_{\rm pl}$ acts about the central axis defined by $l = 0$. The twist imparted onto field lines grows linearly $\propto l$; i.e. field lines threading $l = 0$ receive no net twist, which increases towards the edge of the circular slab. We further define an observer `impact parameter' $0 \leq b \leq 1$, such that $b=1$ corresponds to a time-integrated line of sight grazing the circular slab and $b=0$ represents the diameter chord. 

\subsubsection{Pulse profile}
\begin{figure}
  \centering
{\includegraphics[width=.45\textwidth]{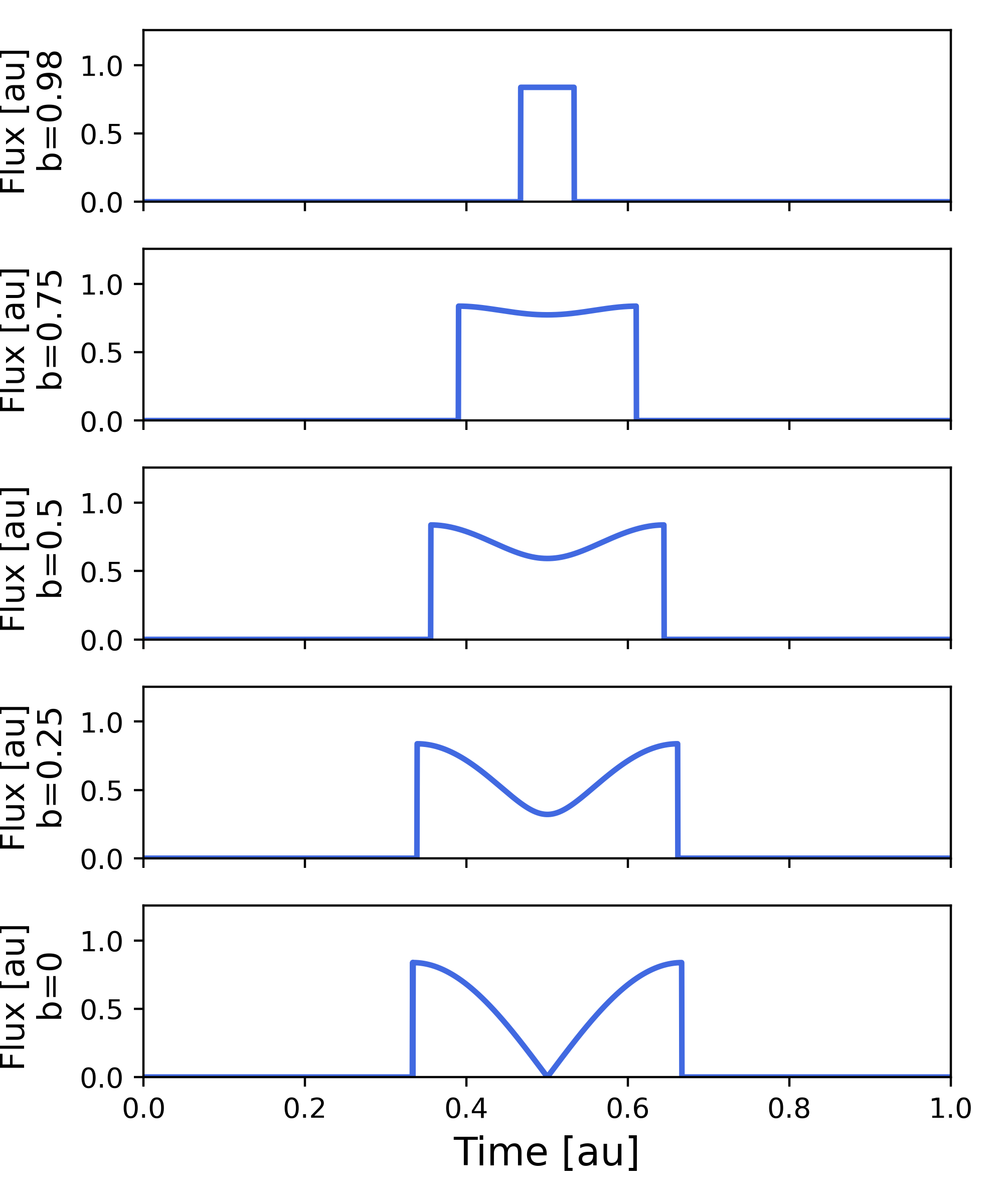}}
\caption{\textit{Simplified} geometric model of synthetic (phase-folded) pulse profiles from a localized plastic motion, with arbitrary flux \& time axes. A uniform, circular, single slab (without smoothing) is assumed, for different values of the `impact parameter' $b$. For high values of $b$, the observer's line of sight does not intersect two distinct regions of the outer ring where most of the dissipation occurs, but we note that lower values of $b$ are more common.}
\label{fig:toy_model_pulses}
\end{figure}

In this simplified geometric picture, dissipation is concentrated within a ring of field lines on the outer edge of the slab. This edge-brightened or ``conal" character expected to be a generic feature of plastic motion or thermoelectrically-driven field twists for a given active region. The twist rate (Eq. \ref{eq:twist_rate}), current requirements, dissipation (Eq. \ref{eq:diss_pl}) \& radio luminosity scale with $l$. In Fig. \ref{fig:toy_model_pulses}, we show pulse profiles produced within this framework, for a range of observers with varying $b$. We find that double peaked pulses are often expected, although rarer, narrower, single peaked pulses are permitted for chords which intersect the region with a larger impact parameter. We caution the reader that this is a \textit{simplified} toy model, and that a variety of variability is anticipated based on the additional complexity of plastic flow morphologies and microphysical intricacies of pair creation, or absorption within the magnetosphere (as is observed for single pulses of radio pulsars). For instance, more than one active region (e.g., with plastic motion) could be operating in these sources.

\subsubsection{Pulse fraction}
Within this toy model the pulse fraction $f_{\rm p}$ (i.e. ratio of the pulse width and the NS period) can inform on the slab size:
\begin{equation}
    f_{\rm p} \approx \frac{b l}{\pi R_{*} \sin(\theta_{\rm fp})}
\end{equation}
For GLEAM-X J1627, where often double-peaked pulse profile may be reproduced by a single slab undergoing plastic motion, the pulse fraction of $\approx 0.03$ can be reproduced for reasonable values of $b$, $l$, and $\theta_{\rm fp}$. The more larger pulse fraction of GPM J1839–10 hints at possibly a larger slab size, or multiple slabs undergoing plastic motion as supported by the higher observed variability. 

\subsubsection{Pulse peak migration}

Plastic motion may cause observed radio/X-ray pulses to migrate across the magnetar period phase on a characteristic timescale:
\begin{equation}
     T_{\rm mig} \approx \frac{2 \sin(\theta) R_{*}}{{\rm v}_{\rm pl}} \approx 10 \; \sin(\theta_{-1}) \, {\rm v}_{\rm pl,4}^{-1} \: {\rm years}
\end{equation}
For fast plastic flows at high co-latitudes this may be observable, as has been suggested by \cite{younes_pulse_2022} to explain pulse peak migration in X-ray observations of SGR 1830-0645. Finally, we note that, pulse migration due to rapid plastic flow could counteract or enhance the derived period derivative $\dot{P}$, depending if net plastic flow is retrograde or prograde.

\subsection{Magnetic Radio Emission in Shorter Period Magnetars}
\label{sect:others}
Although radio emission from shorter period (e.g $P \lesssim 12$ s) canonical magnetars \& high B pulsars is often energetically consistent with being powered by rotational energy \citep{rea_fundamental_2012}, it is possible some radio emission is magnetically powered. \cite{wang_coherent_2019} suggest that radio emission powered by twists may explain observed radio-loud magnetars XTE J1810-197 \citep{camilo_radio_2016} \& PSR J1622-4950 \citep{levin_radio-loud_2010}; and \cite{wang_atypical_2023} suggest that radio emission from SGR 1935+2154 may be of magnetic origin. Finally, we note the newly discovered PSR J1710-3452 \citep{surnis_discovery_2023} appears to bear resemblances to other ULPMs and predictions in this model (longer period, double-peaked structure and intermittent pulsing), and is considered a candidate for this emission model. 
\par
Plastic crustal motion may occur ubiquitously in magnetars, yet be subdominant to rotationally-powered mechanisms or quenched for periods beyond the deathlines and active zones in Fig. \ref{fig:p-pdot_hurley_walker}. In addition, younger active magnetars have not had enough time to build twist to the critical value (Eq. \ref{eq:timescale_critical}) or have undergone crustal slippages or local magnetic field rearrangements (as evidenced by short X-ray bursts) that inhibit the gradual build up of crustal stress through plastic motion. The escape of low-frequency emission may be more difficult in higher density magnetospheres of faster spinning magnetars possibly embedded in evolving post-supernova environments. Finally, higher co-latitude close field line footpoints in more slowly rotating NS may preferentially allow the escape of superluminal O-mode radiation \citep{arons_wave_1986}.

\subsection{Model Limitations}
\label{sect:discussion_pl}
The presented model depends critically on ${\rm v}_{\rm pl}$; a poorly constrained parameter that may sensitively depend on the crust viscosity, {\bf depth}, temperature, field line anchor point, and magnetic field. A better understanding of ${\rm v}_{\rm pl}$ and its distribution function would allow tighter constraints on the predicted luminosity, variability and viable parameter space of this model. In addition, numerical magneto-thermal simulations used to motivate values of ${\rm v}_{\rm pl}$ in this work \citep[e.g.,][]{lander_magnetar_2016,lander_magnetic-field_2019,2024arXiv240214911G} consider younger magnetars whose crust field evolution is active, set by choices of interior field and initial boundary conditions. ULPMs may be much older with cooler outer crusts \citep{potekhin_thermal_2020}, possibly resulting in lower typical values of $\dot{\Psi}$ for both plastic flow or thermoelectric mechanisms.
\par
$\dot{\Psi}_{\rm pl}$ likely depends on the depth of plastic flow \citep[where shallower flows are likely faster, and may be correlated with X-ray activity, e.g.,][]{younes_pulse_2022}, and the nature of late-time crust/core magnetic coupling which drives crustal evolution and stresses. A better understanding of the age of these sources through observations will enable more direct comparison to recent magneto-thermal simulations that consider core field evolution at late times. This will permit a more accurate determination of potential mild twists imparted to the magnetosphere. For the thermoelectric case, $\dot{\Psi}$ will also depend on the core-crust magnetic field coupling, especially if significant core expulsion occurs for older neutron stars. Future studies of the late-time core magnetic field expulsion via the Meissner effect will be crucial to characterise magnetic stresses of old magnetars ({\color{blue} Lander, in prep}).

The nature of efficient twist dissipation very close to the critical value requires further study, especially: whether untwisting occurs with relative stability modulated by pair production, and more detailed (plasma kinetic or detailed Monte Carlo cascade) studies of the gap physics in this key regime.

Finally, throughout this work we have assumed for simplicity a dipolar magnetic field and constant footpoint area. There is strong observational evidence for multi-polar field components \citep[e.g.,][]{bilous_nicer_2019,riley_nicer_2021,2021ApJ...907...63K,2024A&A...685A..70F} in NS which would likely dominate the local magnetic field close to the ULPM surface. The larger local $B$ may dominate magnetic field evolution and increase local plastic motion, and smaller $\rho_{\rm c}$ would influence gap physics (see Fig. \ref{fig:lengthscales_small_rho_c}).

\section{Conclusions}
\label{sect:conclusions}
In this work, we have sought an explanation for coherent radio emission from magnetized neutron stars beyond the death line of rotation-powered mechanisms, motivated by recently discovered ultra-long period radio sources. We present our primary conclusions as follows:

\begin{enumerate}
    \item Crustal plastic motion or thermoelectric effects impart mild (low) twist to the local magnetic field lines
    \item Dissipation of twist occurs only above a characteristic scale  (Eq. \ref{eq:critical_twist}), below which the corotational charge density could suppress particle acceleration
    \item Efficient dissipation of twist usually occurs such that dissipation continues with quasi-stability for the lifetime  of plastic motion or thermoelectric action
    \item Acceleration gaps form close to the surface in the twist zone by either RICS or curvature photons channels, producing electron-positron pairs.
    \item The pair multiplicity of the RICS channel is expected to be lower than the CR channel, and might result in narrower band radio emission
    \item The dissipation luminosity exceeds the pair luminosity only for longer period magnetars leading to a minimum period for radio emission: $P_{\rm RICS} \gtrsim 120 \; (T/10^{6.5} {\rm K})^{-5} \, {\rm sec}$ and $P_{\rm curv} \gtrsim 150 \; ({\rm v_{\rm pl}}/10^{3} {\, \rm cm \, yr^{-1}})^{-7/6} \, {\rm sec}$
    \item A simplified geometric model predicts a pulse fraction \& morphology consistent with some observations, and favors an even number of pulse peaks per active zone
    \item Thermal X-ray/UV return current hotspots ought to be generically present and may be detectable by current instruments
    \item Rotationally-powered RICS pair production is permitted in active zones of B-P space beyond conventional pulsar death lines
    \item Notwithstanding uncertainties of the magnetar period and ${\rm v}_{\rm pl}$ distributions, we may expect many more less luminous, longer period ULPMs
\end{enumerate}

Coupling the radiation model presented here with numerical magneto-thermal evolution simulations may provide more robust predictions for the long-term evolution of radio-loud ULPMs. In addition, further image-plane searches for more ultra-long period radio sources and deep soft X-ray/UV observations of known sources, simultaneous with radio observations where possible would be constructive. 

\section*{Acknowledgements}
A.~J.~C. acknowledges support from the Oxford Hintze Centre for Astrophysical Surveys which is funded through generous support from the Hintze Family Charitable Foundation. Z.~W. acknowledges support by NASA under award number 80GSFC21M0002.
\par
The authors are grateful for fruitful conversations with Sam Lander, Kaustubh Rajwade, James Matthews, Andrey Timokhin, Arthur Suvorov, and Natasha Hurley-Walker which improved this work. We would also like to thank Alice Harding, Constantinos Kalapotharakos, Konstantinos Gourgouliatos, and Paz Beniamini for helpful feedback and discussions on an earlier version of this manuscript. We would also like to thank the anonymous referee whose comments helped improve this work.

%%%%%%%%%%%%%%%%%%%%%%%%%%%%%%%%%%%%%%%%%%%%%%%%%%
\section*{Data Availability}
This theoretical work produced no new data. Figure reproduction packages will be shared upon request to the authors.



\bibliographystyle{mnras}
\bibliography{example_1} % if your bibtex file is called example.bib

\begin{thebibliography}{}
\makeatletter
\relax
\def\mn@urlcharsother{\let\do\@makeother \do\$\do\&\do\#\do\^\do\_\do\%\do\~}
\def\mn@doi{\begingroup\mn@urlcharsother \@ifnextchar [ {\mn@doi@} {\mn@doi@[]}}
\def\mn@doi@[#1]#2{\def\@tempa{#1}\ifx\@tempa\@empty \href {http://dx.doi.org/#2} {doi:#2}\else \href {http://dx.doi.org/#2} {#1}\fi \endgroup}
\def\mn@eprint#1#2{\mn@eprint@#1:#2::\@nil}
\def\mn@eprint@arXiv#1{\href {http://arxiv.org/abs/#1} {{\tt arXiv:#1}}}
\def\mn@eprint@dblp#1{\href {http://dblp.uni-trier.de/rec/bibtex/#1.xml} {dblp:#1}}
\def\mn@eprint@#1:#2:#3:#4\@nil{\def\@tempa {#1}\def\@tempb {#2}\def\@tempc {#3}\ifx \@tempc \@empty \let \@tempc \@tempb \let \@tempb \@tempa \fi \ifx \@tempb \@empty \def\@tempb {arXiv}\fi \@ifundefined {mn@eprint@\@tempb}{\@tempb:\@tempc}{\expandafter \expandafter \csname mn@eprint@\@tempb\endcsname \expandafter{\@tempc}}}

\bibitem[\protect\citeauthoryear{{Adler}}{{Adler}}{1971}]{1971AnPhy..67..599A}
{Adler} S.~L.,  1971, \mn@doi [Annals of Physics] {10.1016/0003-4916(71)90154-0}, \href {https://ui.adsabs.harvard.edu/abs/1971AnPhy..67..599A} {67, 599}

\bibitem[\protect\citeauthoryear{Akgün, Cerdá-Durán, Miralles  \& Pons}{Akgün et~al.}{2018}]{akgun_crustmagnetosphere_2018}
Akgün T.,  Cerdá-Durán P.,  Miralles J.~A.,   Pons J.~A.,  2018, \mn@doi [Monthly Notices of the Royal Astronomical Society] {10.1093/mnras/sty2669}, 481, 5331

\bibitem[\protect\citeauthoryear{Andrade}{Andrade}{1934}]{andrade_theory_1934}
Andrade E. N.~C.,  1934, \mn@doi [The London, Edinburgh, and Dublin Philosophical Magazine and Journal of Science] {10.1080/14786443409462409}, 17, 497

\bibitem[\protect\citeauthoryear{Arons \& Barnard}{Arons \& Barnard}{1986}]{arons_wave_1986}
Arons J.,  Barnard J.~J.,  1986, \mn@doi [The Astrophysical Journal] {10.1086/163978}, 302, 120

\bibitem[\protect\citeauthoryear{{Arzoumanian}, {Chernoff}  \& {Cordes}}{{Arzoumanian} et~al.}{2002}]{2002ApJ...568..289A}
{Arzoumanian} Z.,  {Chernoff} D.~F.,   {Cordes} J.~M.,  2002, \mn@doi [\apj] {10.1086/338805}, \href {https://ui.adsabs.harvard.edu/abs/2002ApJ...568..289A} {568, 289}

\bibitem[\protect\citeauthoryear{{Backer}}{{Backer}}{1970}]{backer_pulsar_1970}
{Backer} D.~C.,  1970, \mn@doi [\nat] {10.1038/228042a0}, \href {https://ui.adsabs.harvard.edu/abs/1970Natur.228...42B} {228, 42}

\bibitem[\protect\citeauthoryear{{Bak}, {Tang}  \& {Wiesenfeld}}{{Bak} et~al.}{1987}]{1987PhRvL..59..381B}
{Bak} P.,  {Tang} C.,   {Wiesenfeld} K.,  1987, \mn@doi [\prl] {10.1103/PhysRevLett.59.381}, \href {https://ui.adsabs.harvard.edu/abs/1987PhRvL..59..381B} {59, 381}

\bibitem[\protect\citeauthoryear{{Baring}}{{Baring}}{1991}]{1991A&A...249..581B}
{Baring} M.~G.,  1991, \aap, \href {https://ui.adsabs.harvard.edu/abs/1991A&A...249..581B} {249, 581}

\bibitem[\protect\citeauthoryear{{Baring} \& {Harding}}{{Baring} \& {Harding}}{1997}]{1997ApJ...482..372B}
{Baring} M.~G.,  {Harding} A.~K.,  1997, \mn@doi [\apj] {10.1086/304152}, \href {https://ui.adsabs.harvard.edu/abs/1997ApJ...482..372B} {482, 372}

\bibitem[\protect\citeauthoryear{{Baring} \& {Harding}}{{Baring} \& {Harding}}{2001}]{2001ApJ...547..929B}
{Baring} M.~G.,  {Harding} A.~K.,  2001, \mn@doi [\apj] {10.1086/318390}, \href {https://ui.adsabs.harvard.edu/abs/2001ApJ...547..929B} {547, 929}

\bibitem[\protect\citeauthoryear{{Baring} \& {Harding}}{{Baring} \& {Harding}}{2007}]{2007Ap&SS.308..109B}
{Baring} M.~G.,  {Harding} A.~K.,  2007, \mn@doi [\apss] {10.1007/s10509-007-9326-x}, \href {https://ui.adsabs.harvard.edu/abs/2007Ap&SS.308..109B} {308, 109}

\bibitem[\protect\citeauthoryear{{Baring}, {Gonthier}  \& {Harding}}{{Baring} et~al.}{2005}]{2005ApJ...630..430B}
{Baring} M.~G.,  {Gonthier} P.~L.,   {Harding} A.~K.,  2005, \mn@doi [\apj] {10.1086/431895}, \href {https://ui.adsabs.harvard.edu/abs/2005ApJ...630..430B} {630, 430}

\bibitem[\protect\citeauthoryear{Baring, Wadiasingh  \& Gonthier}{Baring et~al.}{2011}]{baring_cooling_2011}
Baring M.~G.,  Wadiasingh Z.,   Gonthier P.~L.,  2011, \mn@doi [The Astrophysical Journal] {10.1088/0004-637X/733/1/61}, 733, 61

\bibitem[\protect\citeauthoryear{Beloborodov}{Beloborodov}{2009}]{beloborodov_untwisting_2009}
Beloborodov A.~M.,  2009, \mn@doi [The Astrophysical Journal] {10.1088/0004-637X/703/1/1044}, 703, 1044

\bibitem[\protect\citeauthoryear{Beloborodov \& Li}{Beloborodov \& Li}{2016}]{beloborodov_magnetar_2016}
Beloborodov A.~M.,  Li X.,  2016, \mn@doi [The Astrophysical Journal] {10.3847/1538-4357/833/2/261}, 833, 261

\bibitem[\protect\citeauthoryear{Beloborodov \& Thompson}{Beloborodov \& Thompson}{2007}]{beloborodov_corona_2007}
Beloborodov A.~M.,  Thompson C.,  2007, \mn@doi [The Astrophysical Journal] {10.1086/508917}, 657, 967

\bibitem[\protect\citeauthoryear{{Ben{\'a}{\v{c}}ek}, {Timokhin}, {Mu{\~n}oz}, {Jessner}, {Rievajov{\'a}}, {Pohl}  \& {B{\"u}chner}}{{Ben{\'a}{\v{c}}ek} et~al.}{2024}]{2024arXiv240520866B}
{Ben{\'a}{\v{c}}ek} J.,  {Timokhin} A.,  {Mu{\~n}oz} P.~A.,  {Jessner} A.,  {Rievajov{\'a}} T.,  {Pohl} M.,   {B{\"u}chner} J.,  2024, \mn@doi [arXiv e-prints] {10.48550/arXiv.2405.20866}, \href {https://ui.adsabs.harvard.edu/abs/2024arXiv240520866B} {p. arXiv:2405.20866}

\bibitem[\protect\citeauthoryear{{Beniamini} \& {Kumar}}{{Beniamini} \& {Kumar}}{2020}]{Beniamini+2020}
{Beniamini} P.,  {Kumar} P.,  2020, \mn@doi [\mnras] {10.1093/mnras/staa2489}, \href {https://ui.adsabs.harvard.edu/abs/2020MNRAS.498..651B} {498, 651}

\bibitem[\protect\citeauthoryear{Beniamini, Wadiasingh  \& Metzger}{Beniamini et~al.}{2020}]{beniamini_periodicity_2020}
Beniamini P.,  Wadiasingh Z.,   Metzger B.~D.,  2020, \mn@doi [Monthly Notices of the Royal Astronomical Society] {10.1093/mnras/staa1783}, 496, 3390

\bibitem[\protect\citeauthoryear{Beniamini, Wadiasingh, Hare, Rajwade, Younes  \& van~der Horst}{Beniamini et~al.}{2023}]{beniamini_evidence_2023}
Beniamini P.,  Wadiasingh Z.,  Hare J.,  Rajwade K.~M.,  Younes G.,   van~der Horst A.~J.,  2023, \mn@doi [Monthly Notices of the Royal Astronomical Society] {10.1093/mnras/stad208}, 520, 1872

\bibitem[\protect\citeauthoryear{{Bilous} et~al.,}{{Bilous} et~al.}{2019}]{bilous_nicer_2019}
{Bilous} A.~V.,  et~al., 2019, \mn@doi [\apjl] {10.3847/2041-8213/ab53e7}, \href {https://ui.adsabs.harvard.edu/abs/2019ApJ...887L..23B} {887, L23}

\bibitem[\protect\citeauthoryear{{Blandford}, {Applegate}  \& {Hernquist}}{{Blandford} et~al.}{1983}]{Blandford_1983}
{Blandford} R.~D.,  {Applegate} J.~H.,   {Hernquist} L.,  1983, \mn@doi [\mnras] {10.1093/mnras/204.4.1025}, \href {https://ui.adsabs.harvard.edu/abs/1983MNRAS.204.1025B} {204, 1025}

\bibitem[\protect\citeauthoryear{{Borghese} et~al.,}{{Borghese} et~al.}{2018}]{2018MNRAS.478..741B}
{Borghese} A.,  et~al., 2018, \mn@doi [\mnras] {10.1093/mnras/sty1119}, \href {https://ui.adsabs.harvard.edu/abs/2018MNRAS.478..741B} {478, 741}

\bibitem[\protect\citeauthoryear{Braun, Safi-Harb  \& Fryer}{Braun et~al.}{2019}]{braun_progenitors_2019}
Braun C.,  Safi-Harb S.,   Fryer C.~L.,  2019, \mn@doi [Monthly Notices of the Royal Astronomical Society] {10.1093/mnras/stz2437}, 489, 4444

\bibitem[\protect\citeauthoryear{{Bulik} \& {Miller}}{{Bulik} \& {Miller}}{1997}]{1997MNRAS.288..596B}
{Bulik} T.,  {Miller} M.~C.,  1997, \mn@doi [\mnras] {10.1093/mnras/288.3.596}, \href {https://ui.adsabs.harvard.edu/abs/1997MNRAS.288..596B} {288, 596}

\bibitem[\protect\citeauthoryear{{Bussard}, {Alexander}  \& {Meszaros}}{{Bussard} et~al.}{1986}]{1986PhRvD..34..440B}
{Bussard} R.~W.,  {Alexander} S.~B.,   {Meszaros} P.,  1986, \mn@doi [\prd] {10.1103/PhysRevD.34.440}, \href {https://ui.adsabs.harvard.edu/abs/1986PhRvD..34..440B} {34, 440}

\bibitem[\protect\citeauthoryear{Caleb et~al.,}{Caleb et~al.}{2022}]{caleb_discovery_2022}
Caleb M.,  et~al., 2022, \mn@doi [Nature Astronomy] {10.1038/s41550-022-01688-x}, 6, 828

\bibitem[\protect\citeauthoryear{{Caleb} et~al.,}{{Caleb} et~al.}{2024}]{caleb_2024}
{Caleb} M.,  et~al., 2024, \mn@doi [Nature Astronomy] {10.1038/s41550-024-02277-w}, \href {https://ui.adsabs.harvard.edu/abs/2024NatAs.tmp..107C} {}

\bibitem[\protect\citeauthoryear{{Camilo}, {Ransom}, {Halpern}, {Reynolds}, {Helfand}, {Zimmerman}  \& {Sarkissian}}{{Camilo} et~al.}{2006}]{2006Natur.442..892C}
{Camilo} F.,  {Ransom} S.~M.,  {Halpern} J.~P.,  {Reynolds} J.,  {Helfand} D.~J.,  {Zimmerman} N.,   {Sarkissian} J.,  2006, \mn@doi [\nat] {10.1038/nature04986}, \href {https://ui.adsabs.harvard.edu/abs/2006Natur.442..892C} {442, 892}

\bibitem[\protect\citeauthoryear{{Camilo}, {Ransom}, {Chatterjee}, {Johnston}  \& {Demorest}}{{Camilo} et~al.}{2012}]{camilo_psr_2012}
{Camilo} F.,  {Ransom} S.~M.,  {Chatterjee} S.,  {Johnston} S.,   {Demorest} P.,  2012, \mn@doi [\apj] {10.1088/0004-637X/746/1/63}, \href {https://ui.adsabs.harvard.edu/abs/2012ApJ...746...63C} {746, 63}

\bibitem[\protect\citeauthoryear{Camilo et~al.,}{Camilo et~al.}{2016}]{camilo_radio_2016}
Camilo F.,  et~al., 2016, \mn@doi [The Astrophysical Journal] {10.3847/0004-637X/820/2/110}, 820, 110

\bibitem[\protect\citeauthoryear{Chang, Hsiang, Chu, Chung, Su, Lin  \& Huang}{Chang et~al.}{2023}]{chang_observational_2023}
Chang H.-K.,  Hsiang J.-Y.,  Chu C.-Y.,  Chung Y.-H.,  Su T.-H.,  Lin T.-H.,   Huang C.-Y.,  2023, \mn@doi [Monthly Notices of the Royal Astronomical Society] {10.1093/mnras/stad400}, 520, 4068

\bibitem[\protect\citeauthoryear{Chatterjee et~al.,}{Chatterjee et~al.}{2017}]{chatterjee_direct_2017}
Chatterjee S.,  et~al., 2017, \mn@doi [Nature] {10.1038/nature20797}, 541, 58

\bibitem[\protect\citeauthoryear{Chen \& Beloborodov}{Chen \& Beloborodov}{2017}]{chen_particle--cell_2017}
Chen A.~Y.,  Beloborodov A.~M.,  2017, \mn@doi [The Astrophysical Journal] {10.3847/1538-4357/aa7a57}, 844, 133

\bibitem[\protect\citeauthoryear{{Chen} \& {Ruderman}}{{Chen} \& {Ruderman}}{1993}]{1993ApJ...402..264C}
{Chen} K.,  {Ruderman} M.,  1993, \mn@doi [\apj] {10.1086/172129}, \href {https://ui.adsabs.harvard.edu/abs/1993ApJ...402..264C} {402, 264}

\bibitem[\protect\citeauthoryear{Cooper \& Wijers}{Cooper \& Wijers}{2021}]{cooper_coherent_2021}
Cooper A.~J.,  Wijers R. A. M.~J.,  2021, \mn@doi [Monthly Notices of the Royal Astronomical Society: Letters] {10.1093/mnrasl/slab099}, 508, L32

\bibitem[\protect\citeauthoryear{Cooper et~al.,}{Cooper et~al.}{2022}]{cooper_testing_2022}
Cooper A.~J.,  et~al., 2022, \mn@doi [Monthly Notices of the Royal Astronomical Society] {10.1093/mnras/stac2951}, 517, 5483

\bibitem[\protect\citeauthoryear{Cooper, Gupta, Wadiasingh, Wijers, Boersma, Andreoni, Rowlinson  \& Gourdji}{Cooper et~al.}{2023}]{cooper_pulsar_2023}
Cooper A.~J.,  Gupta O.,  Wadiasingh Z.,  Wijers R. A. M.~J.,  Boersma O.~M.,  Andreoni I.,  Rowlinson A.,   Gourdji K.,  2023, \mn@doi [Monthly Notices of the Royal Astronomical Society] {10.1093/mnras/stac3580}, 519, 3923

\bibitem[\protect\citeauthoryear{Coti~Zelati, Rea, Pons, Campana  \& Esposito}{Coti~Zelati et~al.}{2018}]{coti_zelati_systematic_2018}
Coti~Zelati F.,  Rea N.,  Pons J.~A.,  Campana S.,   Esposito P.,  2018, \mn@doi [Monthly Notices of the Royal Astronomical Society] {10.1093/mnras/stx2679}, 474, 961

\bibitem[\protect\citeauthoryear{Cruces et~al.,}{Cruces et~al.}{2020}]{cruces_repeating_2020}
Cruces M.,  et~al., 2020, \mn@doi [Monthly Notices of the Royal Astronomical Society] {10.1093/mnras/staa3223}, 500, 448

\bibitem[\protect\citeauthoryear{{D'A{\`\i}} et~al.,}{{D'A{\`\i}} et~al.}{2016}]{2016MNRAS.463.2394D}
{D'A{\`\i}} A.,  et~al., 2016, \mn@doi [\mnras] {10.1093/mnras/stw2023}, \href {https://ui.adsabs.harvard.edu/abs/2016MNRAS.463.2394D} {463, 2394}

\bibitem[\protect\citeauthoryear{{Daugherty} \& {Harding}}{{Daugherty} \& {Harding}}{1982}]{1982ApJ...252..337D}
{Daugherty} J.~K.,  {Harding} A.~K.,  1982, \mn@doi [\apj] {10.1086/159561}, \href {https://ui.adsabs.harvard.edu/abs/1982ApJ...252..337D} {252, 337}

\bibitem[\protect\citeauthoryear{Daugherty \& Harding}{Daugherty \& Harding}{1983}]{daugherty_pair_1983}
Daugherty J.~K.,  Harding A.~K.,  1983, \mn@doi [The Astrophysical Journal] {10.1086/161411}, 273, 761

\bibitem[\protect\citeauthoryear{{Daugherty} \& {Harding}}{{Daugherty} \& {Harding}}{1986}]{1986ApJ...309..362D}
{Daugherty} J.~K.,  {Harding} A.~K.,  1986, \mn@doi [\apj] {10.1086/164608}, \href {https://ui.adsabs.harvard.edu/abs/1986ApJ...309..362D} {309, 362}

\bibitem[\protect\citeauthoryear{{De Luca}, {Caraveo}, {Mereghetti}, {Tiengo}  \& {Bignami}}{{De Luca} et~al.}{2006}]{2006Sci...313..814D}
{De Luca} A.,  {Caraveo} P.~A.,  {Mereghetti} S.,  {Tiengo} A.,   {Bignami} G.~F.,  2006, \mn@doi [Science] {10.1126/science.1129185}, \href {https://ui.adsabs.harvard.edu/abs/2006Sci...313..814D} {313, 814}

\bibitem[\protect\citeauthoryear{{Dermer}}{{Dermer}}{1990}]{1990ApJ...360..197D}
{Dermer} C.~D.,  1990, \mn@doi [\apj] {10.1086/169108}, \href {https://ui.adsabs.harvard.edu/abs/1990ApJ...360..197D} {360, 197}

\bibitem[\protect\citeauthoryear{{Dobie} et~al.,}{{Dobie} et~al.}{2024}]{2024arXiv240612352D}
{Dobie} D.,  et~al., 2024, \mn@doi [arXiv e-prints] {10.48550/arXiv.2406.12352}, \href {https://ui.adsabs.harvard.edu/abs/2024arXiv240612352D} {p. arXiv:2406.12352}

\bibitem[\protect\citeauthoryear{{Drake} \& {Craft}}{{Drake} \& {Craft}}{1968}]{drake_second_1968}
{Drake} F.~D.,  {Craft} H.~D.,  1968, \mn@doi [\nat] {10.1038/220231a0}, \href {https://ui.adsabs.harvard.edu/abs/1968Natur.220..231D} {220, 231}

\bibitem[\protect\citeauthoryear{{Duncan}}{{Duncan}}{1998}]{1998ApJ...498L..45D}
{Duncan} R.~C.,  1998, \mn@doi [\apjl] {10.1086/311303}, \href {https://ui.adsabs.harvard.edu/abs/1998ApJ...498L..45D} {498, L45}

\bibitem[\protect\citeauthoryear{{Edgeworth}, {Dalton}  \& {Parnell}}{{Edgeworth} et~al.}{1984}]{edgeworth_pitch_1984}
{Edgeworth} R.,  {Dalton} B.~J.,   {Parnell} T.,  1984, \mn@doi [European Journal of Physics] {10.1088/0143-0807/5/4/003}, \href {https://ui.adsabs.harvard.edu/abs/1984EJPh....5..198E} {5, 198}

\bibitem[\protect\citeauthoryear{{Elfritz}, {Pons}, {Rea}, {Glampedakis}  \& {Vigan{\`o}}}{{Elfritz} et~al.}{2016}]{elfritz_simulated_2016}
{Elfritz} J.~G.,  {Pons} J.~A.,  {Rea} N.,  {Glampedakis} K.,   {Vigan{\`o}} D.,  2016, \mn@doi [\mnras] {10.1093/mnras/stv2963}, \href {https://ui.adsabs.harvard.edu/abs/2016MNRAS.456.4461E} {456, 4461}

\bibitem[\protect\citeauthoryear{{Enoto} et~al.,}{{Enoto} et~al.}{2017}]{2017ApJS..231....8E}
{Enoto} T.,  et~al., 2017, \mn@doi [\apjs] {10.3847/1538-4365/aa6f0a}, \href {https://ui.adsabs.harvard.edu/abs/2017ApJS..231....8E} {231, 8}

\bibitem[\protect\citeauthoryear{{Erber}}{{Erber}}{1966}]{1966RvMP...38..626E}
{Erber} T.,  1966, \mn@doi [Reviews of Modern Physics] {10.1103/RevModPhys.38.626}, \href {https://ui.adsabs.harvard.edu/abs/1966RvMP...38..626E} {38, 626}

\bibitem[\protect\citeauthoryear{{Fan}, {Xu}  \& {Chen}}{{Fan} et~al.}{2024}]{2024arXiv240403882F}
{Fan} Y.-N.,  {Xu} K.,   {Chen} W.-C.,  2024, \mn@doi [arXiv e-prints] {10.48550/arXiv.2404.03882}, \href {https://ui.adsabs.harvard.edu/abs/2024arXiv240403882F} {p. arXiv:2404.03882}

\bibitem[\protect\citeauthoryear{{Fiori}, {Razzano}, {Harding}, {Kerr}, {Mignani}  \& {Saz Parkinson}}{{Fiori} et~al.}{2024}]{2024A&A...685A..70F}
{Fiori} A.,  {Razzano} M.,  {Harding} A.~K.,  {Kerr} M.,  {Mignani} R.~P.,   {Saz Parkinson} P.~M.,  2024, \mn@doi [\aap] {10.1051/0004-6361/202348924}, \href {https://ui.adsabs.harvard.edu/abs/2024A&A...685A..70F} {685, A70}

\bibitem[\protect\citeauthoryear{{Gakis} \& {Gourgouliatos}}{{Gakis} \& {Gourgouliatos}}{2024}]{2024arXiv240214911G}
{Gakis} D.,  {Gourgouliatos} K.~N.,  2024, \mn@doi [arXiv e-prints] {10.48550/arXiv.2402.14911}, \href {https://ui.adsabs.harvard.edu/abs/2024arXiv240214911G} {p. arXiv:2402.14911}

\bibitem[\protect\citeauthoryear{{Goldreich} \& {Reisenegger}}{{Goldreich} \& {Reisenegger}}{1992}]{goldreich_magnetic_1992}
{Goldreich} P.,  {Reisenegger} A.,  1992, \mn@doi [\apj] {10.1086/171646}, \href {https://ui.adsabs.harvard.edu/abs/1992ApJ...395..250G} {395, 250}

\bibitem[\protect\citeauthoryear{{Gonthier}, {Harding}, {Baring}, {Costello}  \& {Mercer}}{{Gonthier} et~al.}{2000}]{2000ApJ...540..907G}
{Gonthier} P.~L.,  {Harding} A.~K.,  {Baring} M.~G.,  {Costello} R.~M.,   {Mercer} C.~L.,  2000, \mn@doi [\apj] {10.1086/309357}, \href {https://ui.adsabs.harvard.edu/abs/2000ApJ...540..907G} {540, 907}

\bibitem[\protect\citeauthoryear{{Gonthier}, {Baring}, {Eiles}, {Wadiasingh}, {Taylor}  \& {Fitch}}{{Gonthier} et~al.}{2014}]{2014PhRvD..90d3014G}
{Gonthier} P.~L.,  {Baring} M.~G.,  {Eiles} M.~T.,  {Wadiasingh} Z.,  {Taylor} C.~A.,   {Fitch} C.~J.,  2014, \mn@doi [\prd] {10.1103/PhysRevD.90.043014}, \href {https://ui.adsabs.harvard.edu/abs/2014PhRvD..90d3014G} {90, 043014}

\bibitem[\protect\citeauthoryear{{Gonz{\'a}lez-Caniulef}, {Zane}, {Turolla}  \& {Wu}}{{Gonz{\'a}lez-Caniulef} et~al.}{2019}]{2019MNRAS.483..599G}
{Gonz{\'a}lez-Caniulef} D.,  {Zane} S.,  {Turolla} R.,   {Wu} K.,  2019, \mn@doi [\mnras] {10.1093/mnras/sty3159}, \href {https://ui.adsabs.harvard.edu/abs/2019MNRAS.483..599G} {483, 599}

\bibitem[\protect\citeauthoryear{Gourgouliatos \& Lander}{Gourgouliatos \& Lander}{2021}]{gourgouliatos_axisymmetric_2021}
Gourgouliatos K.~N.,  Lander S.~K.,  2021, \mn@doi [Monthly Notices of the Royal Astronomical Society] {10.1093/mnras/stab1869}, 506, 3578

\bibitem[\protect\citeauthoryear{{Gourgouliatos}, {Wood}  \& {Hollerbach}}{{Gourgouliatos} et~al.}{2016}]{2016PNAS..113.3944G}
{Gourgouliatos} K.~N.,  {Wood} T.~S.,   {Hollerbach} R.,  2016, \mn@doi [Proceedings of the National Academy of Science] {10.1073/pnas.1522363113}, \href {https://ui.adsabs.harvard.edu/abs/2016PNAS..113.3944G} {113, 3944}

\bibitem[\protect\citeauthoryear{{Harding} \& {Lai}}{{Harding} \& {Lai}}{2006}]{2006RPPh...69.2631H}
{Harding} A.~K.,  {Lai} D.,  2006, \mn@doi [Reports on Progress in Physics] {10.1088/0034-4885/69/9/R03}, \href {https://ui.adsabs.harvard.edu/abs/2006RPPh...69.2631H} {69, 2631}

\bibitem[\protect\citeauthoryear{{Harding} \& {Muslimov}}{{Harding} \& {Muslimov}}{1998}]{1998ApJ...508..328H}
{Harding} A.~K.,  {Muslimov} A.~G.,  1998, \mn@doi [\apj] {10.1086/306394}, \href {https://ui.adsabs.harvard.edu/abs/1998ApJ...508..328H} {508, 328}

\bibitem[\protect\citeauthoryear{{Harding} \& {Muslimov}}{{Harding} \& {Muslimov}}{2011}]{2011ApJ...726L..10H}
{Harding} A.~K.,  {Muslimov} A.~G.,  2011, \mn@doi [\apjl] {10.1088/2041-8205/726/1/L10}, \href {https://ui.adsabs.harvard.edu/abs/2011ApJ...726L..10H} {726, L10}

\bibitem[\protect\citeauthoryear{{Harding}, {Baring}  \& {Gonthier}}{{Harding} et~al.}{1997}]{1997ApJ...476..246H}
{Harding} A.~K.,  {Baring} M.~G.,   {Gonthier} P.~L.,  1997, \mn@doi [\apj] {10.1086/303605}, \href {https://ui.adsabs.harvard.edu/abs/1997ApJ...476..246H} {476, 246}

\bibitem[\protect\citeauthoryear{{Harding}, {Muslimov}  \& {Zhang}}{{Harding} et~al.}{2002}]{2002ApJ...576..366H}
{Harding} A.~K.,  {Muslimov} A.~G.,   {Zhang} B.,  2002, \mn@doi [\apj] {10.1086/341633}, \href {https://ui.adsabs.harvard.edu/abs/2002ApJ...576..366H} {576, 366}

\bibitem[\protect\citeauthoryear{{Harding}, {Wadiasingh}  \& {Baring}}{{Harding} et~al.}{2024}]{2024HEAD...2110725H}
{Harding} A.,  {Wadiasingh} Z.,   {Baring} M.~G.,  2024, in AAS/High Energy Astrophysics Division. p. 107.25

\bibitem[\protect\citeauthoryear{{Hare}, {Pavlov}, {Posselt}, {Kargaltsev}, {Temim}  \& {Chen}}{{Hare} et~al.}{2024}]{2024arXiv240503947H}
{Hare} J.,  {Pavlov} G.~G.,  {Posselt} B.,  {Kargaltsev} O.,  {Temim} T.,   {Chen} S.,  2024, \mn@doi [arXiv e-prints] {10.48550/arXiv.2405.03947}, \href {https://ui.adsabs.harvard.edu/abs/2024arXiv240503947H} {p. arXiv:2405.03947}

\bibitem[\protect\citeauthoryear{{Harko} \& {Cheng}}{{Harko} \& {Cheng}}{2002}]{2002MNRAS.335...99H}
{Harko} T.,  {Cheng} K.~S.,  2002, \mn@doi [\mnras] {10.1046/j.1365-8711.2002.05598.x}, \href {https://ui.adsabs.harvard.edu/abs/2002MNRAS.335...99H} {335, 99}

\bibitem[\protect\citeauthoryear{{Herold}}{{Herold}}{1979}]{1979PhRvD..19.2868H}
{Herold} H.,  1979, \mn@doi [\prd] {10.1103/PhysRevD.19.2868}, \href {https://ui.adsabs.harvard.edu/abs/1979PhRvD..19.2868H} {19, 2868}

\bibitem[\protect\citeauthoryear{{Hewitt} et~al.,}{{Hewitt} et~al.}{2023}]{hewitt_2023}
{Hewitt} D.~M.,  et~al., 2023, \mn@doi [\mnras] {10.1093/mnras/stad2847}, \href {https://ui.adsabs.harvard.edu/abs/2023MNRAS.526.2039H} {526, 2039}

\bibitem[\protect\citeauthoryear{{Hibschman} \& {Arons}}{{Hibschman} \& {Arons}}{2001a}]{2001ApJ...554..624H}
{Hibschman} J.~A.,  {Arons} J.,  2001a, \mn@doi [\apj] {10.1086/321378}, \href {https://ui.adsabs.harvard.edu/abs/2001ApJ...554..624H} {554, 624}

\bibitem[\protect\citeauthoryear{{Hibschman} \& {Arons}}{{Hibschman} \& {Arons}}{2001b}]{2001ApJ...560..871H}
{Hibschman} J.~A.,  {Arons} J.,  2001b, \mn@doi [\apj] {10.1086/323069}, \href {https://ui.adsabs.harvard.edu/abs/2001ApJ...560..871H} {560, 871}

\bibitem[\protect\citeauthoryear{{Hiramatsu} et~al.,}{{Hiramatsu} et~al.}{2023}]{hiramatsu_limits_2023}
{Hiramatsu} D.,  et~al., 2023, \mn@doi [\apjl] {10.3847/2041-8213/acae98}, \href {https://ui.adsabs.harvard.edu/abs/2023ApJ...947L..28H} {947, L28}

\bibitem[\protect\citeauthoryear{{Ho}, {Andersson}  \& {Graber}}{{Ho} et~al.}{2017}]{2017PhRvC..96f5801H}
{Ho} W. C.~G.,  {Andersson} N.,   {Graber} V.,  2017, \mn@doi [\prc] {10.1103/PhysRevC.96.065801}, \href {https://ui.adsabs.harvard.edu/abs/2017PhRvC..96f5801H} {96, 065801}

\bibitem[\protect\citeauthoryear{{Hoffman} \& {Heyl}}{{Hoffman} \& {Heyl}}{2012}]{2012MNRAS.426.2404H}
{Hoffman} K.,  {Heyl} J.,  2012, \mn@doi [\mnras] {10.1111/j.1365-2966.2012.21921.x}, \href {https://ui.adsabs.harvard.edu/abs/2012MNRAS.426.2404H} {426, 2404}

\bibitem[\protect\citeauthoryear{{Horowitz} \& {Kadau}}{{Horowitz} \& {Kadau}}{2009}]{2009PhRvL.102s1102H}
{Horowitz} C.~J.,  {Kadau} K.,  2009, \mn@doi [\prl] {10.1103/PhysRevLett.102.191102}, \href {https://ui.adsabs.harvard.edu/abs/2009PhRvL.102s1102H} {102, 191102}

\bibitem[\protect\citeauthoryear{{Hu}, {Baring}, {Wadiasingh}  \& {Harding}}{{Hu} et~al.}{2019}]{2019MNRAS.486.3327H}
{Hu} K.,  {Baring} M.~G.,  {Wadiasingh} Z.,   {Harding} A.~K.,  2019, \mn@doi [\mnras] {10.1093/mnras/stz995}, \href {https://ui.adsabs.harvard.edu/abs/2019MNRAS.486.3327H} {486, 3327}

\bibitem[\protect\citeauthoryear{Hu, Baring, Harding  \& Wadiasingh}{Hu et~al.}{2022}]{hu_high-energy_2022}
Hu K.,  Baring M.~G.,  Harding A.~K.,   Wadiasingh Z.,  2022, \mn@doi [The Astrophysical Journal] {10.3847/1538-4357/ac9611}, 940, 91

\bibitem[\protect\citeauthoryear{{Hu} et~al.,}{{Hu} et~al.}{2024}]{2024Natur.626..500H}
{Hu} C.-P.,  et~al., 2024, \mn@doi [\nat] {10.1038/s41586-023-07012-5}, \href {https://ui.adsabs.harvard.edu/abs/2024Natur.626..500H} {626, 500}

\bibitem[\protect\citeauthoryear{Hurley-Walker et~al.,}{Hurley-Walker et~al.}{2022a}]{hurley-walker_galactic_2022}
Hurley-Walker N.,  et~al., 2022a, \mn@doi [Publications of the Astronomical Society of Australia] {10.1017/pasa.2022.17}, 39, e035

\bibitem[\protect\citeauthoryear{Hurley-Walker et~al.,}{Hurley-Walker et~al.}{2022b}]{hurley-walker_radio_2022}
Hurley-Walker N.,  et~al., 2022b, \mn@doi [Nature] {10.1038/s41586-021-04272-x}, 601, 526

\bibitem[\protect\citeauthoryear{Hurley-Walker et~al.,}{Hurley-Walker et~al.}{2023}]{hurley-walker_long-period_2023}
Hurley-Walker N.,  et~al., 2023, \mn@doi [Nature] {10.1038/s41586-023-06202-5}, 619, 487

\bibitem[\protect\citeauthoryear{Hyman, Lazio, Kassim, Ray, Markwardt  \& Yusef-Zadeh}{Hyman et~al.}{2005}]{hyman_powerful_2005}
Hyman S.~D.,  Lazio T. J.~W.,  Kassim N.~E.,  Ray P.~S.,  Markwardt C.~B.,   Yusef-Zadeh F.,  2005, \mn@doi [Nature] {10.1038/nature03400}, 434, 50

\bibitem[\protect\citeauthoryear{{Hyman}, {Lazio}, {Roy}, {Ray}, {Kassim}  \& {Neureuther}}{{Hyman} et~al.}{2006}]{hyman_new_2006}
{Hyman} S.~D.,  {Lazio} T. J.~W.,  {Roy} S.,  {Ray} P.~S.,  {Kassim} N.~E.,   {Neureuther} J.~L.,  2006, \mn@doi [\apj] {10.1086/499294}, \href {https://ui.adsabs.harvard.edu/abs/2006ApJ...639..348H} {639, 348}

\bibitem[\protect\citeauthoryear{{Igoshev}, {Hollerbach}, {Wood}  \& {Gourgouliatos}}{{Igoshev} et~al.}{2021}]{igoshev_2021_strong}
{Igoshev} A.~P.,  {Hollerbach} R.,  {Wood} T.,   {Gourgouliatos} K.~N.,  2021, \mn@doi [Nature Astronomy] {10.1038/s41550-020-01220-z}, \href {https://ui.adsabs.harvard.edu/abs/2021NatAs...5..145I} {5, 145}

\bibitem[\protect\citeauthoryear{{Jenet}, {Anderson}, {Kaspi}, {Prince}  \& {Unwin}}{{Jenet} et~al.}{1998}]{jenet_radio_1998}
{Jenet} F.~A.,  {Anderson} S.~B.,  {Kaspi} V.~M.,  {Prince} T.~A.,   {Unwin} S.~C.,  1998, \mn@doi [\apj] {10.1086/305529}, \href {https://ui.adsabs.harvard.edu/abs/1998ApJ...498..365J} {498, 365}

\bibitem[\protect\citeauthoryear{{Johnson} \& {Lippmann}}{{Johnson} \& {Lippmann}}{1949}]{1949PhRv...76..828J}
{Johnson} M.~H.,  {Lippmann} B.~A.,  1949, \mn@doi [Physical Review] {10.1103/PhysRev.76.828}, \href {https://ui.adsabs.harvard.edu/abs/1949PhRv...76..828J} {76, 828}

\bibitem[\protect\citeauthoryear{{Johnston}, {Mitra}, {Keith}, {Oswald}  \& {Karastergiou}}{{Johnston} et~al.}{2024}]{2024arXiv240410254J}
{Johnston} S.,  {Mitra} D.,  {Keith} M.,  {Oswald} L.,   {Karastergiou} A.,  2024, \mn@doi [arXiv e-prints] {10.48550/arXiv.2404.10254}, \href {https://ui.adsabs.harvard.edu/abs/2024arXiv240410254J} {p. arXiv:2404.10254}

\bibitem[\protect\citeauthoryear{Kalapotharakos, Harding, Kazanas  \& Wadiasingh}{Kalapotharakos et~al.}{2019}]{kalapotharakos_fundamental_2019}
Kalapotharakos C.,  Harding A.~K.,  Kazanas D.,   Wadiasingh Z.,  2019, \mn@doi [The Astrophysical Journal] {10.3847/2041-8213/ab3e0a}, 883, L4

\bibitem[\protect\citeauthoryear{{Kalapotharakos}, {Wadiasingh}, {Harding}  \& {Kazanas}}{{Kalapotharakos} et~al.}{2021}]{2021ApJ...907...63K}
{Kalapotharakos} C.,  {Wadiasingh} Z.,  {Harding} A.~K.,   {Kazanas} D.,  2021, \mn@doi [\apj] {10.3847/1538-4357/abcec0}, \href {https://ui.adsabs.harvard.edu/abs/2021ApJ...907...63K} {907, 63}

\bibitem[\protect\citeauthoryear{Kalapotharakos, Wadiasingh, Harding  \& Kazanas}{Kalapotharakos et~al.}{2022}]{kalapotharakos_fundamental_2022}
Kalapotharakos C.,  Wadiasingh Z.,  Harding A.~K.,   Kazanas D.,  2022, \mn@doi [The Astrophysical Journal] {10.3847/1538-4357/ac78e3}, 934, 65

\bibitem[\protect\citeauthoryear{{Kaplan}, {Hyman}, {Roy}, {Bandyopadhyay}, {Chakrabarty}, {Kassim}, {Lazio}  \& {Ray}}{{Kaplan} et~al.}{2008}]{kaplan_search_2008}
{Kaplan} D.~L.,  {Hyman} S.~D.,  {Roy} S.,  {Bandyopadhyay} R.~M.,  {Chakrabarty} D.,  {Kassim} N.~E.,  {Lazio} T.~J.~W.,   {Ray} P.~S.,  2008, \mn@doi [\apj] {10.1086/591436}, \href {https://ui.adsabs.harvard.edu/abs/2008ApJ...687..262K} {687, 262}

\bibitem[\protect\citeauthoryear{{Kaspi} \& {Beloborodov}}{{Kaspi} \& {Beloborodov}}{2017}]{kaspi_magnetars_2017}
{Kaspi} V.~M.,  {Beloborodov} A.~M.,  2017, \mn@doi [\araa] {10.1146/annurev-astro-081915-023329}, \href {https://ui.adsabs.harvard.edu/abs/2017ARA&A..55..261K} {55, 261}

\bibitem[\protect\citeauthoryear{{Katz}}{{Katz}}{2022}]{katz_2022}
{Katz} J.~I.,  2022, \mn@doi [\apss] {10.1007/s10509-022-04146-2}, \href {https://ui.adsabs.harvard.edu/abs/2022Ap&SS.367..108K} {367, 108}

\bibitem[\protect\citeauthoryear{{Kelner}, {Prosekin}  \& {Aharonian}}{{Kelner} et~al.}{2015}]{2015AJ....149...33K}
{Kelner} S.~R.,  {Prosekin} A.~Y.,   {Aharonian} F.~A.,  2015, \mn@doi [\aj] {10.1088/0004-6256/149/1/33}, \href {https://ui.adsabs.harvard.edu/abs/2015AJ....149...33K} {149, 33}

\bibitem[\protect\citeauthoryear{{Kerr}}{{Kerr}}{2022}]{2022ApJ...934...30K}
{Kerr} M.,  2022, \mn@doi [\apj] {10.3847/1538-4357/ac7877}, \href {https://ui.adsabs.harvard.edu/abs/2022ApJ...934...30K} {934, 30}

\bibitem[\protect\citeauthoryear{Kilpatrick et~al.,}{Kilpatrick et~al.}{2021}]{kilpatrick_deep_2021}
Kilpatrick C.~D.,  et~al., 2021, \mn@doi [The Astrophysical Journal] {10.3847/2041-8213/abd560}, 907, L3

\bibitem[\protect\citeauthoryear{{Kilpatrick} et~al.,}{{Kilpatrick} et~al.}{2023}]{kilpatrick_limits_2023}
{Kilpatrick} C.~D.,  et~al., 2023, \mn@doi [arXiv e-prints] {10.48550/arXiv.2311.09316}, \href {https://ui.adsabs.harvard.edu/abs/2023arXiv231109316K} {p. arXiv:2311.09316}

\bibitem[\protect\citeauthoryear{{Krause-Polstorff} \& {Michel}}{{Krause-Polstorff} \& {Michel}}{1985}]{krause-polstorff_electrosphere_1985}
{Krause-Polstorff} J.,  {Michel} F.~C.,  1985, \mn@doi [\mnras] {10.1093/mnras/213.1.43P}, \href {https://ui.adsabs.harvard.edu/abs/1985MNRAS.213P..43K} {213, 43}

\bibitem[\protect\citeauthoryear{{Kuiack}, {Wijers}, {Shulevski}  \& {Rowlinson}}{{Kuiack} et~al.}{2021}]{2021MNRAS.504.4706K}
{Kuiack} M.~J.,  {Wijers} R. A.~M.~J.,  {Shulevski} A.,   {Rowlinson} A.,  2021, \mn@doi [\mnras] {10.1093/mnras/stab1156}, \href {https://ui.adsabs.harvard.edu/abs/2021MNRAS.504.4706K} {504, 4706}

\bibitem[\protect\citeauthoryear{{Kuiper}, {Hermsen}  \& {Mendez}}{{Kuiper} et~al.}{2004}]{2004ApJ...613.1173K}
{Kuiper} L.,  {Hermsen} W.,   {Mendez} M.,  2004, \mn@doi [\apj] {10.1086/423129}, \href {https://ui.adsabs.harvard.edu/abs/2004ApJ...613.1173K} {613, 1173}

\bibitem[\protect\citeauthoryear{Kumar \& Bo{\v s}njak}{Kumar \& Bo{\v s}njak}{2020}]{kumar_frb_2020}
Kumar P.,  Bo{\v s}njak {\v Z}.,  2020, \mn@doi [Monthly Notices of the Royal Astronomical Society] {10.1093/mnras/staa774}, 494, 2385

\bibitem[\protect\citeauthoryear{Kumar, Lu  \& Bhattacharya}{Kumar et~al.}{2017}]{kumar_fast_2017}
Kumar P.,  Lu W.,   Bhattacharya M.,  2017, \mn@doi [Monthly Notices of the Royal Astronomical Society] {10.1093/mnras/stx665}, 468, 2726

\bibitem[\protect\citeauthoryear{{Lai} \& {Ho}}{{Lai} \& {Ho}}{2002}]{2002ApJ...566..373L}
{Lai} D.,  {Ho} W. C.~G.,  2002, \mn@doi [\apj] {10.1086/338074}, \href {https://ui.adsabs.harvard.edu/abs/2002ApJ...566..373L} {566, 373}

\bibitem[\protect\citeauthoryear{Lander}{Lander}{2016}]{lander_magnetar_2016}
Lander S.~K.,  2016, \mn@doi [The Astrophysical Journal] {10.3847/2041-8205/824/2/L21}, 824, L21

\bibitem[\protect\citeauthoryear{Lander}{Lander}{2023}]{lander_game_2023}
Lander S.~K.,  2023, \mn@doi [The Astrophysical Journal Letters] {10.3847/2041-8213/acca1f}, 947, L16

\bibitem[\protect\citeauthoryear{Lander \& Gourgouliatos}{Lander \& Gourgouliatos}{2019}]{lander_magnetic-field_2019}
Lander S.~K.,  Gourgouliatos K.~N.,  2019, \mn@doi [Monthly Notices of the Royal Astronomical Society] {10.1093/mnras/stz1042}, 486, 4130

\bibitem[\protect\citeauthoryear{{Lander}, {Andersson}, {Antonopoulou}  \& {Watts}}{{Lander} et~al.}{2015}]{2015MNRAS.449.2047L}
{Lander} S.~K.,  {Andersson} N.,  {Antonopoulou} D.,   {Watts} A.~L.,  2015, \mn@doi [\mnras] {10.1093/mnras/stv432}, \href {https://ui.adsabs.harvard.edu/abs/2015MNRAS.449.2047L} {449, 2047}

\bibitem[\protect\citeauthoryear{Levin et~al.,}{Levin et~al.}{2010}]{levin_radio-loud_2010}
Levin L.,  et~al., 2010, \mn@doi [The Astrophysical Journal] {10.1088/2041-8205/721/1/L33}, 721, L33

\bibitem[\protect\citeauthoryear{{Lu}, {Beniamini}  \& {Kumar}}{{Lu} et~al.}{2022}]{Lu+2022}
{Lu} W.,  {Beniamini} P.,   {Kumar} P.,  2022, \mn@doi [\mnras] {10.1093/mnras/stab3500}, \href {https://ui.adsabs.harvard.edu/abs/2022MNRAS.510.1867L} {510, 1867}

\bibitem[\protect\citeauthoryear{{Lyne} et~al.,}{{Lyne} et~al.}{2017}]{lyne_two_2017}
{Lyne} A.~G.,  et~al., 2017, \mn@doi [\apj] {10.3847/1538-4357/834/1/72}, \href {https://ui.adsabs.harvard.edu/abs/2017ApJ...834...72L} {834, 72}

\bibitem[\protect\citeauthoryear{{Lyubarsky}}{{Lyubarsky}}{2014}]{lyubarsky_model_2014}
{Lyubarsky} Y.,  2014, \mn@doi [\mnras] {10.1093/mnrasl/slu046}, \href {https://ui.adsabs.harvard.edu/abs/2014MNRAS.442L...9L} {442, L9}

\bibitem[\protect\citeauthoryear{{Lyutikov} \& {Popov}}{{Lyutikov} \& {Popov}}{2020}]{2020arXiv200505093L}
{Lyutikov} M.,  {Popov} S.,  2020, \mn@doi [arXiv e-prints] {10.48550/arXiv.2005.05093}, \href {https://ui.adsabs.harvard.edu/abs/2020arXiv200505093L} {p. arXiv:2005.05093}

\bibitem[\protect\citeauthoryear{{Manchester}, {Hobbs}, {Teoh}  \& {Hobbs}}{{Manchester} et~al.}{2005}]{2005AJ....129.1993M}
{Manchester} R.~N.,  {Hobbs} G.~B.,  {Teoh} A.,   {Hobbs} M.,  2005, \mn@doi [\aj] {10.1086/428488}, \href {https://ui.adsabs.harvard.edu/abs/2005AJ....129.1993M} {129, 1993}

\bibitem[\protect\citeauthoryear{Margalit \& Metzger}{Margalit \& Metzger}{2018}]{margalit_concordance_2018}
Margalit B.,  Metzger B.~D.,  2018, \mn@doi [The Astrophysical Journal] {10.3847/2041-8213/aaedad}, 868, L4

\bibitem[\protect\citeauthoryear{{Margalit}, {Beniamini}, {Sridhar}  \& {Metzger}}{{Margalit} et~al.}{2020}]{margalit_implications_2020}
{Margalit} B.,  {Beniamini} P.,  {Sridhar} N.,   {Metzger} B.~D.,  2020, \mn@doi [\apjl] {10.3847/2041-8213/abac57}, \href {https://ui.adsabs.harvard.edu/abs/2020ApJ...899L..27M} {899, L27}

\bibitem[\protect\citeauthoryear{{Marsh} et~al.,}{{Marsh} et~al.}{2016}]{Marsh_2016}
{Marsh} T.~R.,  et~al., 2016, \mn@doi [\nat] {10.1038/nature18620}, \href {https://ui.adsabs.harvard.edu/abs/2016Natur.537..374M} {537, 374}

\bibitem[\protect\citeauthoryear{{Medin} \& {Lai}}{{Medin} \& {Lai}}{2007}]{2007MNRAS.382.1833M}
{Medin} Z.,  {Lai} D.,  2007, \mn@doi [\mnras] {10.1111/j.1365-2966.2007.12492.x}, \href {https://ui.adsabs.harvard.edu/abs/2007MNRAS.382.1833M} {382, 1833}

\bibitem[\protect\citeauthoryear{{Mereghetti}}{{Mereghetti}}{2008}]{2008A&ARv..15..225M}
{Mereghetti} S.,  2008, \mn@doi [\aapr] {10.1007/s00159-008-0011-z}, \href {https://ui.adsabs.harvard.edu/abs/2008A&ARv..15..225M} {15, 225}

\bibitem[\protect\citeauthoryear{Metzger, Margalit  \& Sironi}{Metzger et~al.}{2019}]{metzger_fast_2019}
Metzger B.~D.,  Margalit B.,   Sironi L.,  2019, \mn@doi [Monthly Notices of the Royal Astronomical Society] {10.1093/mnras/stz700}, 485, 4091

\bibitem[\protect\citeauthoryear{{Mushtukov}, {Nagirner}  \& {Poutanen}}{{Mushtukov} et~al.}{2016}]{2016PhRvD..93j5003M}
{Mushtukov} A.~A.,  {Nagirner} D.~I.,   {Poutanen} J.,  2016, \mn@doi [\prd] {10.1103/PhysRevD.93.105003}, \href {https://ui.adsabs.harvard.edu/abs/2016PhRvD..93j5003M} {93, 105003}

\bibitem[\protect\citeauthoryear{{Nelson} \& {Wasserman}}{{Nelson} \& {Wasserman}}{1991}]{1991ApJ...371..265N}
{Nelson} R.~W.,  {Wasserman} I.,  1991, \mn@doi [\apj] {10.1086/169889}, \href {https://ui.adsabs.harvard.edu/abs/1991ApJ...371..265N} {371, 265}

\bibitem[\protect\citeauthoryear{Niu et~al.,}{Niu et~al.}{2022}]{niu_repeating_2022}
Niu C.-H.,  et~al., 2022, \mn@doi [Nature] {10.1038/s41586-022-04755-5}, 606, 873

\bibitem[\protect\citeauthoryear{{Obenberger} et~al.,}{{Obenberger} et~al.}{2014}]{2014ApJ...785...27O}
{Obenberger} K.~S.,  et~al., 2014, \mn@doi [\apj] {10.1088/0004-637X/785/1/27}, \href {https://ui.adsabs.harvard.edu/abs/2014ApJ...785...27O} {785, 27}

\bibitem[\protect\citeauthoryear{{Olausen} \& {Kaspi}}{{Olausen} \& {Kaspi}}{2014}]{2014ApJS..212....6O}
{Olausen} S.~A.,  {Kaspi} V.~M.,  2014, \mn@doi [\apjs] {10.1088/0067-0049/212/1/6}, \href {https://ui.adsabs.harvard.edu/abs/2014ApJS..212....6O} {212, 6}

\bibitem[\protect\citeauthoryear{{Paczynski}}{{Paczynski}}{1992}]{1992AcA....42..145P}
{Paczynski} B.,  1992, \actaa, \href {https://ui.adsabs.harvard.edu/abs/1992AcA....42..145P} {42, 145}

\bibitem[\protect\citeauthoryear{Parfrey, Beloborodov  \& Hui}{Parfrey et~al.}{2013}]{parfrey_dynamics_2013}
Parfrey K.,  Beloborodov A.~M.,   Hui L.,  2013, \mn@doi [The Astrophysical Journal] {10.1088/0004-637X/774/2/92}, 774, 92

\bibitem[\protect\citeauthoryear{Pastor-Marazuela et~al.,}{Pastor-Marazuela et~al.}{2021}]{pastor-marazuela_chromatic_2021}
Pastor-Marazuela I.,  et~al., 2021, \mn@doi [Nature] {10.1038/s41586-021-03724-8}, 596, 505

\bibitem[\protect\citeauthoryear{{Pelisoli} et~al.,}{{Pelisoli} et~al.}{2024}]{2024arXiv240211015P}
{Pelisoli} I.,  et~al., 2024, arXiv e-prints, \href {https://ui.adsabs.harvard.edu/abs/2024arXiv240211015P} {p. arXiv:2402.11015}

\bibitem[\protect\citeauthoryear{{Perna} \& {Pons}}{{Perna} \& {Pons}}{2011}]{2011ApJ...727L..51P}
{Perna} R.,  {Pons} J.~A.,  2011, \mn@doi [\apjl] {10.1088/2041-8205/727/2/L51}, \href {https://ui.adsabs.harvard.edu/abs/2011ApJ...727L..51P} {727, L51}

\bibitem[\protect\citeauthoryear{{P{\'e}tri}, {Heyvaerts}  \& {Bonazzola}}{{P{\'e}tri} et~al.}{2002}]{petri_global_2002}
{P{\'e}tri} J.,  {Heyvaerts} J.,   {Bonazzola} S.,  2002, \mn@doi [\aap] {10.1051/0004-6361:20020044}, \href {https://ui.adsabs.harvard.edu/abs/2002A&A...384..414P} {384, 414}

\bibitem[\protect\citeauthoryear{Petroff, Hessels  \& Lorimer}{Petroff et~al.}{2022}]{petroff_fast_2022}
Petroff E.,  Hessels J. W.~T.,   Lorimer D.~R.,  2022, \mn@doi [The Astronomy and Astrophysics Review] {10.1007/s00159-022-00139-w}, 30, 2

\bibitem[\protect\citeauthoryear{Philippov, Timokhin  \& Spitkovsky}{Philippov et~al.}{2020}]{philippov_origin_2020}
Philippov A.,  Timokhin A.,   Spitkovsky A.,  2020, \mn@doi [Physical Review Letters] {10.1103/PhysRevLett.124.245101}, 124, 245101

\bibitem[\protect\citeauthoryear{{Pizzolato}, {Colpi}, {De Luca}, {Mereghetti}  \& {Tiengo}}{{Pizzolato} et~al.}{2008}]{2008ApJ...681..530P}
{Pizzolato} F.,  {Colpi} M.,  {De Luca} A.,  {Mereghetti} S.,   {Tiengo} A.,  2008, \mn@doi [\apj] {10.1086/588084}, \href {https://ui.adsabs.harvard.edu/abs/2008ApJ...681..530P} {681, 530}

\bibitem[\protect\citeauthoryear{{Pons} \& {Perna}}{{Pons} \& {Perna}}{2011}]{2011ApJ...741..123P}
{Pons} J.~A.,  {Perna} R.,  2011, \mn@doi [\apj] {10.1088/0004-637X/741/2/123}, \href {https://ui.adsabs.harvard.edu/abs/2011ApJ...741..123P} {741, 123}

\bibitem[\protect\citeauthoryear{{Pons} \& {Vigan{\`o}}}{{Pons} \& {Vigan{\`o}}}{2019}]{2019LRCA....5....3P}
{Pons} J.~A.,  {Vigan{\`o}} D.,  2019, \mn@doi [Living Reviews in Computational Astrophysics] {10.1007/s41115-019-0006-7}, \href {https://ui.adsabs.harvard.edu/abs/2019LRCA....5....3P} {5, 3}

\bibitem[\protect\citeauthoryear{{Pons}, {Miralles}  \& {Geppert}}{{Pons} et~al.}{2009}]{Pons2009}
{Pons} J.~A.,  {Miralles} J.~A.,   {Geppert} U.,  2009, \mn@doi [\aap] {10.1051/0004-6361:200811229}, \href {https://ui.adsabs.harvard.edu/abs/2009A&A...496..207P} {496, 207}

\bibitem[\protect\citeauthoryear{Potekhin, Zyuzin, Yakovlev, Beznogov  \& Shibanov}{Potekhin et~al.}{2020}]{potekhin_thermal_2020}
Potekhin A.~Y.,  Zyuzin D.~A.,  Yakovlev D.~G.,  Beznogov M.~V.,   Shibanov Y.~A.,  2020, \mn@doi [Monthly Notices of the Royal Astronomical Society] {10.1093/mnras/staa1871}, 496, 5052

\bibitem[\protect\citeauthoryear{{Radhakrishnan} \& {Cooke}}{{Radhakrishnan} \& {Cooke}}{1969}]{radhakrishnan_magnetic_1969}
{Radhakrishnan} V.,  {Cooke} D.~J.,  1969, \aplett, \href {https://ui.adsabs.harvard.edu/abs/1969ApL.....3..225R} {3, 225}

\bibitem[\protect\citeauthoryear{Rajwade et~al.,}{Rajwade et~al.}{2020}]{rajwade_possible_2020}
Rajwade K.~M.,  et~al., 2020, \mn@doi [Monthly Notices of the Royal Astronomical Society] {10.1093/mnras/staa1237}, 495, 3551

\bibitem[\protect\citeauthoryear{Rea, Pons, Torres  \& Turolla}{Rea et~al.}{2012}]{rea_fundamental_2012}
Rea N.,  Pons J.~A.,  Torres D.~F.,   Turolla R.,  2012, \mn@doi [The Astrophysical Journal] {10.1088/2041-8205/748/1/L12}, 748, L12

\bibitem[\protect\citeauthoryear{Rea, Borghese, Esposito, Zelati, Bachetti, Israel  \& Luca}{Rea et~al.}{2016}]{rea_magnetar-like_2016}
Rea N.,  Borghese A.,  Esposito P.,  Zelati F.~C.,  Bachetti M.,  Israel G.~L.,   Luca A.~D.,  2016, \mn@doi [The Astrophysical Journal] {10.3847/2041-8205/828/1/L13}, 828, L13

\bibitem[\protect\citeauthoryear{Rea et~al.,}{Rea et~al.}{2022}]{rea_constraining_2022}
Rea N.,  et~al., 2022, \mn@doi [The Astrophysical Journal] {10.3847/1538-4357/ac97ea}, 940, 72

\bibitem[\protect\citeauthoryear{Rea et~al.,}{Rea et~al.}{2023}]{rea_long-period_2023}
Rea N.,  et~al., 2023, A long-period radio transient active for three decades: population study in the neutron star and white dwarf rotating dipole scenarios, \url {http://arxiv.org/abs/2307.10351}

\bibitem[\protect\citeauthoryear{{Rhodes}, {Caleb}, {Stappers}, {Andersson}, {Bezuidenhout}, {Driessen}  \& {Heywood}}{{Rhodes} et~al.}{2023}]{rhodes_frb_2023}
{Rhodes} L.,  {Caleb} M.,  {Stappers} B.~W.,  {Andersson} A.,  {Bezuidenhout} M.~C.,  {Driessen} L.~N.,   {Heywood} I.,  2023, \mn@doi [\mnras] {10.1093/mnras/stad2438}, \href {https://ui.adsabs.harvard.edu/abs/2023MNRAS.525.3626R} {525, 3626}

\bibitem[\protect\citeauthoryear{{Riley} et~al.,}{{Riley} et~al.}{2021}]{riley_nicer_2021}
{Riley} T.~E.,  et~al., 2021, \mn@doi [\apjl] {10.3847/2041-8213/ac0a81}, \href {https://ui.adsabs.harvard.edu/abs/2021ApJ...918L..27R} {918, L27}

\bibitem[\protect\citeauthoryear{Ronchi, Rea, Graber  \& Hurley-Walker}{Ronchi et~al.}{2022}]{ronchi_long-period_2022}
Ronchi M.,  Rea N.,  Graber V.,   Hurley-Walker N.,  2022, \mn@doi [The Astrophysical Journal] {10.3847/1538-4357/ac7cec}, 934, 184

\bibitem[\protect\citeauthoryear{{Ruderman}}{{Ruderman}}{1991a}]{1991ApJ...366..261R}
{Ruderman} M.,  1991a, \mn@doi [\apj] {10.1086/169558}, \href {https://ui.adsabs.harvard.edu/abs/1991ApJ...366..261R} {366, 261}

\bibitem[\protect\citeauthoryear{{Ruderman}}{{Ruderman}}{1991b}]{1991ApJ...382..576R}
{Ruderman} R.,  1991b, \mn@doi [\apj] {10.1086/170744}, \href {https://ui.adsabs.harvard.edu/abs/1991ApJ...382..576R} {382, 576}

\bibitem[\protect\citeauthoryear{{Ruderman}}{{Ruderman}}{1991c}]{1991ApJ...382..587R}
{Ruderman} M.,  1991c, \mn@doi [\apj] {10.1086/170745}, \href {https://ui.adsabs.harvard.edu/abs/1991ApJ...382..587R} {382, 587}

\bibitem[\protect\citeauthoryear{Ruderman \& Sutherland}{Ruderman \& Sutherland}{1975}]{ruderman_theory_1975}
Ruderman M.~A.,  Sutherland P.~G.,  1975, \mn@doi [The Astrophysical Journal] {10.1086/153393}, 196, 51

\bibitem[\protect\citeauthoryear{{Rugg}, {Mahlmann}  \& {Spitkovsky}}{{Rugg} et~al.}{2023}]{2023arXiv231204620R}
{Rugg} N.,  {Mahlmann} J.~F.,   {Spitkovsky} A.,  2023, \mn@doi [arXiv e-prints] {10.48550/arXiv.2312.04620}, \href {https://ui.adsabs.harvard.edu/abs/2023arXiv231204620R} {p. arXiv:2312.04620}

\bibitem[\protect\citeauthoryear{{Skiathas} \& {Gourgouliatos}}{{Skiathas} \& {Gourgouliatos}}{2024}]{skiathas_combined_2024}
{Skiathas} D.,  {Gourgouliatos} K.~N.,  2024, \mn@doi [arXiv e-prints] {10.48550/arXiv.2401.08979}, \href {https://ui.adsabs.harvard.edu/abs/2024arXiv240108979S} {p. arXiv:2401.08979}

\bibitem[\protect\citeauthoryear{Snelders et~al.,}{Snelders et~al.}{2023}]{snelders_microsecond-duration_2023}
Snelders M.~P.,  et~al., 2023, Microsecond-duration bursts from {FRB} {20121102A}, \url {http://arxiv.org/abs/2307.02303}

\bibitem[\protect\citeauthoryear{Spreeuw, Scheers, Braun, Wijers, Miller-Jones, Stappers  \& Fender}{Spreeuw et~al.}{2009}]{spreeuw_new_2009}
Spreeuw H.,  Scheers B.,  Braun R.,  Wijers R. A. M.~J.,  Miller-Jones J. C.~A.,  Stappers B.~W.,   Fender R.~P.,  2009, \mn@doi [Astronomy \& Astrophysics] {10.1051/0004-6361/200810449}, 502, 549

\bibitem[\protect\citeauthoryear{{Stoneham}}{{Stoneham}}{1979}]{1979JPhA...12.2187S}
{Stoneham} R.~J.,  1979, \mn@doi [Journal of Physics A Mathematical General] {10.1088/0305-4470/12/11/028}, \href {https://ui.adsabs.harvard.edu/abs/1979JPhA...12.2187S} {12, 2187}

\bibitem[\protect\citeauthoryear{{Strohmayer}, {Ogata}, {Iyetomi}, {Ichimaru}  \& {van Horn}}{{Strohmayer} et~al.}{1991}]{1991ApJ...375..679S}
{Strohmayer} T.,  {Ogata} S.,  {Iyetomi} H.,  {Ichimaru} S.,   {van Horn} H.~M.,  1991, \mn@doi [\apj] {10.1086/170231}, \href {https://ui.adsabs.harvard.edu/abs/1991ApJ...375..679S} {375, 679}

\bibitem[\protect\citeauthoryear{{Sturner}, {Dermer}  \& {Michel}}{{Sturner} et~al.}{1995}]{1995ApJ...445..736S}
{Sturner} S.~J.,  {Dermer} C.~D.,   {Michel} F.~C.,  1995, \mn@doi [\apj] {10.1086/175736}, \href {https://ui.adsabs.harvard.edu/abs/1995ApJ...445..736S} {445, 736}

\bibitem[\protect\citeauthoryear{Sturrock}{Sturrock}{1971}]{sturrock_model_1971}
Sturrock P.~A.,  1971, \mn@doi [The Astrophysical Journal] {10.1086/150865}, 164, 529

\bibitem[\protect\citeauthoryear{{Surnis} et~al.,}{{Surnis} et~al.}{2023}]{surnis_discovery_2023}
{Surnis} M.~P.,  et~al., 2023, \mn@doi [\mnras] {10.1093/mnrasl/slad082}, \href {https://ui.adsabs.harvard.edu/abs/2023MNRAS.526L.143S} {526, L143}

\bibitem[\protect\citeauthoryear{{Suvorov} \& {Kokkotas}}{{Suvorov} \& {Kokkotas}}{2019}]{2019MNRAS.488.5887S}
{Suvorov} A.~G.,  {Kokkotas} K.~D.,  2019, \mn@doi [\mnras] {10.1093/mnras/stz2052}, \href {https://ui.adsabs.harvard.edu/abs/2019MNRAS.488.5887S} {488, 5887}

\bibitem[\protect\citeauthoryear{{Suvorov} \& {Melatos}}{{Suvorov} \& {Melatos}}{2023}]{Suvorov_evolutionary_2023}
{Suvorov} A.~G.,  {Melatos} A.,  2023, \mn@doi [\mnras] {10.1093/mnras/stad274}, \href {https://ui.adsabs.harvard.edu/abs/2023MNRAS.520.1590S} {520, 1590}

\bibitem[\protect\citeauthoryear{{Tademaru}}{{Tademaru}}{1973}]{1973ApJ...183..625T}
{Tademaru} E.,  1973, \mn@doi [\apj] {10.1086/152252}, \href {https://ui.adsabs.harvard.edu/abs/1973ApJ...183..625T} {183, 625}

\bibitem[\protect\citeauthoryear{{Tendulkar}, {Kaspi}, {Archibald}  \& {Scholz}}{{Tendulkar} et~al.}{2017}]{2017ApJ...841...11T}
{Tendulkar} S.~P.,  {Kaspi} V.~M.,  {Archibald} R.~F.,   {Scholz} P.,  2017, \mn@doi [\apj] {10.3847/1538-4357/aa6d0c}, \href {https://ui.adsabs.harvard.edu/abs/2017ApJ...841...11T} {841, 11}

\bibitem[\protect\citeauthoryear{{The CHIME/FRB Collaboration} et~al.,}{{The CHIME/FRB Collaboration} et~al.}{2020}]{collaboration_periodic_2020}
{The CHIME/FRB Collaboration} et~al., 2020, Periodic activity from a fast radio burst source, \mn@doi{10.11570/20.0002}, \url {https://www.canfar.net/citation/landing?doi=20.0002}

\bibitem[\protect\citeauthoryear{Thompson \& Duncan}{Thompson \& Duncan}{1995}]{thompson_soft_1995}
Thompson C.,  Duncan R.~C.,  1995, \mn@doi [Monthly Notices of the Royal Astronomical Society] {10.1093/mnras/275.2.255}, 275, 255

\bibitem[\protect\citeauthoryear{Thompson, Lyutikov  \& Kulkarni}{Thompson et~al.}{2002}]{thompson_electrodynamics_2002}
Thompson C.,  Lyutikov M.,   Kulkarni S.~R.,  2002, \mn@doi [The Astrophysical Journal] {10.1086/340586}, 574, 332

\bibitem[\protect\citeauthoryear{{Timokhin}}{{Timokhin}}{2010}]{timokhin_time_2010}
{Timokhin} A.~N.,  2010, \mn@doi [\mnras] {10.1111/j.1365-2966.2010.17286.x}, \href {https://ui.adsabs.harvard.edu/abs/2010MNRAS.408.2092T} {408, 2092}

\bibitem[\protect\citeauthoryear{Timokhin \& Arons}{Timokhin \& Arons}{2013}]{timokhin_current_2013}
Timokhin A.~N.,  Arons J.,  2013, \mn@doi [Monthly Notices of the Royal Astronomical Society] {10.1093/mnras/sts298}, 429, 20

\bibitem[\protect\citeauthoryear{{Timokhin} \& {Harding}}{{Timokhin} \& {Harding}}{2015}]{timokhin_polar_2015}
{Timokhin} A.~N.,  {Harding} A.~K.,  2015, \mn@doi [\apj] {10.1088/0004-637X/810/2/144}, \href {https://ui.adsabs.harvard.edu/abs/2015ApJ...810..144T} {810, 144}

\bibitem[\protect\citeauthoryear{{Timokhin} \& {Harding}}{{Timokhin} \& {Harding}}{2019}]{timokhin_multiplicity_2019}
{Timokhin} A.~N.,  {Harding} A.~K.,  2019, \mn@doi [\apj] {10.3847/1538-4357/aaf050}, \href {https://ui.adsabs.harvard.edu/abs/2019ApJ...871...12T} {871, 12}

\bibitem[\protect\citeauthoryear{{Timokhin}, {Eichler}  \& {Lyubarsky}}{{Timokhin} et~al.}{2008}]{2008ApJ...680.1398T}
{Timokhin} A.~N.,  {Eichler} D.,   {Lyubarsky} Y.,  2008, \mn@doi [\apj] {10.1086/587925}, \href {https://ui.adsabs.harvard.edu/abs/2008ApJ...680.1398T} {680, 1398}

\bibitem[\protect\citeauthoryear{{Totani} \& {Tsuzuki}}{{Totani} \& {Tsuzuki}}{2023}]{totani_2023}
{Totani} T.,  {Tsuzuki} Y.,  2023, \mn@doi [\mnras] {10.1093/mnras/stad2532}, \href {https://ui.adsabs.harvard.edu/abs/2023MNRAS.526.2795T} {526, 2795}

\bibitem[\protect\citeauthoryear{{Tsuzuki}, {Totani}, {Hu}  \& {Enoto}}{{Tsuzuki} et~al.}{2024}]{2024MNRAS.530.1885T}
{Tsuzuki} Y.,  {Totani} T.,  {Hu} C.-P.,   {Enoto} T.,  2024, \mn@doi [\mnras] {10.1093/mnras/stae965}, \href {https://ui.adsabs.harvard.edu/abs/2024MNRAS.530.1885T} {530, 1885}

\bibitem[\protect\citeauthoryear{{Urpin} \& {Yakovlev}}{{Urpin} \& {Yakovlev}}{1980}]{1980SvA....24..425U}
{Urpin} V.~A.,  {Yakovlev} D.~G.,  1980, \sovast, \href {https://ui.adsabs.harvard.edu/abs/1980SvA....24..425U} {24, 425}

\bibitem[\protect\citeauthoryear{{Urpin}, {Levshakov}  \& {Iakovlev}}{{Urpin} et~al.}{1986}]{1986MNRAS.219..703U}
{Urpin} V.~A.,  {Levshakov} S.~A.,   {Iakovlev} D.~G.,  1986, \mn@doi [\mnras] {10.1093/mnras/219.3.703}, \href {https://ui.adsabs.harvard.edu/abs/1986MNRAS.219..703U} {219, 703}

\bibitem[\protect\citeauthoryear{{Usov} \& {Shabad}}{{Usov} \& {Shabad}}{1983}]{1983SvAL....9..212U}
{Usov} V.~V.,  {Shabad} A.~E.,  1983, Soviet Astronomy Letters, \href {https://ui.adsabs.harvard.edu/abs/1983SvAL....9..212U} {9, 212}

\bibitem[\protect\citeauthoryear{{Vigan{\`o}}, {Garcia-Garcia}, {Pons}, {Dehman}  \& {Graber}}{{Vigan{\`o}} et~al.}{2021}]{vigano_magneto_2021}
{Vigan{\`o}} D.,  {Garcia-Garcia} A.,  {Pons} J.~A.,  {Dehman} C.,   {Graber} V.,  2021, \mn@doi [Computer Physics Communications] {10.1016/j.cpc.2021.108001}, \href {https://ui.adsabs.harvard.edu/abs/2021CoPhC.26508001V} {265, 108001}

\bibitem[\protect\citeauthoryear{Viganò, Rea, Pons, Perna, Aguilera  \& Miralles}{Viganò et~al.}{2013}]{vigano_unifying_2013}
Viganò D.,  Rea N.,  Pons J.~A.,  Perna R.,  Aguilera D.~N.,   Miralles J.~A.,  2013, \mn@doi [Monthly Notices of the Royal Astronomical Society] {10.1093/mnras/stt1008}, 434, 123

\bibitem[\protect\citeauthoryear{Viganò et~al.,}{Viganò et~al.}{2019}]{vigano_simflowny-based_2019}
Viganò D.,  et~al., 2019, \mn@doi [Computer Physics Communications] {10.1016/j.cpc.2018.11.022}, 237, 168

\bibitem[\protect\citeauthoryear{{Voisin}, {Bonazzola}  \& {Mottez}}{{Voisin} et~al.}{2017}]{2017PhRvD..95j5008V}
{Voisin} G.,  {Bonazzola} S.,   {Mottez} F.,  2017, \mn@doi [\prd] {10.1103/PhysRevD.95.105008}, \href {https://ui.adsabs.harvard.edu/abs/2017PhRvD..95j5008V} {95, 105008}

\bibitem[\protect\citeauthoryear{{Wadiasingh} \& {Chirenti}}{{Wadiasingh} \& {Chirenti}}{2020}]{wadiasingh_fast_2020b}
{Wadiasingh} Z.,  {Chirenti} C.,  2020, \mn@doi [\apjl] {10.3847/2041-8213/abc562}, \href {https://ui.adsabs.harvard.edu/abs/2020ApJ...903L..38W} {903, L38}

\bibitem[\protect\citeauthoryear{Wadiasingh \& Timokhin}{Wadiasingh \& Timokhin}{2019}]{wadiasingh_repeating_2019}
Wadiasingh Z.,  Timokhin A.,  2019, \mn@doi [The Astrophysical Journal] {10.3847/1538-4357/ab2240}, 879, 4

\bibitem[\protect\citeauthoryear{Wadiasingh, Baring, Gonthier  \& Harding}{Wadiasingh et~al.}{2018}]{wadiasingh_resonant_2018}
Wadiasingh Z.,  Baring M.~G.,  Gonthier P.~L.,   Harding A.~K.,  2018, \mn@doi [The Astrophysical Journal] {10.3847/1538-4357/aaa460}, 854, 98

\bibitem[\protect\citeauthoryear{{Wadiasingh} et~al.,}{{Wadiasingh} et~al.}{2019}]{2019BAAS...51c.292W}
{Wadiasingh} Z.,  et~al., 2019, \mn@doi [\baas] {10.48550/arXiv.1903.05648}, \href {https://ui.adsabs.harvard.edu/abs/2019BAAS...51c.292W} {51, 292}

\bibitem[\protect\citeauthoryear{{Wadiasingh}, {Beniamini}, {Timokhin}, {Baring}, {van der Horst}, {Harding}  \& {Kazanas}}{{Wadiasingh} et~al.}{2020}]{wadiasingh_fast_2020}
{Wadiasingh} Z.,  {Beniamini} P.,  {Timokhin} A.,  {Baring} M.~G.,  {van der Horst} A.~J.,  {Harding} A.~K.,   {Kazanas} D.,  2020, \mn@doi [\apj] {10.3847/1538-4357/ab6d69}, \href {https://ui.adsabs.harvard.edu/abs/2020ApJ...891...82W} {891, 82}

\bibitem[\protect\citeauthoryear{{Wang}, {Lai}  \& {Han}}{{Wang} et~al.}{2010}]{wang_polarization_2010}
{Wang} C.,  {Lai} D.,   {Han} J.,  2010, \mn@doi [\mnras] {10.1111/j.1365-2966.2009.16074.x}, \href {https://ui.adsabs.harvard.edu/abs/2010MNRAS.403..569W} {403, 569}

\bibitem[\protect\citeauthoryear{Wang, Zhang, Chen  \& Xu}{Wang et~al.}{2019}]{wang_coherent_2019}
Wang W.,  Zhang B.,  Chen X.,   Xu R.,  2019, \mn@doi [The Astrophysical Journal] {10.3847/1538-4357/ab0e71}, 875, 84

\bibitem[\protect\citeauthoryear{{Wang} et~al.,}{{Wang} et~al.}{2021}]{2021ApJ...920...45W}
{Wang} Z.,  et~al., 2021, \mn@doi [\apj] {10.3847/1538-4357/ac2360}, \href {https://ui.adsabs.harvard.edu/abs/2021ApJ...920...45W} {920, 45}

\bibitem[\protect\citeauthoryear{{Wang} et~al.,}{{Wang} et~al.}{2022}]{2022MNRAS.516.5972W}
{Wang} Z.,  et~al., 2022, \mn@doi [\mnras] {10.1093/mnras/stac2542}, \href {https://ui.adsabs.harvard.edu/abs/2022MNRAS.516.5972W} {516, 5972}

\bibitem[\protect\citeauthoryear{Wang et~al.,}{Wang et~al.}{2023}]{wang_atypical_2023}
Wang P.,  et~al., 2023, Atypical radio pulsations from magnetar {SGR} 1935+2154, \url {http://arxiv.org/abs/2308.08832}

\bibitem[\protect\citeauthoryear{{Wiebicke} \& {Geppert}}{{Wiebicke} \& {Geppert}}{1991}]{1991A&A...245..331W}
{Wiebicke} H.~J.,  {Geppert} U.,  1991, \aap, \href {https://ui.adsabs.harvard.edu/abs/1991A&A...245..331W} {245, 331}

\bibitem[\protect\citeauthoryear{{Wolfson}}{{Wolfson}}{1995}]{wolfson_shear_induced_1995}
{Wolfson} R.,  1995, \mn@doi [\apj] {10.1086/175571}, \href {https://ui.adsabs.harvard.edu/abs/1995ApJ...443..810W} {443, 810}

\bibitem[\protect\citeauthoryear{Younes et~al.,}{Younes et~al.}{2022}]{younes_pulse_2022}
Younes G.,  et~al., 2022, \mn@doi [The Astrophysical Journal Letters] {10.3847/2041-8213/ac4700}, 924, L27

\bibitem[\protect\citeauthoryear{{Zhang}}{{Zhang}}{2023}]{zhang_physics_2023}
{Zhang} B.,  2023, \mn@doi [Reviews of Modern Physics] {10.1103/RevModPhys.95.035005}, \href {https://ui.adsabs.harvard.edu/abs/2023RvMP...95c5005Z} {95, 035005}

\bibitem[\protect\citeauthoryear{{Zhang} \& {Gil}}{{Zhang} \& {Gil}}{2005}]{zhang_gcrt_2005}
{Zhang} B.,  {Gil} J.,  2005, \mn@doi [\apjl] {10.1086/497428}, \href {https://ui.adsabs.harvard.edu/abs/2005ApJ...631L.143Z} {631, L143}

\bibitem[\protect\citeauthoryear{{Zhang}, {Harding}  \& {Muslimov}}{{Zhang} et~al.}{2000}]{2000ApJ...531L.135Z}
{Zhang} B.,  {Harding} A.~K.,   {Muslimov} A.~G.,  2000, \mn@doi [\apjl] {10.1086/312542}, \href {https://ui.adsabs.harvard.edu/abs/2000ApJ...531L.135Z} {531, L135}

\bibitem[\protect\citeauthoryear{{de Ruiter}, {Meyers}, {Rowlinson}, {Shimwell}, {Ruhe}  \& {Wijers}}{{de Ruiter} et~al.}{2023}]{2023arXiv231107394D}
{de Ruiter} I.,  {Meyers} Z.~S.,  {Rowlinson} A.,  {Shimwell} T.~W.,  {Ruhe} D.,   {Wijers} R. A.~M.~J.,  2023, \mn@doi [arXiv e-prints] {10.48550/arXiv.2311.07394}, \href {https://ui.adsabs.harvard.edu/abs/2023arXiv231107394D} {p. arXiv:2311.07394}

\bibitem[\protect\citeauthoryear{{den Hartog}, {Kuiper}, {Hermsen}, {Kaspi}, {Dib}, {Kn{\"o}dlseder}  \& {Gavriil}}{{den Hartog} et~al.}{2008a}]{2008A&A...489..245D}
{den Hartog} P.~R.,  {Kuiper} L.,  {Hermsen} W.,  {Kaspi} V.~M.,  {Dib} R.,  {Kn{\"o}dlseder} J.,   {Gavriil} F.~P.,  2008a, \mn@doi [\aap] {10.1051/0004-6361:200809390}, \href {https://ui.adsabs.harvard.edu/abs/2008A&A...489..245D} {489, 245}

\bibitem[\protect\citeauthoryear{{den Hartog}, {Kuiper}  \& {Hermsen}}{{den Hartog} et~al.}{2008b}]{2008A&A...489..263D}
{den Hartog} P.~R.,  {Kuiper} L.,   {Hermsen} W.,  2008b, \mn@doi [\aap] {10.1051/0004-6361:200809772}, \href {https://ui.adsabs.harvard.edu/abs/2008A&A...489..263D} {489, 263}

\makeatother
\end{thebibliography}

%\begin{thebibliography}{99}
%\end{thebibliography}

%%%%%%%%%%%%%%%%%%%%%%%%%%%%%%%%%%%%%%%%%%%%%%%%%%

%%%%%%%%%%%%%%%%% APPENDICES %%%%%%%%%%%%%%%%%%%%%

\appendix

\section{Quake induced twists}
\label{sect:quake_induced}
Magnetospheric twists ($\Psi_{\rm crit} < \Psi < 1$) may have been imparted due to past crustal failures, which may have been observable as FRBs and/or X-ray bursts. Decaying linear magnetospheric twists have been extensively studied for X-ray loud magnetars (e.g. \citealt{beloborodov_untwisting_2009}), such that the lifetime and luminosity of twist dissipation is theoretically well-understood. Following \cite{beloborodov_untwisting_2009}, we define the magnetosphere in terms of the polodial flux function for a dipole:$f(r, \theta) = \frac{2 \pi \mu \sin^2(\theta)}{r}$ where $f(r, \theta = 0) = 0$ refers to the polar axis and $\mu$ is the magnetic moment of the NS. Flux surfaces are labelled by defining $u \equiv \frac{f}{f_r} = \frac{R}{R_{\rm max}}$ where $R_{\rm max}$ is the maximum altitude reached by magnetic field lines passing through $(r, \theta)$. In this notation $0 < u < 1$ refers to the magnetosphere, $\infty >  u_{\rm c} > 1$ defines the magnetic field anchor point in the inner crust ($\approx 1.1$) and $u = u_{\rm r} = 1$ refers to the neutron star surface.
\par
\cite{beloborodov_untwisting_2009} found that the untwisting magnetosphere comprises of two distinct regions: a `cavity' and a `j-bundle', later supported by axisymmetric PIC simulations of twisted magnetospheres \citep{chen_particle--cell_2017}. The field lines untwist as the cavity slowly expands to larger altitudes `sucking in' electric currents. The process is highly non-linear, depending strongly on the unknown voltage profile across the twist $\Phi(f)$. Dissipation is confined to current-carrying field lines in the magnetosphere ($u < u_{r}$), and specifically concentrated on the cavity/j-bundle boundary, which is labeled as $u_*$. The free energy of the twist can be rewritten as \citep{beloborodov_untwisting_2009}:
\begin{equation}
    E_{\rm tw} = 4 \times 10^{46} B_{15}^2 R_6^3 \Psi^2 u_{*}^3 \: \: {\rm erg s^{-1}}
    \label{eq:energy_u}
\end{equation}
As the twist is greater than $\Psi_{\rm crit}$, we can assume a roughly constant voltage throughout (in time and in flux surface) as implied by the threshold voltage of pair production. Dissipation proceeds on the Ohmic timescale and the luminosity is given by: 
\begin{equation}
\begin{split}
       L_{\rm diss, quake} &= \int_0^{I_{\rm r}} \Phi d I = \frac{c \mu \Psi \Phi u_{*}^2}{4 R^2}  \\
       &= 1.3 \times 10^{37} \; B_{\rm s, 15} \, R_{*, 6} \, \Psi \, \Phi_{10} \, u_{*}^2 \: \: {\rm erg s^{-1}} 
\end{split}
    \label{eq:diss_u}
\end{equation}
Therefore the timescale of the untwist can be determined as:
\begin{equation}
    \tau_{\rm tw} = \frac{E_{\rm tw}}{L_{\rm diss}} = 15 \; \Phi_{10}^{-1} \, B_{\rm s, 15} \, R_{*, 6} \, \Psi \, u_{*} \: \: {\rm years}
    \label{eq:timescale_untwist}
\end{equation}
We note that for a constant target soft photon field the pair discharge voltage $\Phi$ rises commensurately with the magnetic field strength $B$, such that this timescale holds for all magnetic field strengths. This resistive timescale (shown in Fig. \ref{fig:x_rad_lifetime}) is a crude approximation, and depends strongly on how/where the twist is anchored.
\par
The pair production mechanism that limits the untwisting magnetic field produces radio emission (Section \ref{sect:radio}). Assuming a radio efficiency factor $\eta_{\rm r} \ll 1$ from the dissipated energy, such that:
\begin{equation}
    L_{\rm r} = 1.3 \times 10^{31} \; \eta_{\rm r, -4} \, B_{\rm s, 15} \, R_{*, 6} \, u_{*}^2 \, \Psi_{-2} \, \Phi_{\rm pair, 10}  \: \: {\rm erg s^{-1}}
    \label{eq:radio_emission_untwisting}
\end{equation}

The radio luminosity as a function of $u_{*}$ and $\Psi$ is presented in Fig. \ref{fig:x_rad_lifetime}, where we can see that the observed radio luminosities can be easily obtained by even mild twists as expected.

\subsection{X-ray Emission}
\label{sect:x-rays_quake}
As in \S\ref{sect:x-rays}, we may estimate X-ray emission based on the return current. The exact scaling of high-energy luminosity to the dissipation power
$L_{\rm X, therm} \propto \dot{E}$ (e.g., empiricial relations of \citealt{chang_observational_2023}; see \citealt{kalapotharakos_fundamental_2019,kalapotharakos_fundamental_2022} for non-thermal case) is a complicated process that depends on the particle distribution function of backflowing charges on the atmosphere and their relative penetration depths. 

\begin{figure}
  \centering
{\includegraphics[width=.5\textwidth]{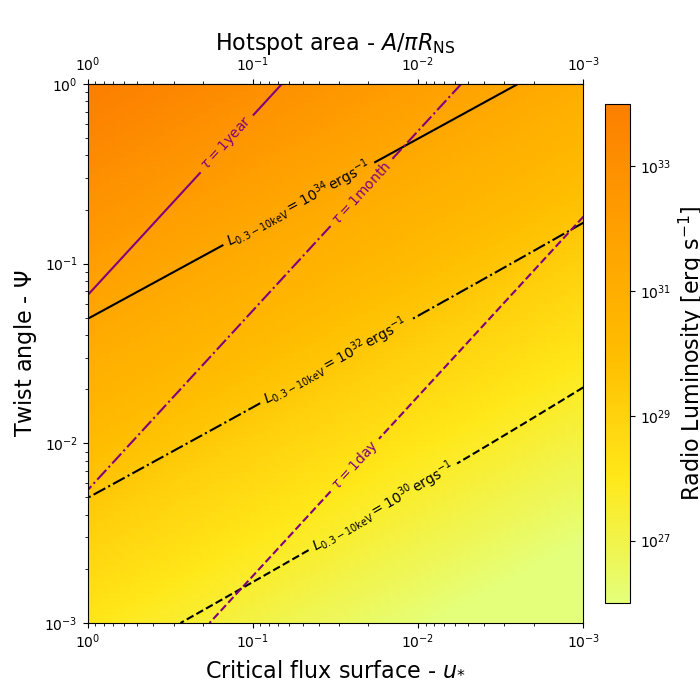}}
\caption{Potential radio luminosity of untwisting field lines in the \protect\cite{beloborodov_untwisting_2009} model (Section \ref{sect:quake_induced}) as a function of twist $\Psi$ and the flux surface of the cavity $u_*$, which scales linearly with the blackbody area $A$ (top axis). Overplotted are the 0.3-10 keV thermal X-ray luminosity associated with the return current, and the approximate expected life-time of the emission via Eq. \ref{eq:timescale_untwist}. }
\label{fig:x_rad_lifetime}
\end{figure}

As in \S\ref{sect:x-rays}, we can naively approximate an X-ray luminosity $L_{\rm X} \sim L_{\rm diss}$ as an upper limit case. In Fig. \ref{fig:x_rad_lifetime} we overplot the expected X-ray luminosity associated with thermal emission from the return current. 
\par

In the untwisting model of \cite{beloborodov_untwisting_2009}, the area of dislodged footprints $A_{\rm fp}$ upon which return currents are confined can be associated with the cavity/j-bundle transition region. 
\begin{equation}
  A_{\rm rc} = \pi R_{*}^2 u_{*}  
\end{equation}
Therefore we can uniquely specify the hotpot blackbody radiation associated with the return current in terms of $u_{*}$:
\begin{equation}
    T_{\rm rc}  = \bigg(\frac{\eta_{\rm X, therm} c \mu \Psi \Phi_{\rm pair} u_{*}}{4 \pi R_{*}^4} \bigg)^{1/4}
\end{equation}
Using this, we can examine the parameter space [$\Psi, u_{*}$] to explore values which are sufficient to produce observed radio luminosity for a given $\eta_{\rm r}$, and compute the expected X-ray component due to the return current. Similarly to Sect. \ref{sect:x-rays}, we integrate thermal radiation from a given hotspot defined by ($A_{\rm rc},T_{\rm rc}$) across the $0.3-10$ keV bandwidth for which X-ray constraints have been reported assuming $\eta_{\rm X} = 1$. Results are presented in Fig. \ref{fig:x_rad_lifetime}.

\subsection{Discussion}
\label{sect:discussion_quake}
\subsubsection{Model predictions}
The evolution of the j-bundle \citep{beloborodov_untwisting_2009} implies that radio emission should decay as a function of time (comparatively, variability in the plastic flow model depends directly on the variability of the plastic flow velocity), although no exact temporal scaling is firmly predicted. Such decay is observed in GLEAM-X J1627 across the two activity epochs, on timescales compatible with mild twists. No such decay is observed in GPM J1839-10, implying dissipation timescales corresponding to much larger twists. Another possibility is that these sources is are still undergoing sudden crustal events capable of imparting twist. This may explain the re-brightening of GLEAM-X J1627, and the erratic pulse profile of GLM J1839-10. In this explanation, one might expect to observe X-ray bursts associated with sudden crustal quake events, and the onset of persistent X-ray emission, none of which have been yet observed.

\subsubsection{Connection to FRBs}
In either interpretation of the ULPM pulsed radio emission considered in this work, radiation is intrinsically linked to crustal behaviour. However for the starquake model, dramatic starquakes are required in the recent past to impart significant twist. Favoured theories of magnetospheric FRBs also rely on crustal slippages to power particle acceleration resulting in FRBs (e.g. \citealt{kumar_fast_2017}). \cite{kumar_frb_2020} suggest that such events may produce Alfvén waves of sufficient amplitude to produce FRBs as they propagate away from the NS surface, to regions in which they are become charge-starved. If repeated crustal displacements which produce FRBs impart twist in the magnetosphere faster than it can be dissipated, then radio-loud ULPMs could represent either current or former FRB sources. The FRB emission is significantly beamed, such that the lack of observations of FRBs from Galactic ULPMs does not necessarily discount them as FRB progenitors. The quasi-stable radio emission predicted in this work may begin during FRB activity (or after it ceases) as the field lines begin to untwist, with luminosities and decay timescales discussed in Section. \ref{sect:quake_induced}. Persistent radio sources have been associated with two FRBs \citep{chatterjee_direct_2017,niu_repeating_2022}, one of which appear to decaying \citep{rhodes_frb_2023}. However the observed luminosities (approximately $10^{38} \, {\rm erg \, s^{-1}}$) are much larger then the luminosities predicted in Eq. \ref{eq:radio_emission_untwisting}, or by even the highest velocity plastic motion. It is more likely that persistent counterparts to FRBs are instead correlated with the magnetar's immediate environment \citep{margalit_concordance_2018}. 

\section{Additional resonant inverse-Compton scattering length scale schematics}
Here we include additional RICS length scale plots, corresponding to different regions in the magnetosphere.

\begin{figure}
  \centering
{\includegraphics[width=.5\textwidth]{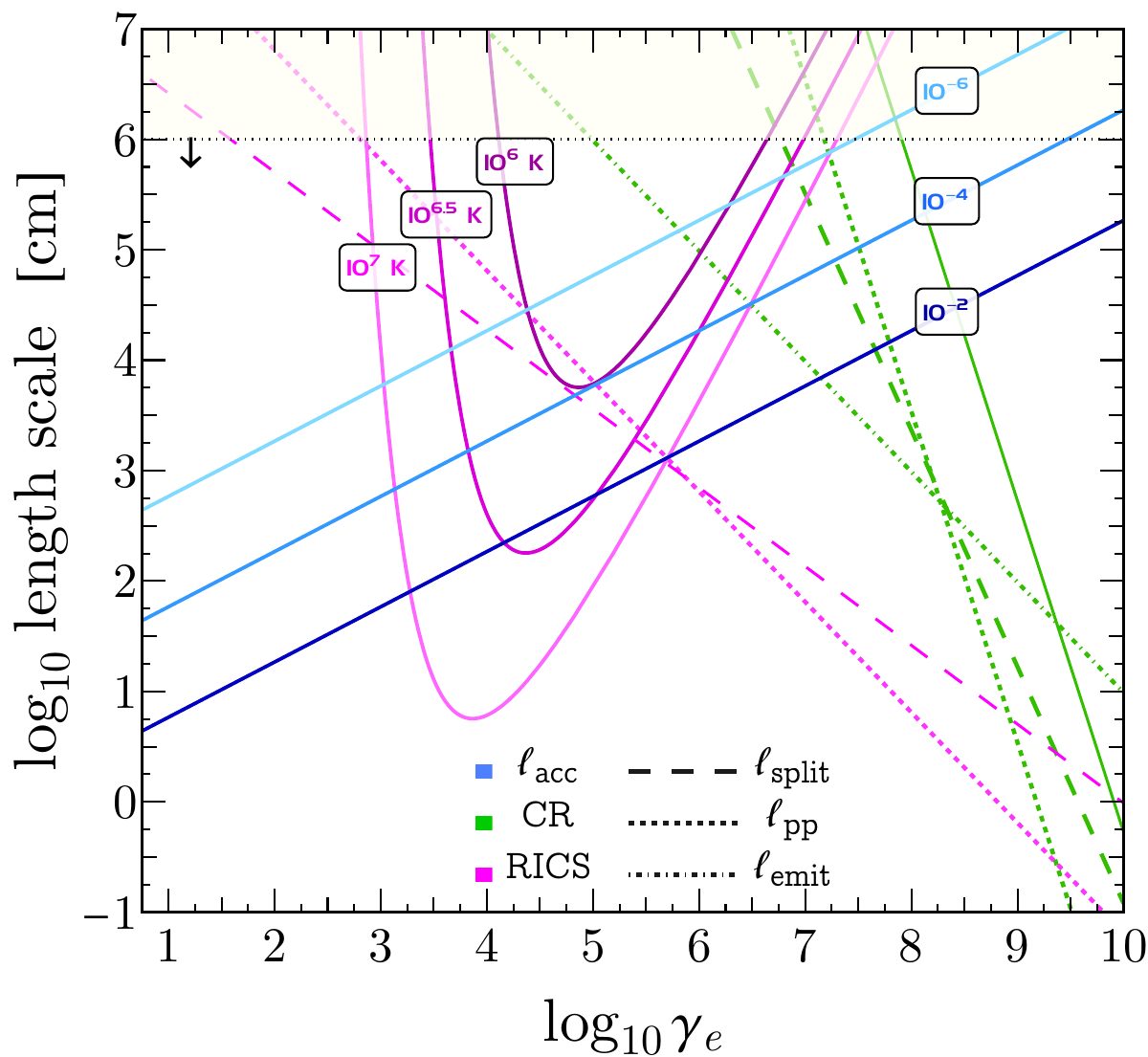}}
\caption{Similar to Fig. \ref{fig:lengthscales_fud}, with the curvature radius increased to $\rho_c = 10^{8.5}$ cm with $\theta_0 = 10^{-2}$ corresponding to situation with locally high curvature radius near a pole with $B=10^{15}$ G. }
\label{fig:lengthscales_higherrhocc}
\end{figure}

\begin{figure}
  \centering
{\includegraphics[width=.5\textwidth]{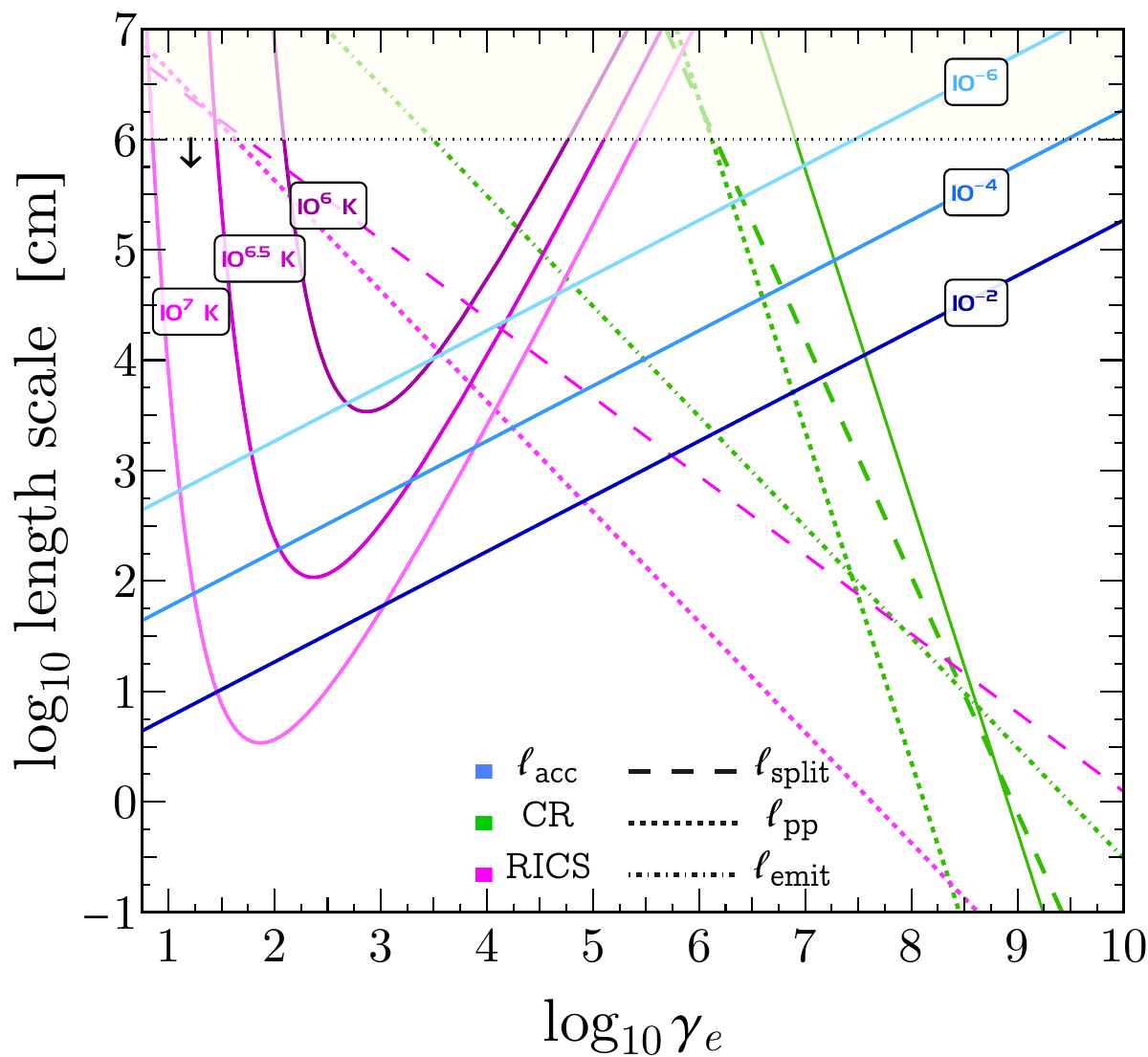}}
\caption{Similar to Fig. \ref{fig:lengthscales_fud}, considering a lower field $B=10^{13}$ G, with above-threshold pair creation ($\chi_0=0.3$) with $\theta_0 = 0.1$ and $\rho_c =10^7$ cm. }
\label{fig:lengthscales_lowB}
\end{figure}

\begin{figure}
  \centering
{\includegraphics[width=.5\textwidth]{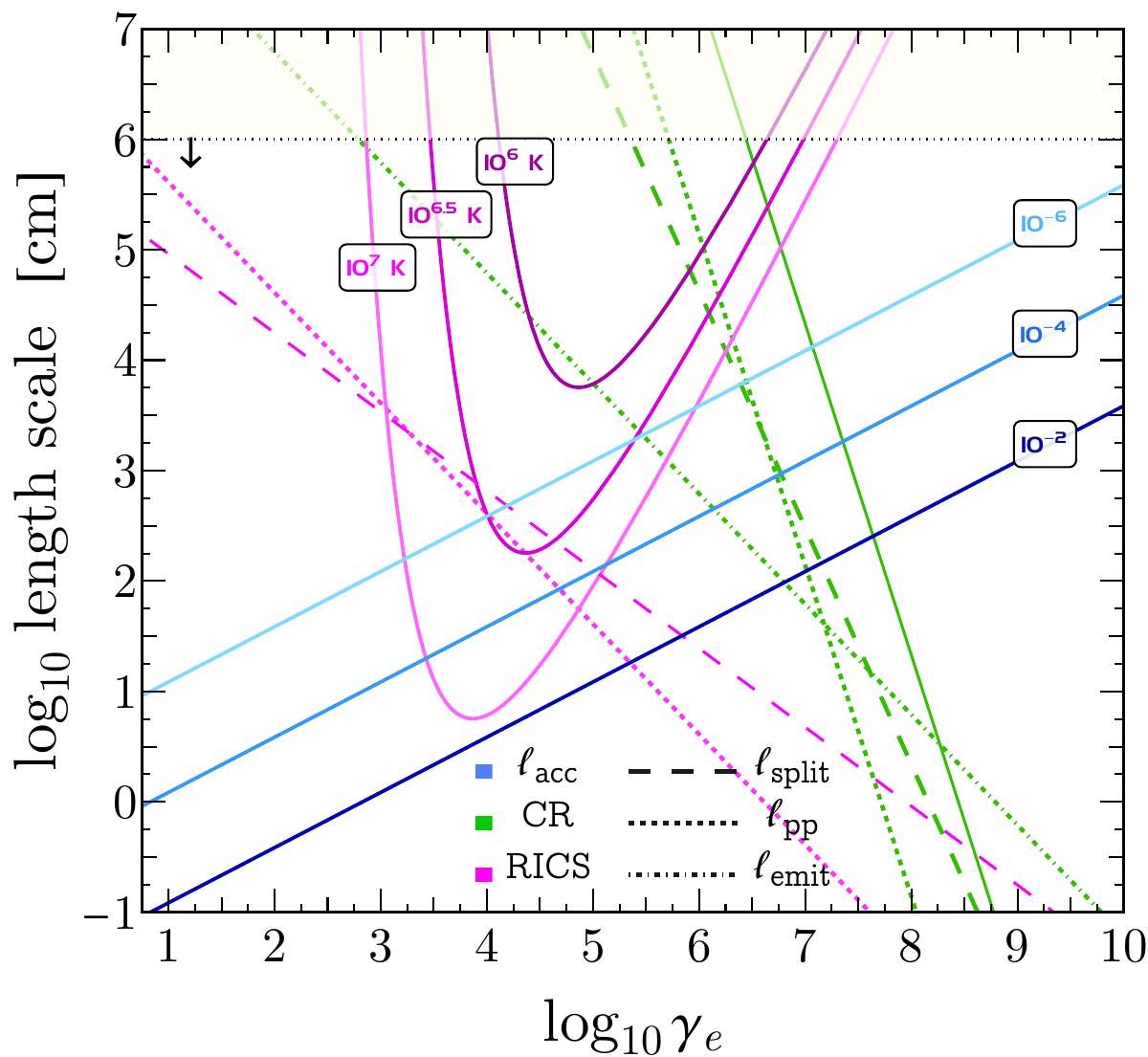}}
\caption{Similar to Fig. \ref{fig:lengthscales_fud}, with smaller curvature radius $\rho_c = 2\times10^6$ cm with $\theta_0 = 0.5$ and $B=10^{15}$ G, relevant for possible multi-polar components.}
\label{fig:lengthscales_small_rho_c}
\end{figure}

%%%%%%%%%%%%%%%%%%%%%%%%%%%%%%%%%%%%%%%%%%%%%%%%%%

% Don't change these lines
\bsp	% typesetting comment
\label{lastpage}
\end{document}